\documentclass[% 
reprint,
superscriptaddress,
nofootinbib,
 amsmath,amssymb,
 aps,
 prd,
longbibliography,
]{revtex4-2}

\usepackage{bm, graphics, graphicx, epsfig, soul, latexsym ,multirow}
\usepackage{comment}
\usepackage{subfigure}
\usepackage{graphicx,color}
\usepackage{dcolumn}
\usepackage{bm,amsmath, amssymb,}
\usepackage{mathrsfs}
\usepackage{bigints}
\usepackage{float}
\usepackage{multirow}
\usepackage{footmisc}
\usepackage{placeins}
\usepackage{subfigure}
\usepackage[normalem]{ulem}
\usepackage{booktabs}
\usepackage{multirow}
\usepackage{array}
\usepackage[colorlinks,urlcolor=blue,citecolor=blue,linkcolor=blue]{hyperref}

\usepackage{amsmath}

\newcommand{\be}{\begin{equation}}
\newcommand{\ee}{\end{equation}}
\begin{document}

%%%%%%%%%%%%%%%%%%%%%%%%%%%%%%%%%%%%%%%%%%%%%%%

\title{Eccentricity-induced systematic error on parametrized tests of general relativity: Hierarchical Bayesian inference applied to a binary black hole population}
\author{Pankaj Saini}
\email{pankajsaini@cmi.ac.in}
\affiliation{Chennai Mathematical Institute, Siruseri 603103, India}
\author{Sajad A. Bhat}
\email{sajad.bhat@iucaa.in}
\affiliation{Chennai Mathematical Institute, Siruseri 603103, India}
\affiliation{Inter-University Centre for Astronomy and Astrophysics, Post Bag 4, Ganeshkhind, Pune - 411007, India}
\author{Marc Favata}
\email{marc.favata@montclair.edu}
\affiliation{Department of Physics \& Astronomy, Montclair State University,
1 Normal Avenue, Montclair, New Jersey 07043, USA}
\author{K.~G.~Arun}
\email{kgarun@cmi.ac.in}
\affiliation{Chennai Mathematical Institute, Siruseri 603103, India}
\affiliation{Institute for Gravitation and the Cosmos, Department of Physics, \\Penn State University, University Park, Pennsylvania 16802, USA}
\date{\today}
 %%%%%%%%%%%%%%%%%%%%%%%%%%%%%%%%%%%%%%%%%%%%%%%%%%%%%%%%%%%%%%%%%%%%%%%%%
\begin{abstract}
One approach to testing general relativity (GR) introduces free parameters in the post-Newtonian (PN) expansion of the gravitational-wave (GW) phase. If systematic errors on these testing GR (TGR) parameters exceed the statistical errors, this may signal a false violation of GR. Here, we consider systematic errors produced by unmodeled binary eccentricity. Since the eccentricity of GW events in ground-based detectors is expected to be small or negligible, the use of quasicircular waveform models for testing GR may be safe when analyzing a small number of events. However, as the catalog size of GW detections increases, more stringent bounds on GR deviations can be placed by combining information from multiple events. In that case, even small systematic biases may become significant. We apply the approach of hierarchical Bayesian inference to model the posterior probability distributions of the TGR parameters inferred from a population of eccentric binary black holes (BBHs). We assume each TGR parameter value varies across the BBH population according to a Gaussian distribution. The means and standard deviations that parametrize these Gaussians are related to the statistical and systematic errors measured for each BBH event. We compute the posterior distributions for these Gaussian \emph{hyperparameters}, characterizing the BBH population via three possible eccentricity distributions. This is done for LIGO and Cosmic Explorer (CE, a proposed third-generation detector). We find that systematic biases from unmodeled eccentricity can signal false GR violations for both detectors when considering constraints set by a catalog of events. We also compute the projected bounds on the $10$ TGR parameters when eccentricity is included as a parameter in the waveform model. This is done via multiplying the individual likelihoods for each event in the catalog, and also by combining them via hierarchical inference. We find that the first four dimensionless TGR deformation parameters can be bounded at $90\%$ confidence to $\delta \hat{\varphi}_i \lesssim 10^{-2}$ for LIGO and $\lesssim 10^{-3}$ for CE [where $i=(0,1,2,3)$; GR predicts zero for all values of the $\delta \hat{\varphi}_i$]. The most stringent bound applies to the $-1$PN (dipole) parameter: it is constrained to $|\delta \hat{\varphi}_{-2}|\lesssim 10^{-5}$ ($ \lesssim 4 \times 10^{-7}$) by LIGO (CE). In comparison to the circular orbit case, the combined bounds on the TGR parameters worsen by a modest factor of $\lesssim 2$ when eccentricity is included in the waveform. (The dipole parameter bound degrades by a factor of $\sim 3 \mbox{--} 4$ in this case.) 

\end{abstract}

\maketitle

\section{Introduction}
General relativity (GR) stands out as the most successful theory for explaining the nature of gravity~\citep{Will:2005va, Sathyaprakash:2009xs, Yunes:2013dva, Berti:2015itd, Krishnendu:2021fga}. The detection of gravitational waves (GWs) by LIGO~\cite{LIGOScientific:2014pky} and Virgo~\cite{virgo:2014yos} enabled tests of GR in the strong-field regime~\cite{LIGOScientific:2016lio,LIGOScientific:2019fpa,LIGOScientific:2018dkp,LIGOScientific:2020tif,LIGOScientific:2021sio} associated with compact binary mergers. Around $90$ GW events have been detected through the third observing run~\cite{LIGOScientific:2021psn}; this number will continue to grow in the coming years as detector sensitivity improves and new detectors come online~\cite{KAGRA:2013rdx, KAGRA:2020tym,Saleem:2021iwi}. Combining information from a growing number of GW observations will enable us to perform more stringent tests of GR~\cite{DelPozzo:2011pg}.

Accurate waveforms in modified theories of gravity that cover the inspiral, merger, and ringdown phases of binary evolution are currently lacking (see, however, Refs.~\cite{Sennett:2016klh, Bernard:2018ivi, Bernard:2018hta, Tahura:2018zuq, Khalil:2018aaj,Okounkova:2019zjf,Okounkova:2020rqw, Cayuso:2020lca, East:2020hgw, East:2021bqk, Shiralilou:2021mfl}). Hence, current approaches of testing GR mostly rely on {\it null} tests. A null test introduces one or more parameters that are expected to capture some beyond-GR effects; their values, by definition, are zero in GR. These parameters are constrained using the GW data to deduce the region in the non-GR parameter space that is allowed by the data. Hence, it is crucial that the model we use to perform these null tests is accurate and free of systematic errors.

Owing to the complexity of the two-body problem, gravitational waveforms used for data analysis typically involve approximations and assumptions. These include truncation of the post-Newtonian (PN) series, or not accounting for physical effects like orbital eccentricity, spin, precession, and nonquadrupolar modes.
These approximations in the waveform model introduce systematic biases in the estimated parameters. When testing GR, these biases may manifest as a false violation of GR (see, for example, Refs~\cite{Saini:2022igm, Hu:2022bji,Bhat:2022amc, Narayan:2023vhm}). Hence, it is critical to understand the nature and magnitude of waveform systematics before claiming a violation of GR.  Here we address the systematic biases on parametrized tests of GR due to the neglect of eccentricity. This is applied to non-GR parameter constraints determined from observations of a binary black hole (BBH) population by both current and third-generation (3G) GW detectors. This work extends our previous analysis~\cite{Saini:2022igm} which considered these biases at the level of individual events.

\subsection{Parametrized tests of GR}
Parametrized tests of GR introduce one or more free parameters into a mathematical function or object that can be related to observational quantities. The free parameters are frequently expressed in terms of small deviations from their GR values. Examples include the parametrized post-Newtonian formalism~\cite{Arun:2004hn,Arun:2006hn}, the post-Einsteinian formalism~\cite{Yunes:2009ke,PPE:2011ys}, or the testing GR (TGR) formalism~\cite{TIGER:2013upa, GW150914TGR:2016lio}. This paper focuses on the latter.

In the TGR formalism the primary observable is the Fourier transform of the GW phase. This is analytically approximated as
\be
\tilde{h}(f) \propto f^{-{7/6}} e^{i\Psi(f)}, 
\ee
where the phase function is expanded in a PN series in the relative orbital speed $v= (\pi M f)^{1/3}$ for circular orbits. Here $M=m_1+m_2$ is the binary total mass, $m_{1,2}$ are the component masses, and $f$ is the observed GW frequency. To 3.5 PN order the phase expansion has the form~\cite{Blanchet:2002av}
\begin{align}
\label{eq:PsiSPA}
\Psi(f) &\sim \frac{3}{128 \eta v^5} \left[ 1 + \varphi^{\rm GR}_2 v^2 + \varphi^{\rm GR}_3 v^3 + \cdots \right] \\
& \sim \frac{3}{128 \eta v^5} \sum_{i=0}^{7} (\varphi_{i}^{\rm GR}v^{i} 
      +  \varphi_i^{\rm log, GR}  v^{i} \log v) \,.
\end{align}
The coefficients depend on the binary reduced mass ratio $\eta=m_1 m_2/M^2$, the dimensionless component spin vectors $\bm{\chi}_{1,2}$, and possibly other parameters characterizing the structure of the bodies. For example, $\varphi_0^{\rm GR}=1$, $\varphi_1^{\rm GR}=0$, $\varphi_2^{\rm GR}=\frac{3715}{756} +\frac{55}{9}\eta$, and $\varphi_3^{\rm GR}=-16\pi + 4\beta$, where $\beta$ is the spin-orbit term~\cite{Kidder:1992fr,Poisson:1993zr,Kidder:1995zr,Buonanno:2009zt}
\begin{multline}
\beta = \left[ \frac{113}{24} \left(1+\sqrt{1-4\eta}\right) -\frac{19}{6}\eta \right] {\bm \chi}_{1} \cdot \hat{\bm L}_{\rm N} \\
+ \left[ \frac{113}{24} \left(1-\sqrt{1-4\eta}\right) -\frac{19}{6}\eta \right] {\bm \chi}_{2} \cdot \hat{\bm L}_{\rm N} \;,
\end{multline}
and ${\bm \chi}_{1,2} \cdot \hat{\bm L}_{\rm N}$ are the components of the dimensionless spin vectors along the Newtonian orbital angular momentum direction $\hat{\bm L}_{\rm N}$. (For aligned spins these components are the same as the dimensionless spin parameters ${\bm \chi}_{1,2} \cdot \hat{\bm L}_{\rm N} = \chi_{1,2} \in [-1,1]$. In this work, we consider only the case where $\chi_{1,2} \in [0,1]$.)

In the TGR formalism, the PN phase coefficients $(\varphi_i^{\rm GR}, \varphi_i^{\rm log, GR})$ are modified by introducing fractional deviation parameters $(\delta \hat{\varphi}_i, \delta \hat{\varphi}_i^{\rm log})$ via
\begin{subequations}
\label{deformationA}
\begin{align}
\varphi_{i}^{\rm GR} &\xrightarrow{} \varphi_{i} = \varphi_{i}^{\rm GR} (1 + \delta \hat{\varphi}_{i}) \,, \\
\varphi_i^{\rm log, GR} &\xrightarrow{} \varphi_i^{\rm log} = \varphi_i^{\rm log, GR} (1 + \delta \hat{\varphi}_i^{\rm log}) \,.
\end{align}
\end{subequations}
The summation in Eq.~\eqref{eq:PsiSPA} is also extended to selected values $i<0$; in particular, we consider $i=-2$ corresponding to dipole gravitational radiation. (See Sec.~\ref{waveform model} for details of our waveform model.) 

The deviation parameters $\delta \hat{\varphi_i}$ are treated as new degrees of freedom that capture a particular class of deviations from GR.\footnote{For simplicity, we drop references to the parameters $\varphi_i^{\rm log, GR}$, $\varphi_i^{\rm log}$, or $\delta \hat{\varphi}_i^{\rm log, GR}$ associated with logarithmic-dependent terms in the phase expansion; they are implied throughout when mentioning the $\varphi_i^{\rm GR}$, $\varphi_i$, and $\delta \hat{\varphi}_i$ parameters.} In the limit $\delta\hat{\varphi}_i \rightarrow 0$, GR is recovered. Precise measurement of the $\delta\hat{\varphi}_i$ is crucial for these tests. If detected GW signals have $\delta\hat{\varphi}_i$ parameter measurements consistent with zero, the signals are consistent with GR. If the test shows clear deviations from zero (or the GR value), that is an indicator of a potential GR violation or the presence of systematic errors due to inaccuracies or missing physics in the waveform model.

Inaccurate waveform modeling will start affecting a TGR parameter $\delta \hat{\varphi}_i$ when the systematic errors are larger than the corresponding statistical uncertainties associated with that parameter~\cite{Vallisneri:2013rc}. Given the sensitivity of current detectors, the systematic bias may be hidden by the large statistical uncertainty due to detector noise. However, as the sensitivity of current detectors improves and more sensitive 3G detectors become operational, more events with a large signal-to-noise ratio (SNR) will be detected. As the statistical errors scale as $\text{SNR}^{-1}$ while systematic errors are independent of SNR, systematic biases are likely to be dominant for loud events. For those events, even a small shift in the “true” value of $\delta\hat{\varphi}_i$ may significantly bias the test of GR.

For an individual event, the systematic bias may be sufficiently small (relative to the statistical uncertainty) to avoid appearing as a false GR violation. In other words, the waveform model may be sufficiently accurate to analyze individual events but not for a catalog of events.

\subsection{Biases due to the neglect of eccentricity from individual events and a population}
{\it In this paper, we focus on the systematic bias in the TGR parameters from unmodeled orbital eccentricity in a population of eccentric BBHs.} The emission of GWs carries energy and angular momentum away from the binary. This energy and angular momentum loss causes the orbit to shrink and circularize on a (relatively) rapid timescale~\cite{PetersMathews:1963ux,Peters:1964zz}. For example, consider a black hole binary with masses $20 M_{\odot}$ and $15 M_{\odot}$ and an eccentricity of $0.9$. The eccentricity is specified at a reference GW frequency of $0.01$ Hz (twice the orbital frequency). The two black holes are initially separated by $\sim 12 d_{\oplus}$ ($d_{\oplus}=\text{Earth diameter}$) and moving with characteristic speed $\sim 0.01 c$, where $c$ is the speed of light. For this system it takes $\sim 60$ days for the eccentricity to reduce to $e\sim 0.1$ at $1$ Hz. The separation has reduced to $\sim 0.55 d_{\oplus}$ and the characteristic speed has increased to $0.07 c$. Further, it only takes an additional $\sim 2$ hours for the eccentricity to reduce to $e \sim 0.01$ at $10$ Hz. At this point, the binary's separation is $\sim 0.11 d_{\oplus}$ and the speed has reached up to $\sim 0.15 c$. Hence, for long-lived binaries it is expected that the orbit is highly circularized when observed in the frequency band of ground-based detectors. This motivates the use of quasicircular waveforms for the detection and parameter estimation of compact binaries with the current LIGO-Virgo-KAGRA detectors.

A binary's eccentricity depends on its formation history. Binaries formed through isolated formation channels in the Galactic field are expected to be in circular orbits due to mass transfer episodes between the binary constituents. On the other hand, dynamically formed binaries inside dense stellar environments (e.g., in globular clusters, nuclear star clusters, or near supermassive black holes) may have highly eccentric orbits due to multibody interactions~\cite{Postnov:2014tza, Samsing:2017xmd, Mapelli:2021taw}. Binaries formed with high eccentricity may retain residual eccentricity when they enter the frequency bands of ground-based GW detectors~\cite{Wen:2002km,OLeary:2008myb,Antonini:2012ad,Antonini:2013tea, Antonini:2015zsa, Rodriguez:2018pss}. In fact, there are already claims of detected eccentricity in BBHs observed by LIGO/Virgo in the GWTC-3 catalog~\cite{Romero-Shaw:2020thy,Gayathri:2020coq,Romero-Shaw:2021ual,OShea:2021ugg,Romero-Shaw:2022xko}. However, it is hard to distinguish between the eccentricity and spin-precession effects, especially for the short duration signals such as GW190521~\cite{Romero-Shaw:2022fbf}.     

Favata~\cite{Favata:2013rwa} showed that the systematic biases due to the neglect of eccentricity for binary neutron stars can become greater than the statistical errors even for small eccentricities ($e_0 \sim 10^{-3} \mbox{--} 10^{-2}$; here and throughout $e_0$ will refer to the binary eccentricity at a reference GW frequency of $10$ Hz). Reference~\cite{Favata:2021vhw} extended those results and compared the systematic biases predicted from the Fisher matrix approach with Bayesian estimates, finding them to be in good agreement. Recently, Ref.~\cite{Divyajyoti:2023rht} showed that neglecting eccentricity in the GW signals can significantly bias the chirp mass of the binary. Reference~\cite{Narikawa:2016uwr} studied the effect of neglecting orbital eccentricity, spin, and tidal deformation as possible sources of systematic bias using the parametrized post-Einsteinian formalism \cite{Yunes:2009ke}. And Ref.~\cite{Pang:2018hjb} showed that neglecting higher modes in the waveform model can also bias tests of GR.

More recently, Saini~\textit{et~al.}~\cite{Saini:2022igm} studied the effect of eccentricity-induced systematic bias on parametrized tests of GR. They found that the systematic bias on the leading-order TGR parameter typically exceeds the statistical errors for a BBH with eccentricity $e_0\sim 0.04$ in LIGO and $e_0\sim 0.005$ in Cosmic Explorer (CE). A similar study~\cite{Bhat:2022amc} considered the inspiral-merger-ringdown consistency test of GR~\cite{Ghosh:2016qgn}, showing that an eccentricity of $e_0 \sim 0.1$ ($\sim 0.01$) in LIGO (CE) causes a significant systematic bias in the final mass and spin of the remnant BH. Reference~\cite{Narayan:2023vhm} also studied the effect of neglecting eccentricity on various tests of GR. However, all the above studies were done for individual events and not at the level of an observed population.
  
Recently, Ref.~\cite{Moore:2021eok} showed how a systematic bias on GR tests can accumulate for a catalog of GW events. Even if these biases are less dominant when analyzing individual events, a systematic bias can accumulate when combining the individual bounds on $\delta\hat{\varphi}_i$ across multiple events, mimicking a GR deviation. Reference~\cite{Hu:2022bji} studied the effect of overlapping signals and inaccurate waveforms on parametrized tests of GR for 3G GW detectors; they also found that systematic biases could accumulate when combining multiple events, leading to an apparent GR violation. 

%%%%%%%%%%%%%%%%%%%%%%%%%%%%%%%%%%%%%%%%%%%%%%%%%%%%%%%%%%%%
%%%%%%%%%%%%%%%%%%%%%%%%%%%%%%%%%%%%%%%%%%%%%%%%%%%%%%%%%%%%
\begin{figure*}
   \centering
    \begin{subfigure}{\includegraphics[width=0.49\textwidth]{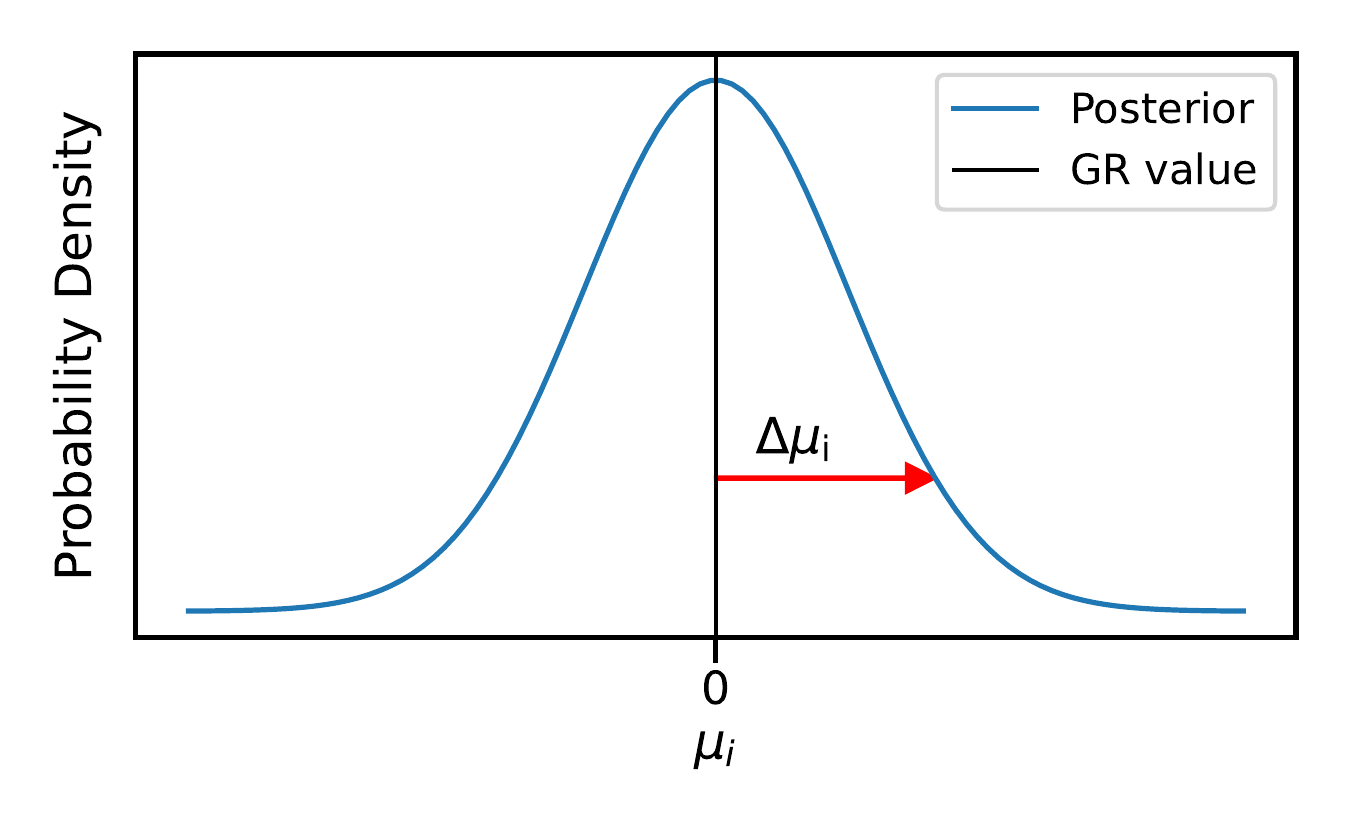}}
    \end{subfigure}
    \vspace{-0.7cm}
    \begin{subfigure}{\includegraphics[width=0.49\textwidth]{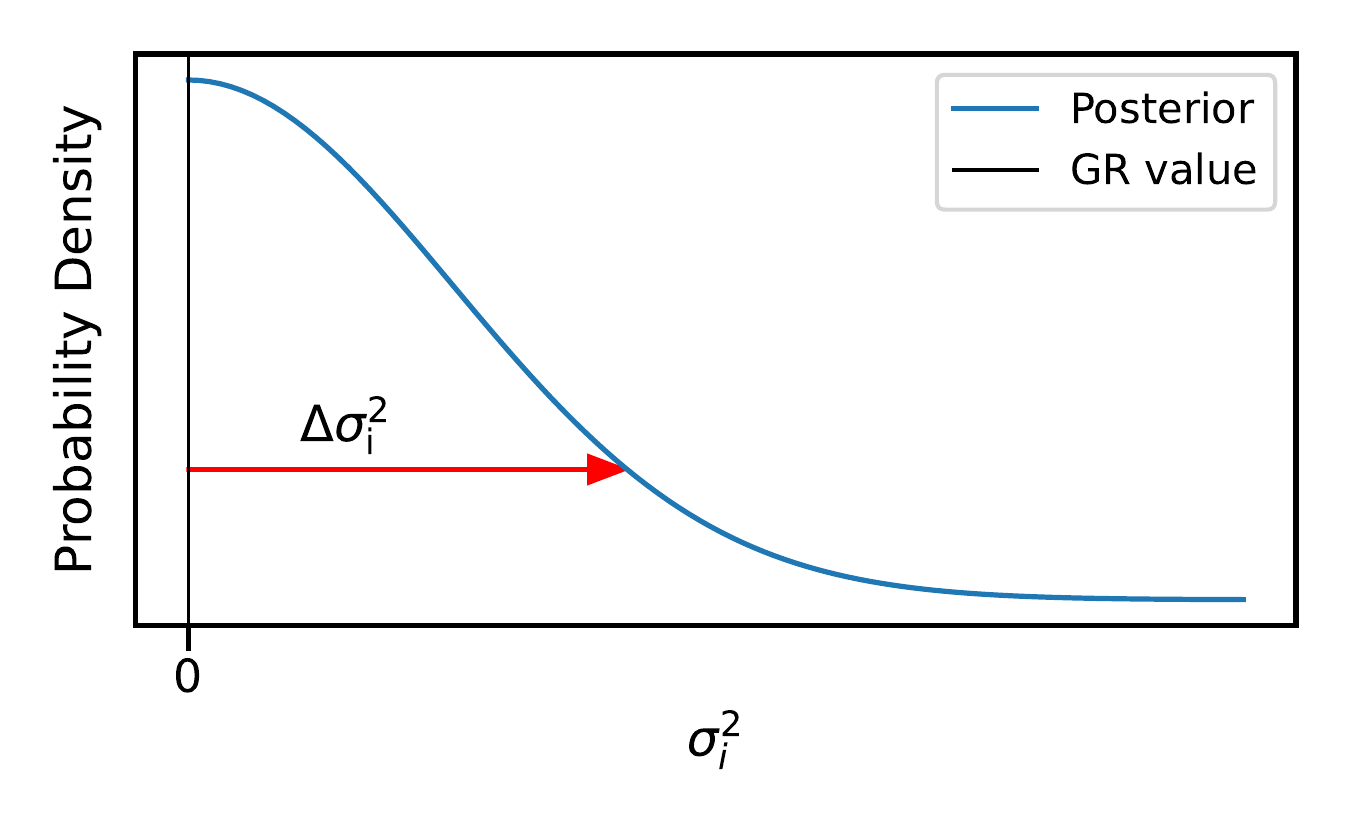}}
    \end{subfigure}
    \begin{subfigure}{\includegraphics[width=0.49\textwidth]{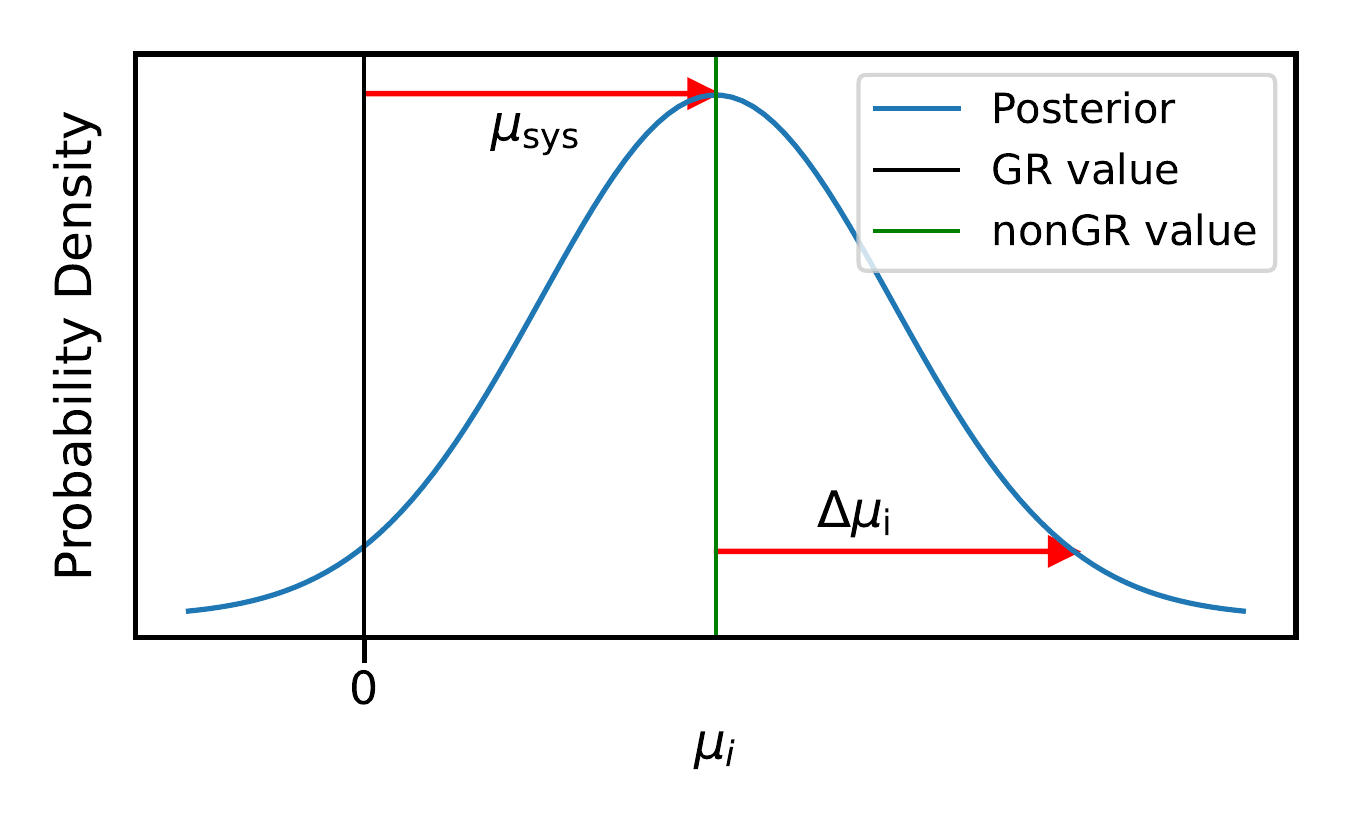}}
    \end{subfigure}
    \vspace{-0.7cm}
    \begin{subfigure}{\includegraphics[width=0.49\textwidth]{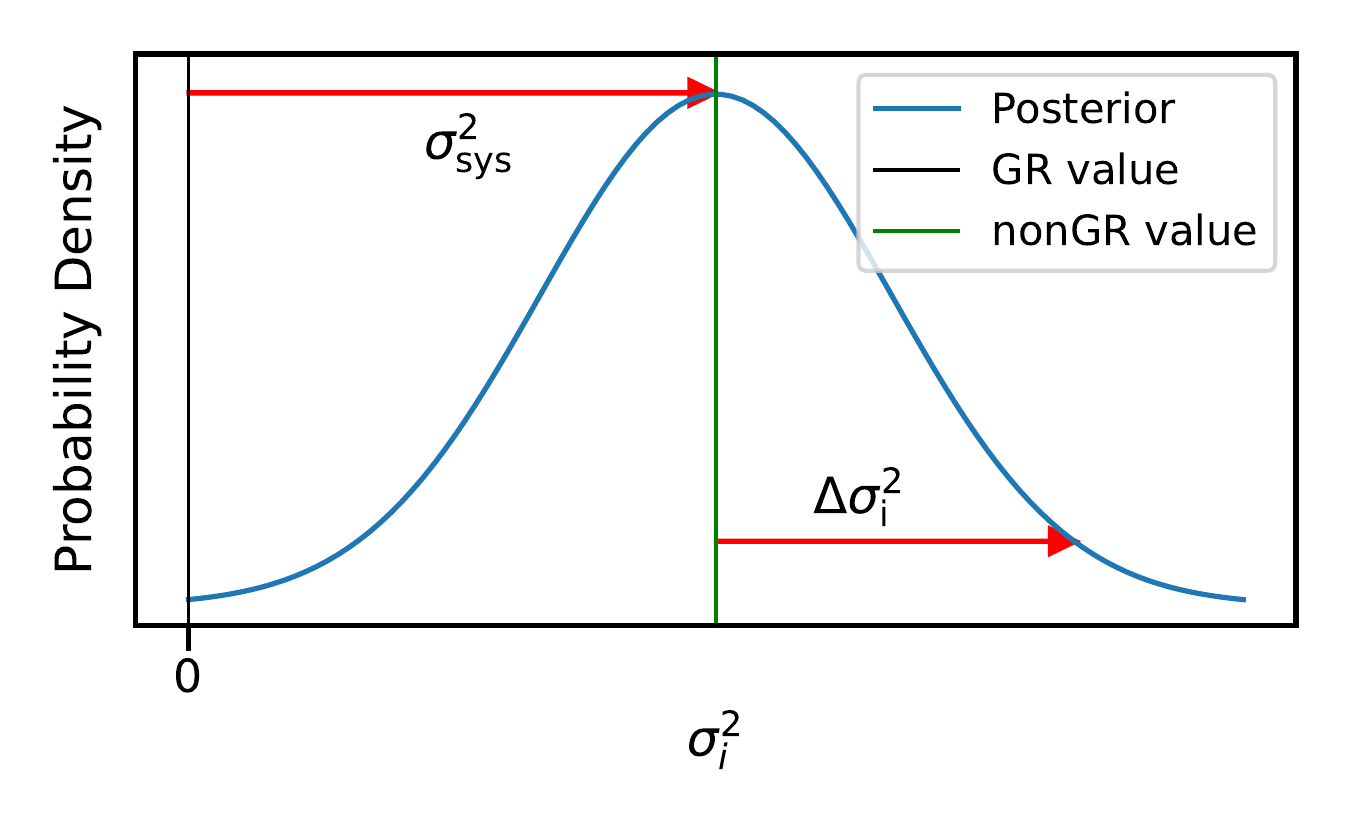}}
    \end{subfigure}
    \caption{\label{schematic figure} Schematic diagram depicting the expected behavior of the population hyperparameters $\mu_i$ and $\sigma_i$ for the unbiased (upper panels) and biased (lower panels) cases. The blue curves represent the one-dimensional marginalized posterior probability distributions for the hyperparameters $\mu_i$ and $\sigma_i$. For simplicity, we represent the posteriors via Gaussian distributions. In the absence of any bias (upper panel), the peaks of the probability distributions are exactly at the GR values (zeros) for both $\mu_i$ and $\sigma_i$. In this case, the width of the distributions depends only on the statistical width $(\tilde{\sigma}_i^{j})$ of the individual $\delta\hat{\varphi}_i$ posteriors [$p(\delta\hat{\varphi}_i^{j})$] in the population. The use of an inaccurate waveform model shifts the peak of the individual posteriors $p(\delta\hat{\varphi}_i^{j})$ away from zero. The shift in the $\mu_i$ posterior is given by the mean of the systematic bias ($\mu_{\rm sys}$) in the $\delta\hat{\varphi}_i^{j}$ posteriors (assuming the statistical errors on the $\delta\hat{\varphi}_i^{j}$ are the same for all events, $\tilde{\sigma}_i^{j} = \tilde{\sigma}_{0,i}$). The shift in the $\sigma_i^2$ posterior is proportional to the variance in the systematic biases ($\sigma_{\rm sys}^2$). The widths of the $\mu_i$ and $\sigma_i$ posteriors in the biased case depend on $\sigma_{\rm sys}$ and $\tilde{\sigma}_{0,i}$ [see Eqs.~\eqref{eq:mu} and~\eqref{eq:sigma}].
    }
\end{figure*}
%%%%%%%%%%%%%%%%%%%%%%%%%%%%%%%%%%%%%%%%%%%%%%%%%%%%%%%%%%%%
%%%%%%%%%%%%%%%%%%%%%%%%%%%%%%%%%%%%%%%%%%%%%%%%%%%%%%%%%%%%

\subsection{Inference of TGR parameters from a BBH population}
In addition to a formalism for computing the systematic waveform bias, this problem requires understanding how individual bounds on GR deviation parameters are combined for a population of sources. To assess the systematic biases in a population, one constructs a population likelihood function by combining the likelihoods from individual events. There are primarily three ways of combining the information from different GW events, with the combined likelihood depending on assumptions about the individual TGR parameters $\delta\hat{\varphi}_i$ ~\cite{Zimmerman:2019wzo}:
\begin{enumerate}
    \item {\it Shared common value:} The TGR parameters $\delta\hat{\varphi}_i$ are assumed to share a common value across all events in the population. In this case, multiplying the likelihoods of the individual events will give the combined likelihood.
    \item  {\it Unrelated TGR parameters:} Each event can have an independent value of the TGR parameter. This implies that there is no relation between the measured values of a particular $\delta\hat{\varphi}_i$ for any two events. This is equivalent to assuming that the non-GR theory has a coupling parameter that varies from source to source. While this would be an unnatural assumption for a modified gravity theory, it would be a more natural assumption for black hole mimickers that are parametrized by their tidal deformability~\cite{Cardoso:2017cfl,Johnson-Mcdaniel:2018cdu} or spin-induced multipole moments~\cite{Krishnendu:2017shb,Krishnendu:2019tjp,Saleem:2021vph,Saini:2023gaw}. In this case, the natural way to combine information is to multiply the Bayes factors of the GR versus non-GR hypotheses across events (assuming flat priors on $\delta \hat{\varphi}_i$)~\cite{Zimmerman:2019wzo}.
    \item  {\it Common distribution:} In this scenario, the TGR parameters $\delta\hat{\varphi}_i$ are assumed to follow a common (unknown) underlying distribution that can depend on the system parameters (such as the BH masses and spins)~\cite{Isi:2019asy, LIGOScientific:2020tif, LIGOScientific:2021sio}. This is more natural to expect in a physically meaningful theory of gravity and is the assumption we make in this work. Hierarchical inference~\cite{Zimmerman:2019wzo, Isi:2019asy,Isi:2022cii} is the method proposed to deal with this scenario; it is introduced below and discussed in further detail in Sec.~\ref{hierarchical inference}.
\end{enumerate}
Here, we closely follow the prescription of Ref.~\cite{Isi:2019asy}. The assumption about the nature of the GR deviations depends on the particular theory of gravity. In the case of parametrized tests of GR we expect the TGR parameters $\delta\hat{\varphi}_i$ to be functions of the individual source properties like the masses and spins. This implies that the $\delta\hat{\varphi}_i$ are expected to be related across events through some common distribution. 

In the absence of any knowledge about the distribution of the TGR parameters or their functional dependence on the source parameters, we assume (for simplicity) that the population likelihood for TGR parameters follows a Gaussian distribution with unknown mean and standard deviation (to be estimated from the data). In other words, we assume the underlying distribution of $\delta\hat{\varphi}_i$ to be a Gaussian $p(\delta\hat{\varphi}_i)$ characterized by two hyperparameters $\mu_i$ (mean) and $\sigma_i$ (standard deviation), with the latter to be inferred by observing a population of sources:
\begin{equation}
    p(\delta\hat{\varphi}_i|\mu_i,\sigma_i) = \mathcal{N}(\mu_i,\sigma_i)\,.
\end{equation}
(Recall that $i$ labels the $i^{\rm th}$ TGR parameter, which corresponds to modifications of the GW phasing at different PN orders.) It is worth stressing that, {\it a priori}, both hyperparameters parametrize a theoretical model of modified gravity; in the GR limit $(\mu_i, \sigma_i) \rightarrow 0$. Hence, if GR is the true theory of gravity, the posteriors on $\mu_i$ and $\sigma_i$ should be consistent with zero.
We further assume that the posterior probability distribution on the individual $\delta\hat{\varphi}_i$ for the $j^{\rm th}$ event follows a Gaussian distribution ${\mathcal N}$ with mean $\Tilde{\mu}_i^j$ and spread $\Tilde{\sigma}_i^j$:
\begin{equation}
     p(\delta\hat{\varphi}_i^{j}|d_j) = \mathcal{N}(\Tilde{\mu}_i^{j}, \Tilde{\sigma}_i^{j})\,.
\end{equation}

To gain better intuition, it is instructive to examine relevant limiting cases. Consider first the case where the standard deviation of the distribution of $\delta\hat{\varphi}_i$ across events vanishes, $\sigma_i\to 0$. This means the value of $\delta\hat{\varphi}_i$ is identical for all events. In this case, $p(\delta\hat{\varphi}_i)$ will be centered around $\mu_i$ (which is same for all the events and $\neq 0$ if GR is violated). The width of the $p(\delta\hat{\varphi}_i)$ then arises solely from statistical errors in the measurement of the remaining hyperparameter $\mu_i$; this depends on the widths of the individual Gaussians ($\tilde{\sigma}_i^j$). This is similar to the scenario of shared common values discussed in Case 1 earlier. 

Consider another extreme scenario where $\sigma_i\to \infty$. This would mean that none of the values of $\delta\hat{\varphi}_i$ is the same for any two events, and therefore $p(\delta\hat{\varphi}_i)$ will be a flat distribution (as described in Case 2 above). These examples put the three scenarios discussed in perspective. In this paper we will only consider the methods in Cases 1 and 3 when combining events from a population of BBHs.

To better understand the meaning of the hyperparameters $\mu_i$ and $\sigma_i$, we show two cases schematically in Fig.~\ref{schematic figure}. When GR is correct and a perfect (unbiased) GR waveform model is used to analyze the GW signal, the inferred $\delta\hat{\varphi}_i^j$ posteriors from each event are free of any systematic bias and peak at zero. In this case (upper panels), we expect the posterior distributions for the $\mu_i$ and $\sigma_i^2$ (or $\sigma_i$) to be consistent with zero. If the detector noise is additionally zero ($\tilde{\sigma}_i^{j}\to 0$), the $p(\mu_i)\to\delta(\mu_i)$ and $p(\sigma_i^2)\to\delta(\sigma_i^2)$, where $\delta$ is the Dirac $\delta$ function.

Even in the case where GR is correct, if the GR waveform model has inaccuracies or missing physics, the corresponding $\delta\hat{\varphi}_i^j$ posteriors will peak away from zero. The amount of this shift from zero depends on the mismatch between the waveform model and the GW signal. At the level of population inference, the biases in the individual $\delta\hat{\varphi}_i^j$ posteriors translate into the posterior distributions of the hyperparameters $\mu_i$ and $\sigma_i^2$. The $\mu_i$ and $\sigma_i^2$ posteriors show the inconsistency with GR (lower panels of Fig.~\ref{schematic figure}). If the widths of all the individual Gaussians are the same (i.e.~$\tilde{\sigma}_i^j= \tilde{\sigma}_{0,i}$), the shifts in the $\mu_i$ posteriors from zero are equal to the {\it mean of the systematic biases} of the $\delta\hat{\varphi}_i^j$ posteriors ($\mu_{\rm sys}$). The shift in the $\sigma_i^2$ posteriors is equal to the {\it variance of the systematic biases} $(\sigma_{\rm sys}^{2})$. Note that the peak of the $\sigma_i^2$ posteriors is also affected by the individual widths $\tilde{\sigma}_{i}^{j}$~\cite{Zimmerman:2019wzo, Isi:2022cii}. Those $\delta\hat{\varphi}_i^{j}$ posteriors that are uninformative can cause the $p(\mu_i)$ and $p(\sigma_i^2)$ [or equivalently $p(\sigma_i)$] to shift from the true values of $\mu_i$ and $\sigma_i^2$. For the biased case and in the limit where the statistical errors for the $i^{\rm th}$ TGR parameter are the same for each event ($\tilde{\sigma}_i^j= \tilde{\sigma}_{0,i}$), the widths (or standard deviations) of the $\mu_i$ and $\sigma_i^2$ posteriors are given by~\cite{Isi:2022cii}
\begin{align}\label{eq:mu}
    \Delta\mu_i &= \sqrt{\frac{\sigma_{\rm sys}^2+\tilde{\sigma}_{0,i}^2}{N}} \,, \\
\label{eq:sigma}
    \Delta(\sigma_i^2) &= \sqrt{\frac{2}{N}}(\sigma_{\rm sys}^2+\tilde{\sigma}_{0,i}^2) \,.
\end{align}
As the number of events $N$ grows, the spread in $p(\mu_i)$ and $p(\sigma_i^2)$ reduces as $\sim 1/\sqrt{N}$~\cite{Isi:2019asy, Isi:2022cii}\footnote{Reference~\cite{Pacilio:2023uef} recently pointed out that the variance arising from the finite number of catalog events causes the $p(\mu_i)$ and $p(\sigma_i^2)$ to exclude the null hypothesis even if it is correct.}. In the unbiased case ($\sigma_{\rm sys}\to 0$), the width  $\Delta\mu_i$ of $p(\mu_i)$ is $\tilde{\sigma}_{0,i}/\sqrt{N}$. This is similar to the scaling that statistical errors follow when the events are combined by multiplying the likelihoods.

If either the $\mu_i$ or $\sigma_i$ posterior excludes zero with confidence, this suggests a deviation from GR or a systematic bias. If a true GR deviation or a systematic bias is symmetrically distributed around zero, the $\mu_i$ posteriors will be consistent with zero, but the $\sigma_i$ posteriors will exclude zero. If the values of $\delta\hat{\varphi}_i^j$ for individual events are centered around some common value (different from GR but with little to no spread), the posterior of $\sigma_i$ will be centered around zero showing no inconsistency; but the $\mu_i$ posterior will peak around the common value of the individual $\delta\hat{\varphi}_i^j$. (This corresponds to Case 1 above.)

\subsection{Present work}
We study the cumulative effect of systematic bias on the $\delta\hat{\varphi}_i$ due to unmodeled eccentricity. This assumes that a population of events is observed, combining the individual posteriors of $\delta\hat{\varphi}_i$ hierarchically, and asking under what circumstances they mimic a GR violation. We do this for current-generation LIGO-like detectors, as well as for Cosmic Explorer type third-generation (3G) detectors. We consider a population of BBHs with mass and spin distributions inferred via population inference using data in the third GW transient catalog (GWTC-3)~\cite{LIGOScientific:2021psn,GWTC3:2021sio}. The eccentricities of these binaries are drawn from three different eccentricity distributions (discussed in Sec.~\ref{eccentricity distribution} below). We find that the eccentricity-induced systematic biases implied by these modeled populations cause the posteriors on the hyperparameters $\mu_i$ and $\sigma_i$ to exclude zero, indicating a deviation from GR.

In addition to quantifying the systematic biases, we compute the combined bounds on the $\delta\hat{\varphi}_i$ from the simulated population, {\it including the effect} of eccentricity. We find that including eccentricity in the parameter space has a mild effect on the combined bounds. The bounds on the lower PN order $\delta\hat{\varphi}_i$ ($i=0, 1, 2, 3$) degrade by only a factor of $\lesssim 2$ due to the inclusion of eccentricity in the parameter space. The bound on the dipole TGR parameter degrades by a factor of $\sim 3\mbox{--}4$. The bounds on the higher PN order $\delta\hat{\varphi}_i$ are not significantly affected and are comparable to the bounds set if one assumes a circularized population of BBHs.

Section~\ref{waveform model} of this paper briefly discusses the waveform model used for studying the eccentricity-induced systematic bias on the TGR parameters. Section~\ref{statistical uncertainities} explains how statistical errors are calculated using the Fisher matrix framework, while Sec.~\ref{systematic bias} discusses the Cutler-Vallisneri formalism~\cite{Cutler:2007mi} for calculating the systematic errors. Section~\ref{bayesian analysis} explains the basic formalism of Bayesian analysis, while Sec.~\ref{hierarchical inference} applies it to the hierarchical Bayesian inference method for combining individual likelihoods. Section~\ref{population analysis results} discusses our BBH population model, our methodology, and the results of our hierarchical inference study. In Sec.~\ref{sec:combined bound} we directly compute the (statistical) TGR parameter bounds set by the observed BBH population, examining the extent to which eccentric waveforms can worsen the parameter bounds relative to quasicircular waveforms. We obtain these bounds using both the multiplication of likelihoods and hierarchical Bayesian inference methods. Section~\ref{conclusions} presents our conclusions. We use units with $G=c=1$ henceforth.

\section{Waveform model and parametrization of non-GR effects}\label{waveform model}
\subsection{Post-Newtonian waveform model}
The time-domain strain measured by an interferometric GW detector can be expressed as a linear combination of plus ($h_+$) and cross ($h_{\times}$) polarizations of GWs,
\begin{equation}
    h(t) = F_{+}(\theta, \phi,\psi) h_+(\iota, \beta,t) +  F_{\times}(\theta, \phi,\psi) h_{\times}(\iota, \beta, t)\,,
\end{equation}
where $F_{+,\times}(\theta,\phi,\psi)$ are the antenna pattern functions. The angles $\theta$ and $\phi$ describe the sky position of the source with respect to the detector, $\psi$ is the polarization angle of the incoming GW, $\iota$ is the inclination angle of the binary, and $\beta$ defines the initial orientation of the orbital ellipse; in the circular limit $\beta$ can be absorbed into the phase constant.

The {\it frequency-domain} strain can be calculated using the stationary phase approximation (SPA)~\cite{Droz:1999qx} as
\begin{equation}\label{waveform}
    \Tilde{h}(f) = \mathcal{A} e^{i\Psi(f)} =\hat{\mathcal{A}} f^{-7/6} e^{i \Psi(f)} \,,
\end{equation}
where $\mathcal{A}$ is the amplitude of the waveform. Averaging over the angles $(\theta, \phi, \psi, \iota, \beta)$ in the quadrupole approximation,  
\begin{equation}
    \hat{\mathcal{A}} = \frac{1}{\sqrt{30} \pi^{2/3}} \frac{\mathcal{M}^{5/6}}{D_L}\,,
\end{equation}
where $\mathcal{M}= (m_1 m_2)^{3/5}/M^{1/5}$ is the chirp mass, $m_1$ and $m_2$ are the component masses of the binary (in the source frame), $M = m_1 + m_2$ is the source-frame total mass, and $D_L$ is the source luminosity distance. The luminosity distance and redshift for a flat universe are related by ~\cite{Hogg:1999ad}
\begin{equation}
\label{eq:dLz}
D_L(z) = \frac{1+z}{H_0} \int_0^z \frac{dz'}{\sqrt{\Omega_M (1+z')^3 + \Omega_{\Lambda}}}\,,
\end{equation} 
where $\Omega_{m}$ and $\Omega_{\Lambda}$ denote the matter density and dark energy density parameters (respectively), and $H_{0}$ is the Hubble constant.
The cosmological parameters for a flat universe are taken from Planck observations~\cite{Planck:2015fie}: $H_{0}=67.90$(km/s)/Mpc, $\Omega_{m}=0.3065$, and $\Omega_{\Lambda}=0.6935$. 

The inspiral phase of the binary is well described by the {\it post-Newtonian }(PN) approximation~\cite{Blanchet:2002av}. This approach solves the Einstein equations for the two-body problem in the limit of small speeds and weak gravity. Physical quantities during the inspiral phase of the orbit are usually expressed as power series in the relative orbital speed parameter $v$~\cite{Blanchet:1995ez,Blanchet:1995fg,Kidder:1995zr,Blanchet:2002av,Blanchet:2006gy,Arun:2008kb,Marsat:2012fn,Mishra:2016whh}.
For a binary with a small orbital eccentricity ($\lesssim 0.2$), the SPA phase $\Psi(f)$ in Eq.~\eqref{waveform} can be decomposed into a circular phase $\Psi^{\rm circ.}_{\rm 3.5 PN}$ and an eccentric phase $\Delta\Psi^{\rm ecc.}_{\rm 3PN}$~\cite{Moore:2016qxz},
\begin{equation}
\label{SPA}
    \Psi(f) = \phi_c + 2\pi f t_c + \frac{3}{128 \eta v^5}\bigg(\Psi^{\rm circ.}_{\rm 3.5 PN} + \Delta\Psi^{\rm ecc.}_{\rm 3PN}\bigg) \, ,
\end{equation}
where $\phi_c$ and $t_c$ are the phase and time of coalescence (respectively), $\eta = (m_1 m_2)/M^2$ is the symmetric mass ratio, and $v= (\pi M f)^{1/3}$ is the PN orbital velocity parameter. Note that $\mathcal{M}$ and $M$ are the chirp mass and total mass in the {\it source} frame of the binary. When observing binaries at large distances and interpreting $f$ as the GW frequency measured by Earth-based detectors, the total mass and chirp mass in the above equations should be replaced by the observed (redshifted) total and chirp masses:
\begin{equation}
    M_{\rm obs} = (1+z) M,\,\,\,\,\,\,  \mathcal{M}_{\rm obs} = (1+z) \mathcal{M}\,.
\end{equation}
The {\it quasicircular} piece of the frequency-domain phase can be expanded in powers of the orbital velocity parameter $v$ as
\begin{equation} \label{PN phase}
     \Psi^{\rm circ.}_{\rm 3.5 PN} = \sum_{i=0}^{7} (\varphi_{i}^{\rm GR}v^{i} 
      +  \varphi_i^{\rm log, GR}  v^{i} \log v) \,.
\end{equation}
Here the $\varphi_i^{\rm GR}$ and $\varphi_i^{\rm log, GR}$ are the PN coefficients. In the case of binaries with component spins that are aligned (or antialigned) with the orbital angular momentum, these coefficients are functions of the masses and the magnitude of the individual spin angular momenta. The values of $(\varphi_i, \varphi_i^{\rm log})$ up to 3.5PN order can be found in Refs.~\cite{Arun:2004hn,Arun:2008kb,Buonanno:2009zt,Wade:2013hoa,Mishra:2016whh, Blanchet:2023bwj,Blanchet:2023sbv}. In a PN series, a term proportional to $v^{2 n}$ relative to the {\it Newtonian term} (proportional to $v^{-5}$) is called the $n$PN order term.

The term $\Delta\Psi^{\rm ecc.}_{\rm 3PN}$ in Eq.~\eqref{SPA} is the leading order [$\mathcal{O}(e_0^2)$] eccentric correction up to 3PN order given by the {\tt TaylorF2Ecc} waveform model \cite{Moore:2016qxz}, 
\begin{align}
    \label{eccentric phase}
    \Delta\Psi^{\rm ecc.}_{\rm 3PN} &= -\frac{2355}{1462} e_{0}^{2} \nonumber \Big(\frac{v_{0}}{v}\Big)^{19/3}\Bigg[1+  \nonumber \bigg(\frac{ 299076223}{81976608} \\                     \nonumber
     &+ \frac{18766963}{2927736} \eta \bigg) v^{2} + \bigg(\frac{2833}{1008} - \frac{197}{36} \eta \bigg)v_{0}^{2} \\ 
     &- \frac{2819123}{282600} \pi v^{3} + \frac{377}{72}\pi v_{0}^{3} + \cdots + \mathcal{O}(v^{7})\Bigg]\,. 
\end{align}
Here $v_0 = (\pi M f_0)^{1/3}$ and $e_0$ is the orbital eccentricity defined at a reference frequency $f_0$. The value of $f_0$ is often chosen to be $10$ Hz, corresponding to the low-frequency limit of LIGO detectors; we adopt that value in this work. \footnote{Different waveform models often use different definitions of orbital eccentricity. One should ideally use a model-agnostic definition of eccentricity when comparing eccentricity measurements from different analyses~\cite{PhysRevD.107.064024, Shaikh:2023ypz}.}

The $\tt TaylorF2Ecc$ waveform model is valid for small eccentricities $e_0\lesssim 0.2$ for comparable mass systems. Since the orbital eccentricity of the binary decays rapidly due to the emission of GWs, ground-based detectors are expected to observe most binaries as quasicircular or with very low values of eccentricity. Hence, the $\tt TaylorF2Ecc$ waveform model is sufficient for the purposes of this study. The waveform model accounts for the dominant second harmonic of the GW signal and ignores the eccentricity-induced higher harmonics that are subdominant in the small-eccentricity limit. Spin contributions are accounted for only in the circular part of the phase ($\Delta \Psi_{\rm 3.5PN}^{\rm circ.}$) and neglected in the eccentric piece of the phasing.

\subsection{Parametrizing deviations from GR}
This work focuses on the parametrized test of GR~\cite{Arun:2004hn,2006PhRvD..74b4006A, Cornish:2011ys, Agathos:2013upa}. In parametrized tests of GR, PN coefficients are deformed around their GR values by introducing TGR parameters at each PN order. These free parameters encapsulate the GR deviation via the following replacements in the GW frequency-domain SPA phasing [Eq.~\eqref{PN phase}]:
\begin{subequations}
\label{deformation1}
\begin{align}
\varphi_{i}^{\rm GR} &\xrightarrow{} \varphi_{i} = \varphi_{i}^{\rm GR} + \delta\varphi_{i} \,, \\
\varphi_{i}^{\rm log, GR} &\xrightarrow{} \varphi_{i}^{\rm log} = \varphi_{i}^{\rm log, GR} + \delta \varphi_{i}^{\rm log} \, ,
\end{align}
\end{subequations}
where GR denotes the PN coefficient value in general relativity.
It is more convenient to work with an alternative formulation in terms of dimensionless deviation coefficients:
\begin{subequations}
\label{deformation2}
\begin{align}
\varphi_{i}^{\rm GR} &\xrightarrow{} \varphi_{i} = \varphi_{i}^{\rm GR} (1 + \delta \hat{\varphi}_{i}) \,, \\
\varphi_i^{\rm log, GR} &\xrightarrow{} \varphi_i^{\rm log} = \varphi_i^{\rm log, GR} (1 + \delta \hat{\varphi}_i^{\rm log}) \,,
\end{align}
\end{subequations}
where $\delta\hat{\varphi}_{i}$ and $\delta\hat{\varphi}_{i}^{\rm log}$ are the fractional deviation coefficients. We also extend the summation index in Eq.~\eqref{PN phase} to $i=-2$ to allow for a dipole radiation term (the $i=-1$ term vanishes). GR deviations are only introduced in the circular part of the SPA phase; the eccentric part (which will source the systematic bias) is left as is. 

The following deviation parameters are introduced in Eq.~\eqref{PN phase}:
\begin{equation}
\{\delta\hat{\varphi}_{-2},\delta\hat{\varphi}_{0}, \delta\hat{\varphi}_{1}, \delta\hat{\varphi}_{2},\delta\hat{\varphi}_{3}, \delta\hat{\varphi}_{4}, \delta\hat{\varphi}_{5}^{\rm log}, \delta\hat{\varphi}_{6}, \delta\hat{\varphi}_{6}^{\rm log}, \delta\hat{\varphi}_{7}\} \,.
\end{equation}
There is no deviation to the nonlogarithmic frequency-independent term at 2.5PN order, as this term is degenerate with the coalescence phase and can be absorbed into the redefinition of $\phi_c$. The deviations at $-1$PN ($\delta\hat{\varphi}_{-2}$) and 0.5PN ($\delta\hat{\varphi}_{1}$) orders are absolute deviations as the GR values for those coefficients are zero:
\begin{align}
\varphi_{-2} & = \delta \hat{\varphi}_{-2}\,,\\
\varphi_{1} & = \delta \hat{\varphi}_{1} \,.
\end{align}
The presence of a $-1$PN term is associated with the emission of dipole radiation, which is forbidden in GR~\cite{Will:2014kxa,Will:2004xi}. The excitation of additional fields can give rise to dipole radiation, resulting in a faster decay of the binary's orbit relative to GR~\cite{Yagi:2011xp,Lang:2014osa}. Since the leading-order dipole term modifies the GW phase at $-1$PN order, this term has a larger influence at lower GW frequencies (when the binary separation is larger).  

\section{Estimation of statistical and systematic errors}\label{statistical uncertainities}
\subsection{Statistical errors}
In the large SNR limit, the statistical errors on binary parameters can be estimated using the {\it Fisher information matrix} framework~\cite{Finn:1992wt,Cutler_Flanagan,Clifford_Will}. This framework readily estimates the $1\sigma$ errors in the binary parameters, centered around the injected values. (See Refs.~\cite{vallisneri2008use,Rodriguez:2013mla} for caveats associated with the use of the Fisher matrix.) 

The inner product between two frequency domain signals ${\tilde h}_1(f)$ and ${\tilde h}_2(f)$ is defined as 
\begin{equation}\label{innerproduct}
    (h_{1}|h_{2}) = 2 \int_{f_{\rm low}}^{f_{\rm high}}\frac{\Tilde{h}_{1}^{*}(f) \Tilde{h}_{2}(f) + \Tilde{h}_{1}(f)\Tilde{h}_{2}^{*}(f) }{S_{n}(f)}\, df,
\end{equation}
where $S_{n}(f)$ is the one-sided noise power spectral density (PSD) of the detector, and $\ast$ represents complex conjugation. The limits of integration in Eq.~\eqref{innerproduct} are fixed by the sensitivity of the detector and the properties of the source.

For stationary, Gaussian noise and in the large-SNR limit, the probability distribution of the waveform parameters $\bm \theta$ given the data $d(t)$ can be approximated as
\begin{equation}\label{probability}
 p({\bm \theta}|d) \propto p^{0}({\bm \theta}) \exp\left[ -\frac{1}{2} \Gamma_{ab} (\theta_{a} - \hat{\theta}_{a}) (\theta_{b} - \hat{\theta}_{b}) \right]\,, \end{equation}
where $p^{0}(\bm{\theta})$ is the prior probability of the parameters ${\bm {\theta}} =\{ \theta_a\}$. The $\hat{\theta}_{a}$ are the best-fit values that maximize the Gaussian likelihood function and correspond to the “true'' values of the source parameters in the absence of any bias. In Eq.~\eqref{probability}, $\Gamma_{ab}$ is the Fisher information matrix, which is defined by the inner product
\begin{equation}\label{fisher}
    \Gamma_{ab} = \Bigg(\frac{\partial h}{\partial \theta_{a}}\Bigg|\frac{\partial h}{\partial \theta_{b}}\Bigg)\,.
\end{equation}
The SNR $\rho$ is defined via
\begin{equation}\label{snr}
    \rho^{2} = (h|h) = 4\int_{0}^{\infty}\frac{|\Tilde{h}(f)|^{2}}{S_{n}(f)}\,df\,.
\end{equation}
Since we are not interested in the source distance and sky localization in this study, the parameter $\hat{\mathcal{A}}$ (which sets the signal's SNR) is not a parameter of interest; it decouples completely from the rest of the Fisher matrix~\cite{Clifford_Will}. 

Using Eqs.~\eqref{innerproduct} and ~\eqref{snr}, the Fisher matrix becomes~\cite{Favata:2021vhw}
\begin{equation}
    \Gamma_{ab} = \frac{\rho^2}{I_{7/3}} \int_{f_{\rm low}}^{f_{\rm high}} \frac{f^{-7/3}}{S_n(f)} \partial_a \Psi(f) \partial_b \Psi(f) df \,,
\end{equation}
where $I_{7/3}$ is 
\begin{equation}
    I_{7/3} = \int_{f_{\rm low}}^{f_{\rm high}}\frac{f^{-7/3}}{S_{n}(f)} df\,.
\end{equation}
If the prior probability $p^{0}(\boldsymbol{\theta})$ follows a Gaussian distribution, then the covariance matrix is 
\begin{equation}\label{covariance}
 \Sigma_{ab}=(\Gamma_{ab}+ \Gamma_{ab}^{0})^{-1}\;,   
\end{equation}
where $\Gamma_{ab}^{0}$ is the prior matrix. The 1$\sigma$ width of the projected posterior probability distribution on the binary parameter $\theta_{a}$ is given by the square root of the relevant diagonal element of the covariance matrix:
\begin{equation}\label{error}
    \sigma_a = \sqrt{\Sigma_{aa}} \;.
\end{equation}

The parameters that describe our quasicircular, nonprecessing waveform model are 
\begin{equation}
    \theta_a= \{t_{c},\phi_{c},\log\mathcal{M},\log\eta,\chi_{1},\chi_{2},\delta\hat{\varphi}_{i}\}\, .
\end{equation}
Here $\chi_{1}$ and $\chi_{2}$ are the dimensionless spin parameters of the primary and secondary BHs, respectively; we assume that the spins are aligned with the orbital angular momentum ($\chi_{1,2} \in[0,1]$). We use Gaussian priors on $\phi_c$ and $\chi_{1,2}$ with zero means. The 1$\sigma$ widths of the ($\phi_c$, $\chi_{1,2}$) prior distributions are given by $\delta\phi_c=\pi$ and $\delta\chi_{1,2}=1$, respectively. They are incorporated into the Fisher matrix via the addition of the diagonal elements $\Gamma_{aa}^{0} = 1/\delta{\theta}_a^2$. These priors on $\phi_c$ and $\chi_{1,2}$ improve the conditioning of the Fisher matrix. No priors are used on the $\delta\hat{\varphi}_i$ or the other parameters.

We perform a {\it one parameter test} in which only one of the $\delta\hat{\varphi}_{i}$ is varied along with the other system parameters. If the GW signal is described by a specific modified theory of gravity, a deviation could appear in more than one PN order. Ideally, all $\delta\hat{\varphi}_{i}$ should be measured simultaneously---along with other system parameters---to test the true nature of the deviations~\cite{Arun:2006yw}. Tests in which all the $\delta\hat{\varphi}_{i}$ are measured simultaneously are called {\it multiparameter tests}. Considering current detector sensitivities, multiparameter tests give uninformative posteriors due to correlations between the $\delta\hat{\varphi}_{i}$ and the system parameters~\cite{Datta:2020vcj}. However, multiparameter tests will be possible via multiband observations involving 3G ground-based detectors (Cosmic Explorer~\cite{Reitze:2019iox} and Einstein Telescope~\cite{Punturo:2010zz}) and space-based detectors (LISA)~\cite{Gupta:2020lxa}. Moreover, principal component analysis can solve the degeneracy problem by constructing certain linear combinations of the PN deviation parameters that are better measured~\cite{Saleem:2021nsb, Datta:2022izc}. However, as our goal is to assess the role of systematic errors on GR tests, one-parameter tests provide a good representative set for our study.
  
\subsection{Systematic bias} \label{systematic bias}
In addition to the statistical errors due to detector noise, the parameters $\theta_a$ are also subject to systematic bias due to modeling errors in the gravitational waveforms. Here we investigate the impact of eccentricity-induced systematic biases on the deviation parameters $\delta\hat{\varphi}_{i}$. To calculate this systematic bias we use the Cutler-Vallisneri formalism \cite{Cutler:2007mi}. Systematic biases are defined as the difference between the “true” value of the parameter $\theta^{\rm T}_a$ (in the absence of any systematic bias) and the “best-fit” value of the parameter $\hat{\theta}_a$ (the peak of the recovered Gaussian probability distribution obtained using an approximate waveform):
\begin{equation}
    \Delta\theta_a = \theta^{\rm T}_a - \hat{\theta}_a\,.
\end{equation}
If the approximate waveform is written in terms of an approximate amplitude ($\mathcal{A_{\rm AP}}$) and approximate phase ($\Psi_{\rm AP}$),
\begin{equation}
    \Tilde{h}_{\rm AP}(f)=\mathcal{A}_{\rm AP}  e^{i\Psi_{\rm AP}} \;,
\end{equation}
and if the true waveform differs from the approximate waveform by $\Delta \mathcal{A}$ in amplitude and $\Delta\Psi$ in phase via
\begin{equation}
     \Tilde{h}_{\rm T}(f)=(\mathcal{A}_{\rm AP} + \Delta \mathcal{A}) e^{i(\Psi_{\rm AP}+\Delta\Psi)} \;,
\end{equation}
then the systematic error on the parameter $\theta_{a}$ is approximated as \cite{Cutler:2007mi}
\begin{equation}\label{systematic errors}
    \Delta \theta_{a} \approx \Sigma_{ab}\bigg(\big[\Delta\mathcal{A} + i \mathcal{A}_{\rm AP} \Delta\Psi \big] e^{i \Psi_{\rm AP}} \bigg| \partial^{b} \Tilde{h}_{\rm AP}\bigg) \,.
\end{equation}
We neglect eccentric corrections to the amplitude by setting ($\Delta\mathcal{A}=0$) in Eq.~\eqref{systematic errors}. Small eccentricity corrections to the waveform amplitude will be less important compared to phase corrections, as GW detectors are more sensitive to the GW phase than to the amplitude. Using Eq.~\eqref{snr}, the systematic error $\Delta\theta_a$ simplifies to \cite{Favata:2021vhw}
\begin{equation} 
    \Delta\theta_a = \frac{\rho^2}{I_{7/3}} \Sigma_{ab}  \int_{f_{\rm low}}^{f_{\rm high}} \frac{f^{-7/3}}{S_n(f)} \Delta\Psi \, \partial_b \Psi_{\rm AP} \,df \,,
\end{equation}
where the covariance matrix $\Sigma_{ab}$ in Eq.~\eqref{systematic errors} is calculated using the approximate (quasicircular) waveform. Our approximate waveform is the standard {\tt TaylorF2} quasicircular waveform up to 3.5PN order [Eq.~\eqref{PN phase}], with the $\delta\hat{\varphi}_{i}$ introduced at each PN order. The $\Delta\Psi$ are the eccentric corrections to the phase ($\Delta\Psi^{\rm ecc.}_{\rm 3PN}$) defined in Eq.~\eqref{eccentric phase}. 

The above statistical and systematic error formalism is applied to the LIGO \cite{LIGOScientific:2014pky} and CE \cite{Reitze:2019iox} detectors. The LIGO noise PSD is taken from Eq.~(4.7) of \cite{Ajith:2011ec}. For CE we use the noise PSD from Eq.~(3.7) of \cite{Kastha:2018bcr}. The lower frequency cutoff is chosen to be $10$ Hz for LIGO and $5$ Hz for CE. Since we use the inspiral waveform, the upper frequency cutoff is chosen as the frequency corresponding to the innermost stable circular orbit $(f_{\rm isco})$ of the remnant BH \cite{1972ApJ...178..347B,Husa:2015iqa,Hofmann:2016yih}. This frequency depends on the two component masses ($m_{1}$ and $m_{2}$) of the BHs, their dimensionless spins ($\chi_{1}$  and $\chi_{2}$), and the source redshift $z$. See Appendix C of \cite{Favata:2021vhw} for the full expression. The reference frequency $f_0$ (at which the eccentricity $e_0$ is defined) is chosen to be $10$ Hz for both LIGO and CE.

\section{Bayesian parameter estimation and Hierarchical inference}\label{bayesian analysis}
\subsection{A review of Bayesian parameter estimation}
In Bayesian inference~\cite{2019PASA...36...10T}, we infer the posterior probability density on model parameters $\bm{\theta}$ given GW data. In Sec.~\ref{statistical uncertainities}, we used $\bm\theta$ to represent the waveform parameters that include both system parameters and TGR parameters. For clarity, in this section we represent system and TGR parameters separately. Let us assume $\vec{\theta}_j$ represents a set of system parameters for the $j^{\rm th}$ event in a BBH population. A quasicircular aligned binary system is described by $11$ parameters: two mass parameters, two dimensionless spin parameters, luminosity distance, two sky position angles, two binary orientation parameters, and the coalescence time and phase. The $\delta\hat{\varphi}_i$ are the TGR null parameters. The parametric model for the GW signal therefore is described by $h(\bm\theta) \equiv h(\vec{\theta}_j, \delta\hat{\varphi}_{i})$. Note that in this section we assume the $\delta \hat{\varphi}_i$ have universal values and do not vary by GW source.

Using Bayes' theorem, the posterior probability distribution for parameters $\vec{\theta}_j$ and $\delta\hat{\varphi_i}$ given GW data from the $j^{\rm th}$ event $d_j$ can be written as
\begin{equation}\label{posterior1}
    p(\vec{\theta}_j, \delta\hat{\varphi_i}|d_j) = \frac{\mathcal{\pi}(\vec{\theta}_j, \delta\hat{\varphi_i}) \mathcal{L}(d_j|\vec{\theta}_j, \delta\hat{\varphi_i})}{p(d_j)}\,,
\end{equation}
where $\mathcal{\pi}(\vec{\theta}_j, \delta\hat{\varphi}_i)$ is the prior probability of the parameters $\vec{\theta}_j$ and $\delta\hat{\varphi}_i$, and $\mathcal{L}(d_j|\vec{\theta}_j, \delta\hat{\varphi}_i)$ is the likelihood function of the data $d_j$ given the model parameters $\Vec{\theta}_j$, $\delta\hat{\varphi}_i$. The term $p(d_j)$ in the denominator is called the \emph{evidence}; it is obtained by marginalizing the likelihood over the prior 
\begin{equation}\label{evidence}
    p(d_j) = \int \mathcal{\pi}(\vec{\theta}_j, \delta\hat{\varphi}_i) \mathcal{L}(d_j|\vec{\theta}_j, \delta\hat{\varphi}_i)\,d\vec{\theta}_j \, d\delta\hat{\varphi}_i\,.
\end{equation}
Since we are interested only in the shape of the posterior, Eq.~\eqref{posterior1} can be expressed as
\begin{equation}
        p(\vec{\theta}_j, \delta\hat{\varphi_i}|d_j) \propto \mathcal{\pi}(\vec{\theta}_j, \delta\hat{\varphi_i}) \mathcal{L}(d_j|\vec{\theta}_j, \delta\hat{\varphi_i}) \,.
\end{equation}
If $d_j \rightarrow \bm d = \{d_1, d_2, \cdots d_N \}$ is the set of data from $N$ independent observations and $\vec{\Theta}$ represents the set of individual event system parameters $\vec{\theta}_j$ from those same observations, i.e., $\vec{\Theta} = \{\Vec{\theta}_1, \Vec{\theta}_2, \cdots \Vec{\theta}_N$\}, the combined likelihood is the product of the individual likelihoods:
\begin{equation}
    \mathcal{L}(\bm{d}|\Vec{\Theta}, \delta\hat{\varphi}_i) = \prod_{j=1}^{N}\mathcal{L}(d_{j}|\Vec{\theta}_j, \delta\hat{\varphi}_i)\,,
\end{equation}
The posterior probability on $\delta\hat{\varphi}_i$ given $\bm{d}$ can be computed by marginalizing over the system parameters $\vec{\Theta}$:
\begin{align}\label{dphi_posterior}
    p(\delta\hat{\varphi}_i|\bm{d}) &= \int p(\vec{\Theta}, \delta\hat{\varphi}_i|\bm{d}) \,d\vec{\Theta} \,, \nonumber\\
    & \propto \prod_{j=1}^{N} \int\bigg[\mathcal{\pi}(\Vec{\theta}_j,\delta\hat{\varphi}_i) \mathcal{L}(d_{j}|\Vec{\theta}_j, \delta\hat{\varphi}_i) \,d\Vec{\theta}_j\bigg]\,,
\end{align}
where $d\Vec{\Theta} = d\Vec{\theta}_{1} d\Vec{\theta}_{2} \cdots d\Vec{\theta}_{N}$. Equation~\eqref{dphi_posterior} is only valid if the $\delta\hat{\varphi}_i$ have the same value for all $N$ events. If the $\delta\hat{\varphi}_i$ do not share common values across events but are drawn from a common distribution for all the events, then the population distribution of the $\delta\hat{\varphi}_i$ is described by a more general {\it hierarchical} Bayesian inference discussed next.

\subsection{Hierarchical Bayesian inference}\label{hierarchical inference}
In the hierarchical method~\cite{Zimmerman:2019wzo,Isi:2019asy,Isi:2022cii}, the TGR parameters $\delta\hat{\varphi}_i$ are assumed to follow an underlying distribution. In a particular modified theory of gravity, the various $\delta\hat{\varphi}_i$ will be functions of the system masses and spins; so the values of the $\delta\hat{\varphi}_i$ will vary from system to system. Not knowing this functional dependence for a generic modification of GR, the $\delta\hat{\varphi}_i$ parameters are approximated as following a Gaussian distribution that is characterized by the hyperparameters $\mu_i$ and $\sigma_i$; these correspond to the mean and standard deviation computed from $N$ measurements of the $\delta \hat{\varphi}_i$. In principle, the true probability distribution of the $\delta\hat{\varphi}_i$ might be more complicated than a Gaussian and might require, apart from $\mu_i$ and $\sigma_i$, the measurement of higher moments. But, for simplicity, we assume the underlying distribution to be a Gaussian. If the GW signal is consistent with GR, the values of $\mu_i$ and $\sigma_i$ should be consistent with zero. 

Given the data $\bm {d} = \{d_1, d_2, \cdots d_N\}$ from $N$ observations, the joint posterior probability on $\mu_i$ and $\sigma_i$ can be written as
\begin{equation}\label{posterior2}
    p(\mu_i, \sigma_i | \bm{d}) \propto \pi(\mu_i, \sigma_i) \mathcal{L}(\bm{d}|\mu_i, \sigma_i) \,.
\end{equation}
Assuming no prior knowledge of the underlying distribution (aside from the Gaussian assumption), the prior $\pi(\mu_i, \sigma_i)$ on $\mu_i$ and $\sigma_i$ is taken to be uniform. The prior on $\mu_i$ is symmetric around zero, while the prior for $\sigma_i$ is positive. The priors for each $\mu_i$ and $\sigma_i$ are wide enough to include the range of individual systematic biases. The likelihood function $\mathcal{L}(\bm{d}|\mu_i, \sigma_i)$ is the product of the individual event likelihoods 
\begin{equation}\label{likelihood}
   \mathcal{L}(\bm{d}|\mu_i, \sigma_i) = \prod_{j=1}^{N}\mathcal{L}(d_{j}|\mu_i, \sigma_i)\,.
\end{equation}
Equation~\eqref{likelihood} can further be expanded as
\begin{equation}\label{likelihood_expansion1}
    \mathcal{L}(\bm{d}|\mu_i, \sigma_i) = \prod_{j=1}^{N} \bigg[\int \mathcal{L}(d_j|\delta\hat{\varphi}_i^{j}) p(\delta\hat{\varphi}_i|\mu_i,\sigma_i) \,d\delta\hat{\varphi}_i\bigg]\,.
\end{equation}
The term $\mathcal{L}(d_j|\delta\hat{\varphi}_i^{j})$ is the likelihood of $\delta\hat{\varphi}_i$ for the $j^{\rm th}$ event. Since we assume no prior on the $\delta\hat{\varphi}_i$ for individual events, the likelihood $\mathcal{L}(d_j|\delta\hat{\varphi}_i^{j})$ is the same as the posterior probability $p(\delta\hat{\varphi}_i^{j}|d_j)$. The posterior probability $p(\delta\hat{\varphi}_i^{j}|d_j)$ corresponds to the individual event posteriors for $\delta\hat{\varphi}_i$ which can be obtained by marginalizing over system parameters.

The term $p(\delta\hat{\varphi}_i|\mu_i,\sigma_i)$ in Eq.~\eqref{likelihood_expansion1} is the predicted population distribution of $\delta\hat{\varphi}_i$ across events given the hyperparameters $\mu_i$ and $\sigma_i$. Ideally, the functional form of the population distribution $p(\delta\hat{\varphi}_i|\mu_i,\sigma_i)$ will depend on the theory of gravity considered. We assume $p(\delta\hat{\varphi}_i|\mu_i,\sigma_i)$  follows a Gaussian distribution,
\begin{equation}\label{population gaussian}
    p(\delta\hat{\varphi}_i|\mu_i,\sigma_i) = \mathcal{N}(\mu_i,\sigma_i)\,.
\end{equation}

Obtaining the posterior probability distribution $p(\delta\hat{\varphi}_i^{j}|d_j)$ for a large number of events is computationally expensive and time-consuming. To avoid these difficulties, we assume that the posterior probability $p(\delta\hat{\varphi}_i^{j}|d_j)$ for individual events follows a Gaussian distribution ${\mathcal N}$ with mean $\Tilde{\mu}_i^j$ and spread $\Tilde{\sigma}_i^j$: 
\begin{equation}\label{individual gaussian}
     p(\delta\hat{\varphi}_i^{j}|d_j) = \mathcal{N}(\Tilde{\mu}_i^{j}, \Tilde{\sigma}_i^{j})\,.
\end{equation}
A Gaussian likelihood is a valid assumption for high SNR signals; but for weak signals, it may not be true. We use the Fisher information matrix to calculate the 1$\sigma$ widths $\Tilde{\sigma}_i^{j}$, following the procedure described in Sec.~\ref{statistical uncertainities}. The mean $\Tilde{\mu}_i^{j}$ is fixed by the systematic bias shift from zero. We use the Cutler-Vallisneri formalism to calculate the systematic bias on $\delta\hat{\varphi}_i$ as explained in Sec.~\ref{systematic bias}. Note that we do not add additional scatter to the $\tilde{\mu}_{i}^{j}$ due to different noise realizations. The individual posteriors $\delta\hat{\varphi}_i^{j}$ are built only considering the biases caused by mismodeling. This is equivalent to zero-noise injection and recovery studies. In reality, the ``best-fit" value will not be the ``true value" of the parameter due to the different noise realizations. This simplification will not alter our qualitative conclusions, as the scatter in $\tilde{\mu}_{i}^{j}$ is much greater than what is expected from the additional scatter due to different noise realizations.

The integral in Eq.~\eqref{likelihood_expansion1} can be solved using Monte-Carlo summation: 
\begin{equation}\label{monte carlo}
    \mathcal{L}(\bm{d}|\mu_i, \sigma_i) = \prod_{j=1}^{N} \frac{1}{N_i^{\rm sample}}\sum_{k=1}^{N_i^{\rm sample}} p(\delta\hat{\varphi}_i^{j,k}|\mu_i,\sigma_i)\,,
\end{equation}
where $\delta\hat{\varphi}_i^{j,k}$ refers to the $k^{\rm th}$ posterior sample for $\delta\hat{\varphi}_i$ drawn from the posterior $p(\delta\hat\varphi_i^{j}|d_j)$ for the $j^{\rm th}$ event. For each event, we draw $N_i^{\rm sample}=10000$ samples of $\delta\hat{\varphi}_i^j$.

The joint posterior probability $p(\mu_i, \sigma_i|\bm{d)}$ in Eq.~\eqref{posterior2} can be calculated using Eq.~\eqref{likelihood_expansion1} together with Eq.~\eqref{population gaussian} and Eq.~\eqref{individual gaussian}. The individual posteriors on $\mu_i$ or $\sigma_i$ can be calculated by marginalizing over $\sigma_i$ or $\mu_i$ (respectively). 
The population distribution of $p(\delta\hat{\varphi}_i|\bm{d})$ given data $\bm{d}$ can be reconstructed by marginalizing over the inferred distributions of hyperparameters $\mu_i$, $\sigma_i$:
\begin{equation}
    p(\delta\hat{\varphi}_i|\bm{d}) = \int p(\delta\hat{\varphi}_i|\mu_i,\sigma_i) p(\mu_i,\sigma_i|\bm{d}) \,d\mu_i \,d\sigma_i\,.
\end{equation}

\section{Hierarchical population inference of the TGR parameters}\label{population analysis results}
\subsection{Binary black hole population}\label{bbh population}
\subsubsection{Mass and spin distribution}
We consider a population of BBHs analogous to that observed in GWTC-3. We assume that the masses of the primary BHs in the population follow the {\tt Power Law + Peak} model~\cite{LIGOScientific:2021psn} (a mixture model of power law and Gaussian distributions \cite{Talbot:2017yur}). The model consists of seven parameters: power law slope $\alpha$, mixture fraction $\lambda_{\rm peak}$, minimum ($m_{\rm min}$) and maximum ($m_{\rm max}$) BH masses, mean ($\mu_m$) and standard deviation ($\sigma_m$) of the Gaussian, and a smoothing factor $\delta_{m}$. Defining $\vec{\Lambda}_m = \{\alpha, m_{\rm min}, m_{\rm max}, \delta_m, \mu_m, \sigma_m, \lambda_{\rm peak}\}$,
\begin{equation}
    p(m_1|\vec{\Lambda}_m) = (1-\lambda_{\rm peak}) \mathcal{P} + \lambda_{\rm peak} \mathcal{G}\,,
\end{equation}
where $\mathcal{P}$ is the power law distribution
\begin{equation}
    \mathcal{P} \propto m_1^{-\alpha} S(m_1, m_{\rm min}, \delta_m) \Theta(m_{\rm max}- m_1)\,,
\end{equation}
and $\mathcal{G}$ is the Gaussian distribution
\begin{equation}
    \mathcal{G} \propto \exp\bigg(-\frac{(m_1-\mu_m)^2}{2 \sigma_m^2}\bigg)  S(m_1, m_{\rm min}, \delta_m)\,.
\end{equation}
Here $S(m_1, m_{\rm min}, \delta_{m})$ is a smoothing function that rises from $0$ to $1$ as mass increases from $m_{\rm min}$ to $m_{\rm min} + \delta_m$, 
\begin{equation}
    S(m_1, m_{\rm min}, \delta_m) = 
    \frac{1}{1+e^{f(m-m_{\rm min}, \delta_m)}} \,,
\end{equation}
where
\begin{equation}
    f(m^{\prime}, \delta_m) = \frac{\delta_{m}}{m^{\prime}} - \frac{\delta_m}{m^{\prime}- \delta_m}\,.
\end{equation}
We take the values of the $\vec{\Lambda}_m$ parameters to be the median values for the GWTC-3 population~\cite{LIGOScientific:2021psn}: $\lambda_{\rm peak}=0.04$, $\alpha=3.4$, $m_{\rm min} = 5.08$, $m_{\rm max}=86.85$, $\mu_m=33.73$, $\sigma_m = 3.56$, $\delta_m=4.83$. 

The mass ratio of BBHs in the population is assumed to follow a power law distribution,
\begin{equation}
        p(q|\beta, m_1, m_{\rm min}, \delta_m) \propto q^{\beta}  S(q m_1, m_{\rm min}, \delta_m) \,.
\end{equation}
The value of $\beta$ is taken to be $1.08$. The spin directions of both BHs are assumed to be aligned with the orbital angular momentum of the binary.
The spin magnitudes are drawn from the {\tt Default spin model} distribution~\cite{Talbot:2017yur,LIGOScientific:2021psn}. In that model, the magnitudes of the two dimensionless spin parameters $\chi_{1,2}$ are drawn from a {\tt Beta distribution}:
\begin{equation}
    p(\chi_{1,2}|\alpha_\chi, \beta_\chi) \propto \chi_{1,2}^{\alpha_\chi -1} (1-\chi_{1,2})^{\beta_\chi-1}\,,
\end{equation}
where $\alpha_\chi$ and $\beta_\chi$ are the shape parameters that determine the mean and variance of the distribution. The values of these parameters are restricted to $\alpha_\chi>1$ and $\beta_\chi>1$, to ensure nonsingular component spin distributions. Beta distributions are generally convenient distributions for the range $[0,1]$, which is the allowed parameter range for BH spins. The values are $\alpha_\chi= 1.6$, $\beta_\chi=4.11$.

%%%%%%%%%%%%%%%%%%%%%%%%%%%%%%%%%%%%%%%%%%%%%%%%%%%%%%%%%%%%%%%%%%%%%
\begin{figure}
    \centering
    \includegraphics[width=0.49\textwidth]{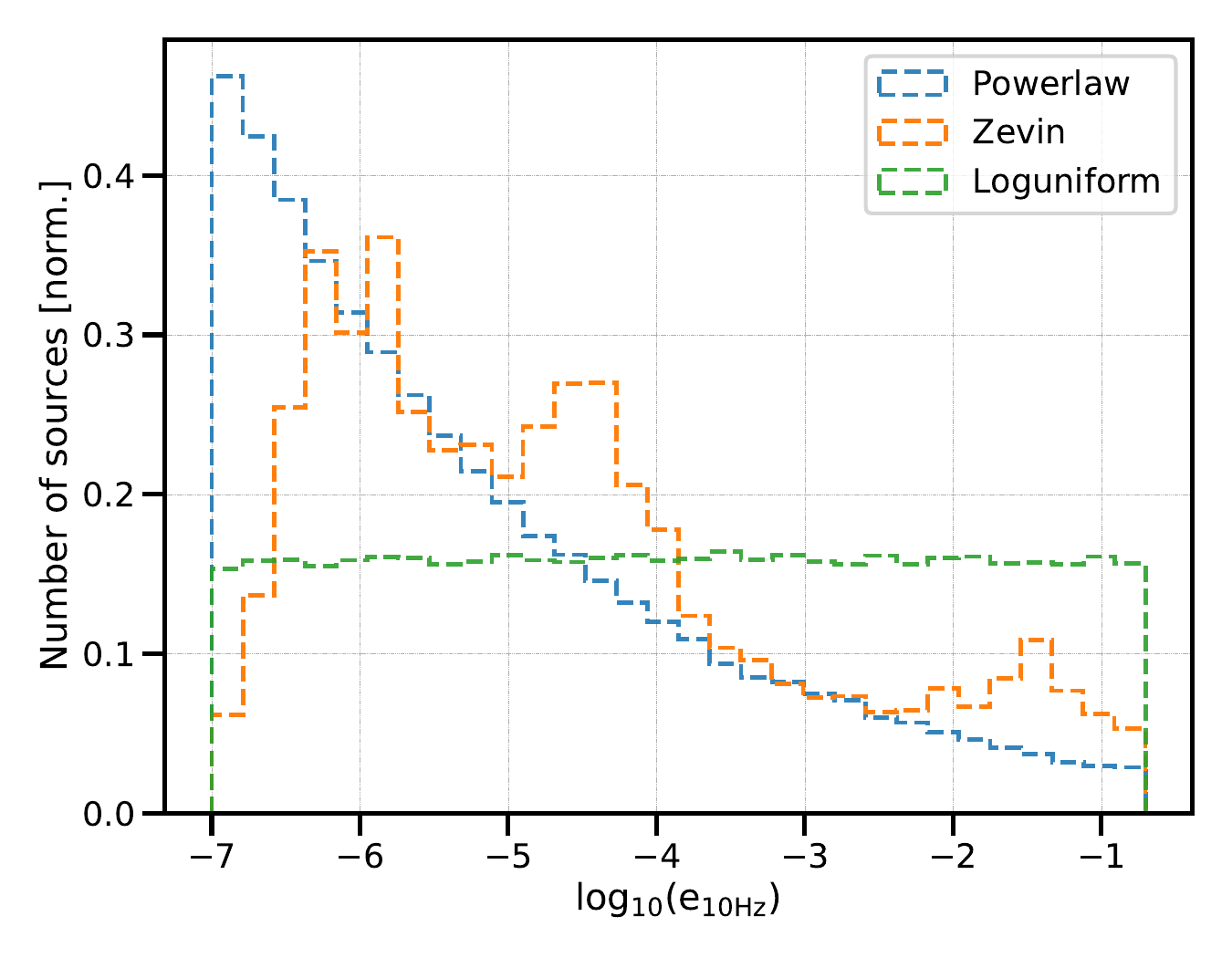}
    \caption{Eccentricity distributions for the BBH populations considered in this study. The {\tt Powerlaw} distribution assumes a PDF with distribution $e_0^{-\alpha}$ with power law index $\alpha=1.2$. The {\tt Loguniform} ($\alpha=-1$) distribution has the largest number of sources with higher eccentricities $e_0>10^{-3}$ (measurable in CE), followed by the {\tt Zevin}~\cite{Zevin:2021rtf} and  {\tt Powerlaw} distributions.}
    \label{eccentricity_distribution}
\end{figure}
%%%%%%%%%%%%%%%%%%%%%%%%%%%%%%%%%%%%%%%%%%%%%%%%%%%%%%%%%%%%%%%%%%%%%

\subsubsection{\label{subsec:ecc}Eccentricity distribution}\label{eccentricity distribution}
At design sensitivity LIGO is expected to detect $\sim 220$-$360$ BBHs per year~\cite{Baibhav:2019gxm, Borhanian:2022czq}. 
Most of the detected BBHs are expected to have very small eccentricities at $10$ Hz. However, a handful may still possess small to moderate eccentricities. The future third-generation detector CE is expected to detect a very large number of BBHs, $\sim 10^{4}\mbox{--}10^{5}$ per year~\cite{Baibhav:2019gxm,Borhanian:2022czq}. Additionally, CE is more sensitive at lower frequencies, where binaries can retain larger eccentricities. Hence, more eccentric BBHs are likely to be detected by CE.

Applying the minimum-information assumption to the underlying astrophysical eccentricity distribution, we draw the eccentricity samples from one of the following eccentricity distributions (shown in Fig.~\ref{eccentricity_distribution}):
\begin{enumerate}
    \item {\tt Loguniform}: The probability density function (PDF) of the eccentricity is proportional to $e_0^{-1}$. The range of the distribution is $e_0\in[10^{-7},0.2]$. \footnote{The upper limit of the eccentricity distribution is restricted to $e_0\leq0.2$ because our waveform model becomes less accurate for $e_0>0.2$.}
    \item {\tt Zevin}: An expected astrophysically motivated eccentricity distribution based on cluster simulations is given in Fig.~1 of Zevin $et$ al.~\cite{Zevin:2021rtf}. Note that Fig.~1 of Ref.~\cite{Zevin:2021rtf} contains two eccentricity distributions (at $10$ Hz) plotted by dashed and solid lines. We use the eccentricity distribution shown by the dashed lines. It represents the intrinsic eccentricity distribution of the considered population. We will refer to this as the {\tt Zevin} eccentricity distribution from here onward. Similar to the {\tt Loguniform} and {\tt Powerlaw} distributions, we restrict the {\tt Zevin} distribution to the range $e_0\in[10^{-7},0.2]$. Note that this distribution is based on cluster simulations that are subject to multiple assumptions. We draw the eccentricity samples from the marginalized one-dimensional eccentricity distribution and do not account for the correlations between eccentricity and other source parameters, assuming them to be small. Note that the ratio of systematic to statistical errors strongly depends on the eccentricity value and is weakly dependent on other binary parameters [see, e.g., Eq.~\eqref{eq:ratio}]. Therefore, the assumption of small correlations (if any) is unlikely to affect our results significantly.
    \item {\tt Powerlaw}: The PDF is proportional to $e_0^{-\alpha}$ with the eccentricity range $e_0\in[10^{-7},0.2]$. We choose $\alpha=1.2$; this produces a power law model that (i) does not deviate significantly from the other two and (ii) preserves at least a few high eccentricity sources (with $e_0>10^{-3}$). 

\end{enumerate}

\subsubsection{Redshift distribution}
Sources are distributed uniformly in comoving volume according to the following redshift distribution:
\begin{equation}
    p(z) \propto \frac{1}{1+z} \frac{dV_c}{dz}\,,
\end{equation}
where $\frac{dV_c}{dz}$ is the comoving volume element at redshift $z$. The factor $(1+z)^{-1}$ converts the detector frame time to the source frame time. The observed population of BBHs shows evidence for the evolution of the merger rate with redshift~\cite{LIGOScientific:2021psn}. However, those results are valid for $z\lesssim1$; in our study we distribute sources up to $z=2.5$, where those estimates may not apply. The choice of redshift distribution is likely to have a negligible effect on our analysis.

\subsection{Analysis} 
We draw a population of $10^5$ BBHs from the {\tt Powerlaw + Peak} distribution, the {\tt Beta} distribution, and one of the three eccentricity distributions: {\tt Loguniform}, {\tt Powerlaw}, {\tt Zevin}. These BBHs are distributed uniformly in a comoving volume up to redshift $z=2.5$. We calculate the SNR for these sources using the LIGO design sensitivity curve. These sources have SNR $\sim\mathcal{O}(1\mbox{--}50)$ in the LIGO band. Since LIGO and CE at design sensitivities are expected to detect $220-360$ or $8.6\times10^4 - 5.4\times10^5$ BBHs per year (respectively)~\cite{Baibhav:2019gxm}, we set an SNR threshold of $12$ as a detection criterion to ensure that we are in a high SNR regime where Fisher matrix estimates are expected to be reasonably accurate. \footnote{We consider single detector configurations throughout this paper as bounds on GR deviations (of the type discussed here) scale straightforwardly with the number of detectors.} Out of $10^5$ BBHs, $458$ sources pass this SNR criterion for LIGO. Since we introduce TGR parameters in the inspiral part of the waveform, we select only inspiral-dominated events that have sufficient numbers of GW cycles in the detector band. To enforce this, we impose an additional condition that the total mass of each event satisfies $M\leq60 M_{\odot}$. We find that $326$ (out of $458$) sources pass the mass criteria. Only these BBHs take part in the further analysis. We use the same population for CE. In CE, these sources have SNR $\mathcal{O}(400\mbox{--}1500)$. Since CE will detect many more events, its population will be different than LIGO's~\cite{Vitale:2016aso}. However, for ease of comparison between the LIGO and CE posteriors---and to reduce computational costs---we have chosen to analyze the same BBH population for both LIGO and CE.

It is computationally challenging to obtain the posteriors on the individual $\delta\hat{\varphi}_i^j$ using a full Bayesian analysis. Hence, we resort to a semianalytical computation of these statistical and systematic errors (via Fisher matrix methods). Note that the Fisher analysis shows very good agreement with Monte Carlo simulations for low-mass systems with high SNR~\cite{2008CQGra..25r4007C}, but can sometimes overestimate the parameter uncertainties compared to Bayesian estimates for higher-mass systems~\cite{Rodriguez:2013mla}. Therefore, one should in principle use full Bayesian analyses for more robust parameter inference, especially for LIGO systems where SNRs are $\mathcal{O}(12 \mbox{--}50)$. For each source in the BBH population, we calculate the statistical and systematic errors on $\delta\hat{\varphi}_i$ according to the methods in Secs.~\ref{statistical uncertainities} and \ref{systematic bias} for both LIGO and CE. A few outlier sources with very large statistical errors were removed from the analysis. This left $303$ sources in our BBH population. For each source we construct one-dimensional Gaussian posteriors for $\delta\hat{\varphi}_i$, $p(\delta\hat{\varphi}_i^{j}|d_j) = \mathcal{N}(\Tilde{\mu}_i^{j}, \Tilde{\sigma}_i^{j})$. The Fisher matrix determines $\Tilde{\sigma}_i^{j}$, and the Cutler-Vallisneri formalism determines $\Tilde{\mu}_i^{j}$. If there is no systematic bias, the means $\Tilde{\mu}_i^j$ of the individual Gaussians should peak at zero. The systematic bias shifts the means of the individual Gaussians away from the GR values (zero).

Next, our task is to obtain the combined likelihood over these events. To calculate the population level likelihood---which is assumed to be a Gaussian $\mathcal{N}(\mu_i, \sigma_i)$---we use hierarchical Bayesian inference as explained in Sec.~\ref{hierarchical inference}. We use {\tt GWPopulation} \cite{Talbot:2019okv} for the hierarchical Bayesian framework and the {\tt Dynesty} sampler~\cite{speagle2020dynesty} to sample over $\mu_i$ and $\sigma_i$. As mentioned earlier the priors on $\mu_i$ are assumed to be uniform and symmetric around zero, and the priors on $\sigma_i$ are assumed to be uniform and positive.

%%%%%%%%%%%%%%%%%%%%%%%%%%%%%%%%%%%%%%%%%%%%%%%%%%%%%%%%%%%%%%%%%%%%%%%%%%%%%
%%%%%%%%%%%%%%%%%%%%%%%%%%%%%%%%%%%%%%%%%%%%%%%%%%%%%%%%%%%%%%%%%%%%%%%%%%%%%
\begin{figure*}[hpt!]
    \centering
    \begin{subfigure}{\includegraphics[width=0.49\textwidth]{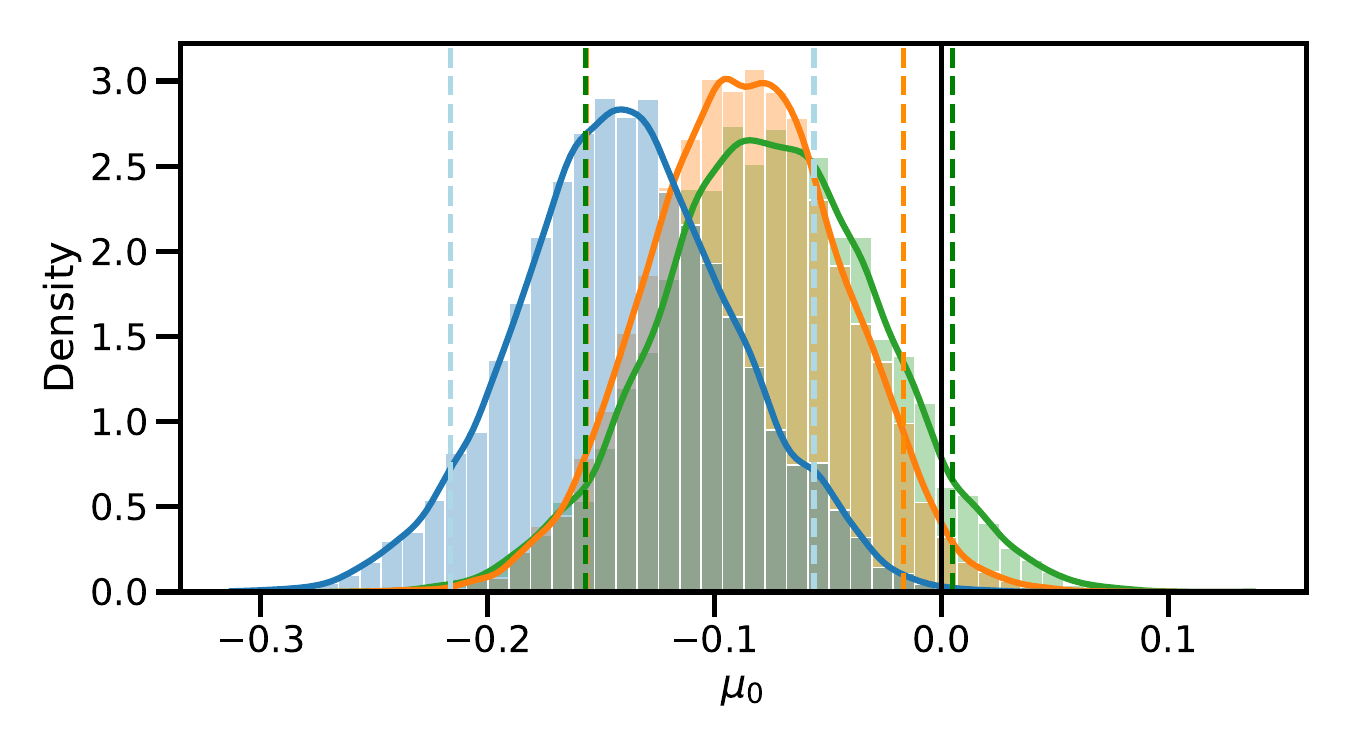}}
    \end{subfigure}
    \vspace{-0.6cm}
    \begin{subfigure}{\includegraphics[width=0.49\textwidth]{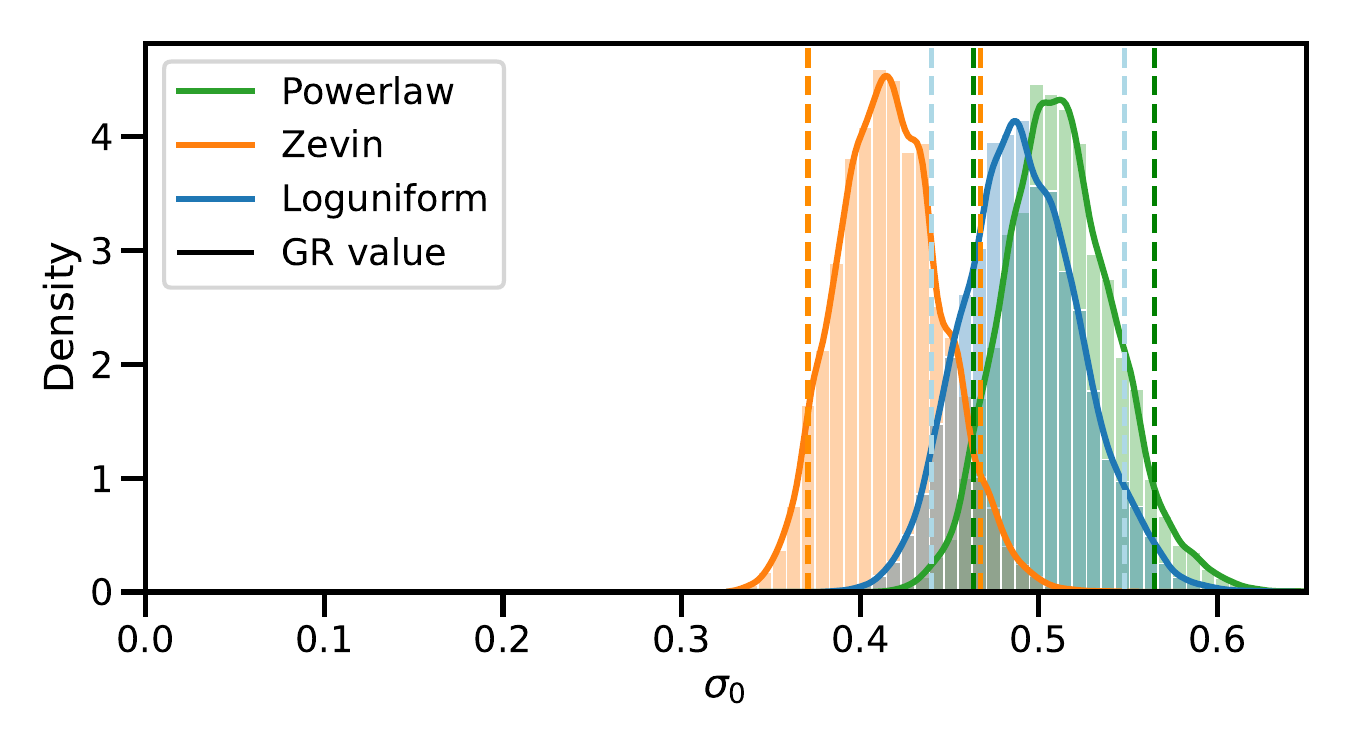}}
    \end{subfigure}

    \begin{subfigure}{\includegraphics[width=0.49\textwidth]{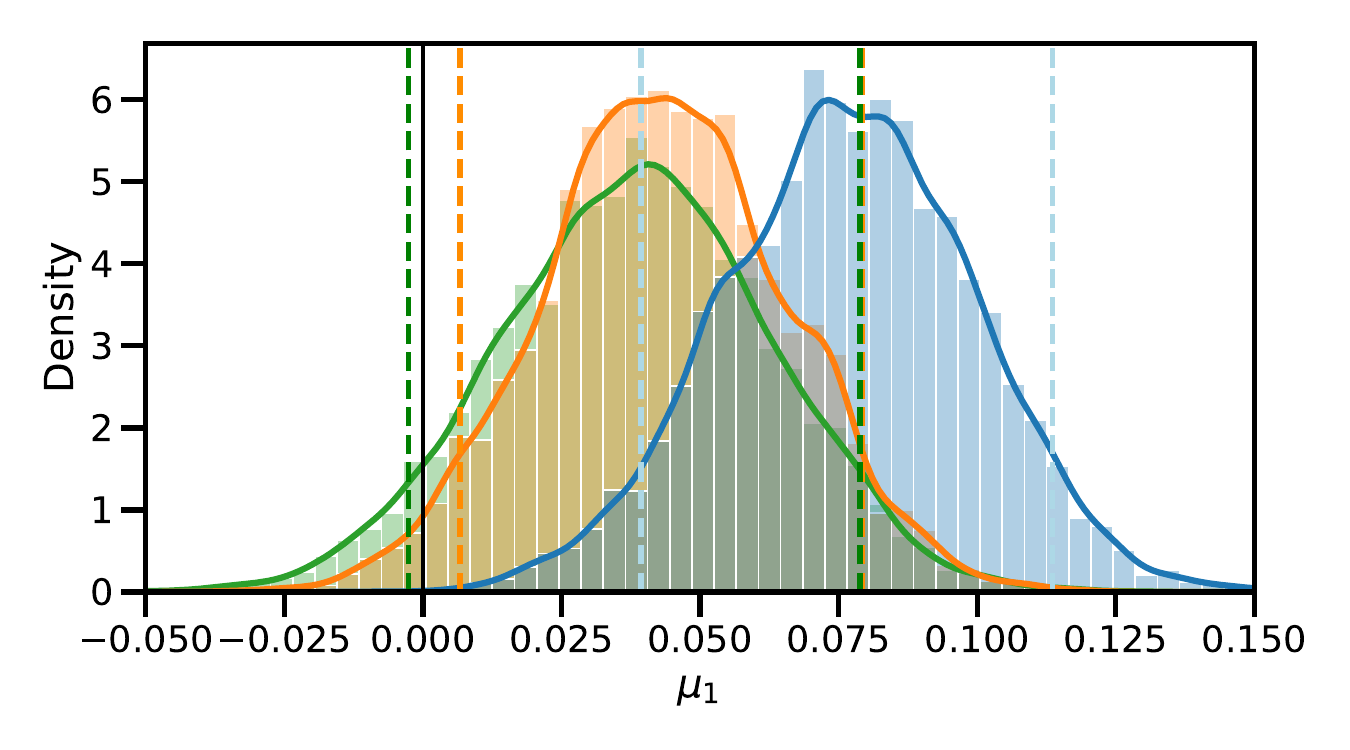}}
    \end{subfigure}
    \vspace{-0.6cm}
    \begin{subfigure}{\includegraphics[width=0.49\textwidth]{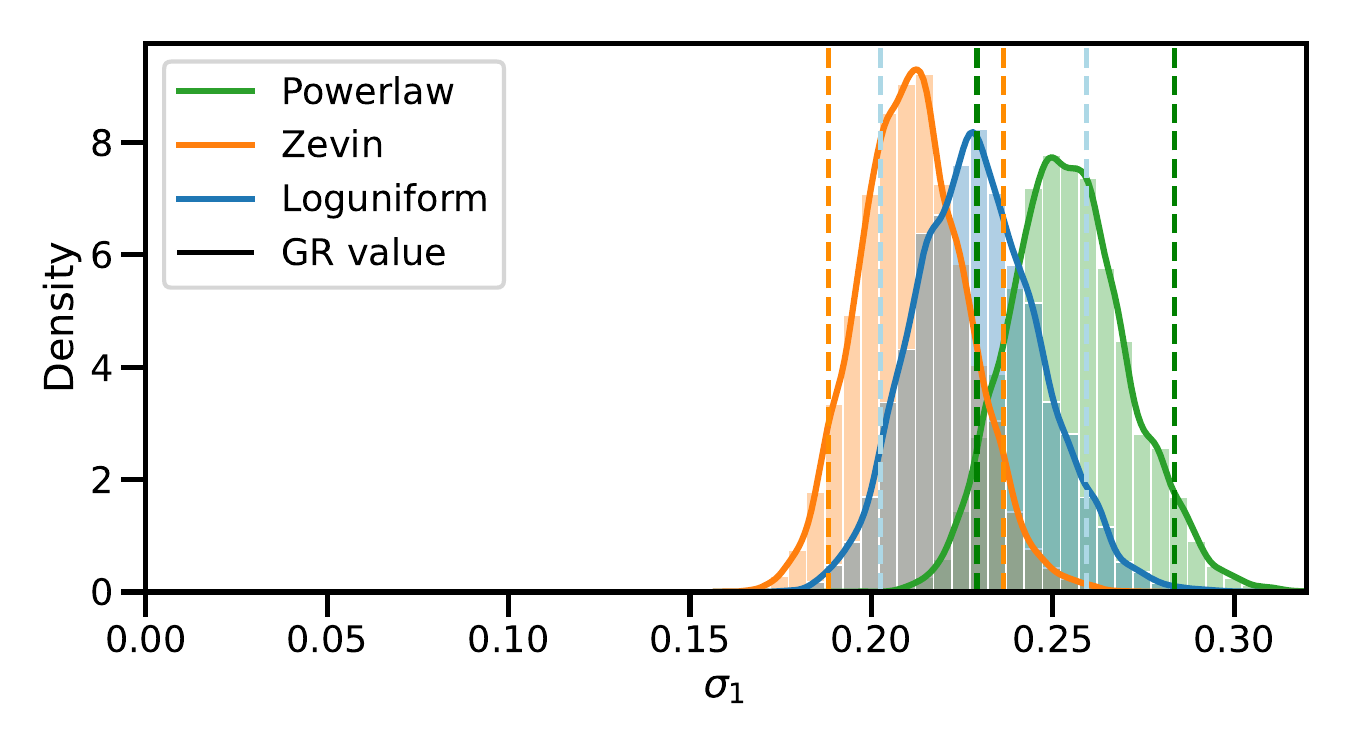}}
    \end{subfigure}

    \begin{subfigure}{\includegraphics[width=0.49\textwidth]{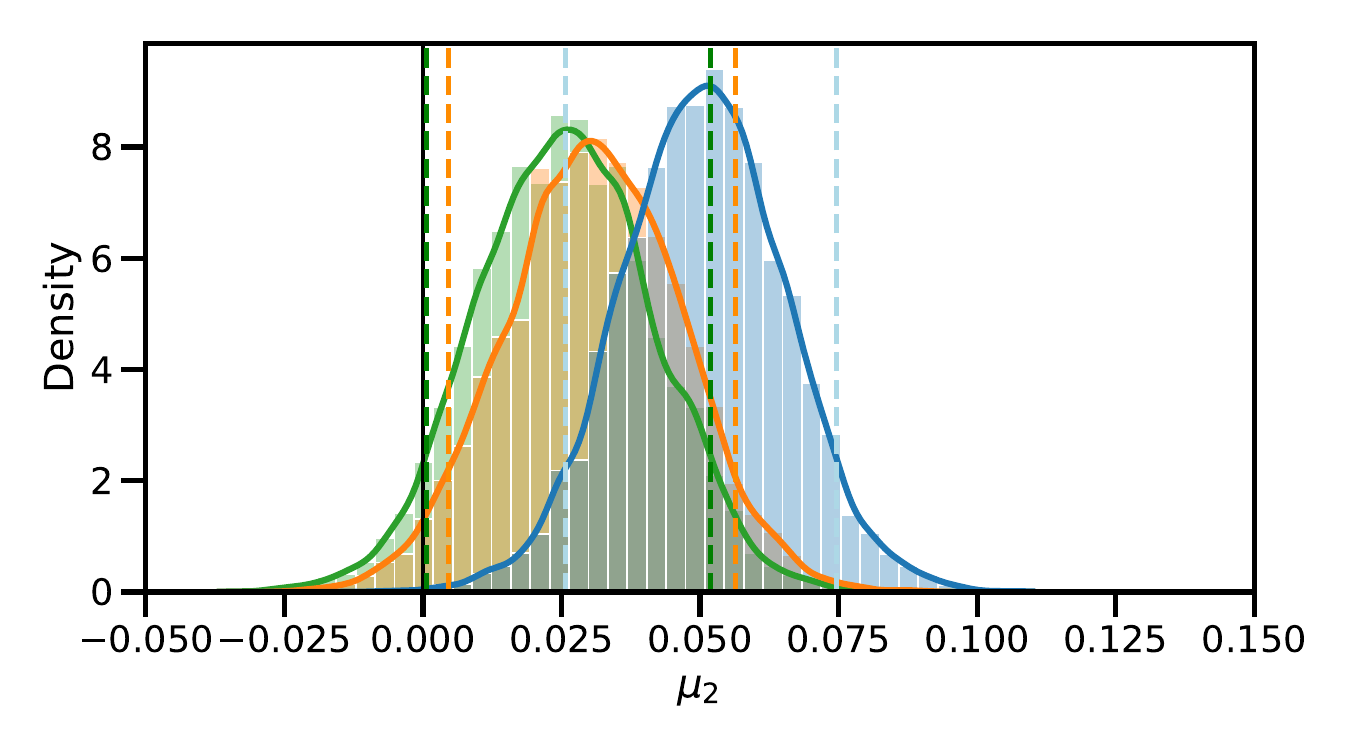}}
    \end{subfigure}
    \vspace{-0.6cm}
    \begin{subfigure}{\includegraphics[width=0.49\textwidth]{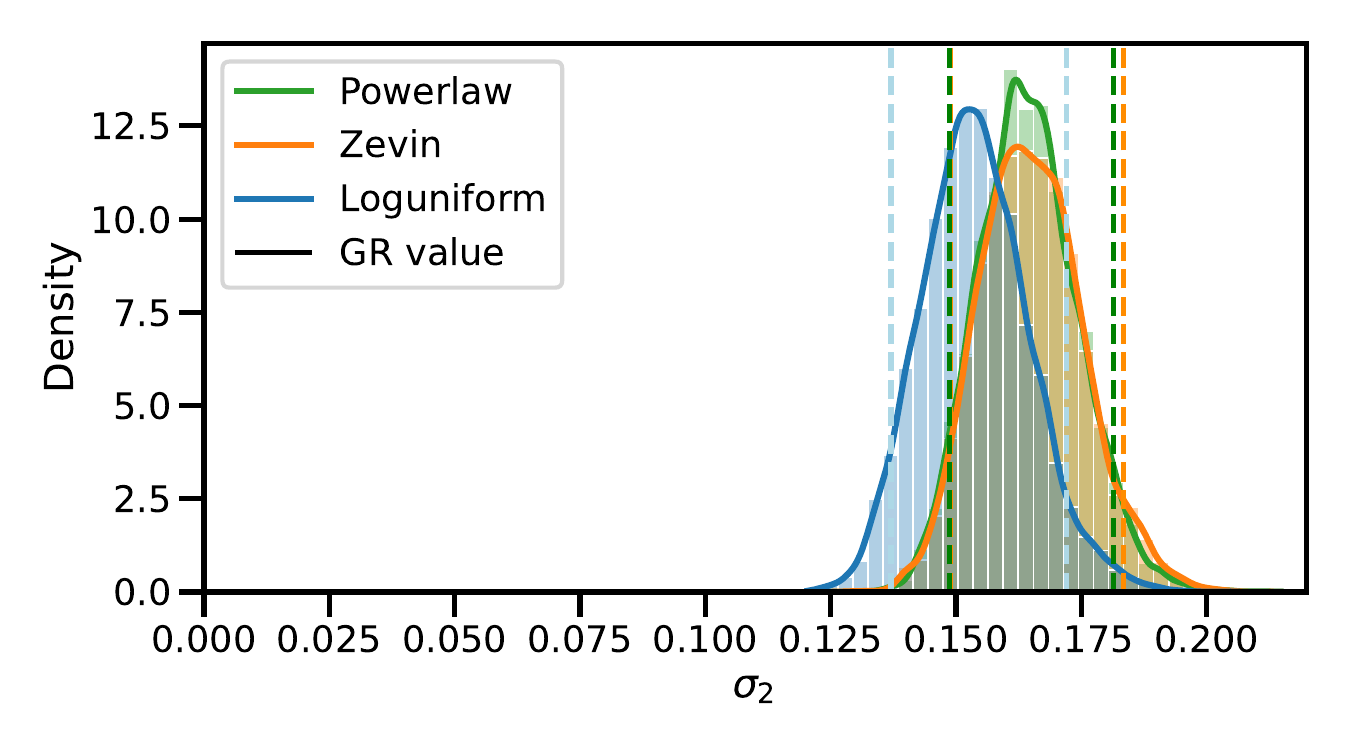}}
    \end{subfigure}

    \begin{subfigure}{\includegraphics[width=0.49\textwidth]{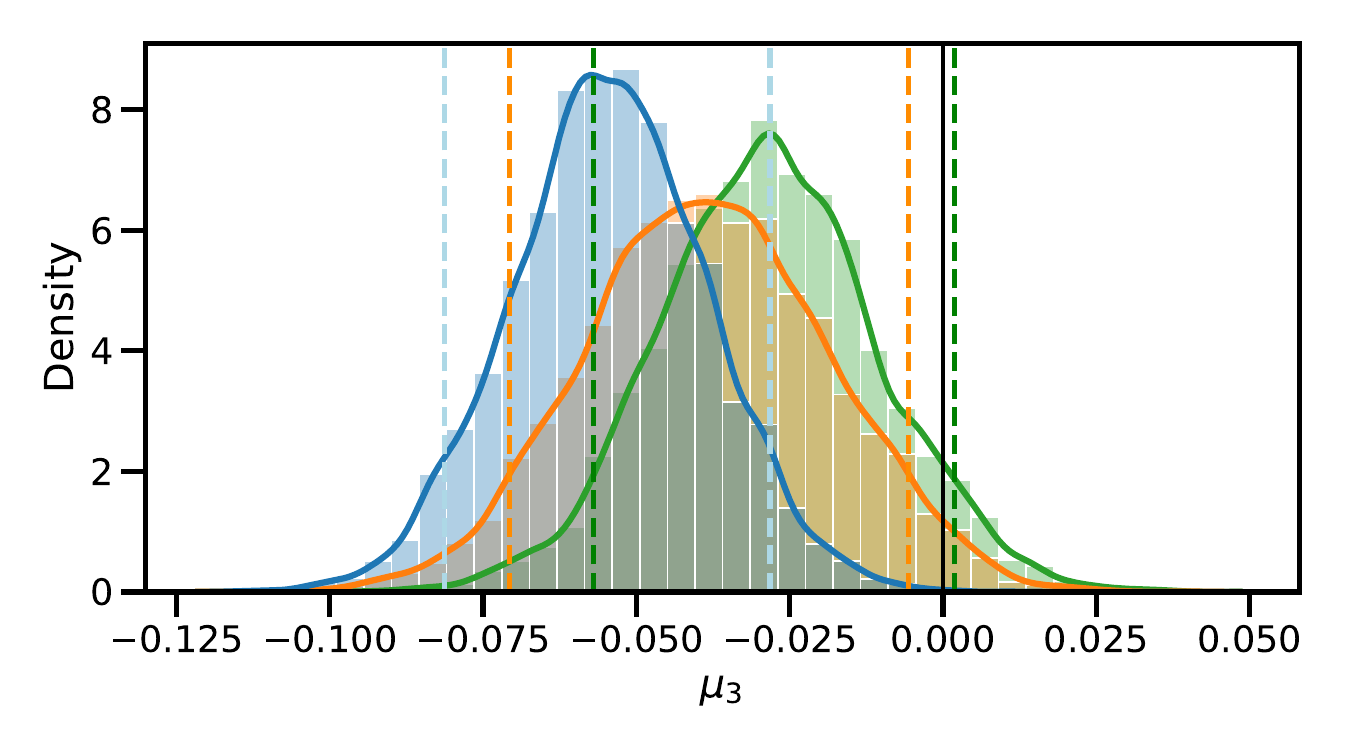}}
    \end{subfigure}
    \vspace{-0.6cm}
    \begin{subfigure}{\includegraphics[width=0.49\textwidth]{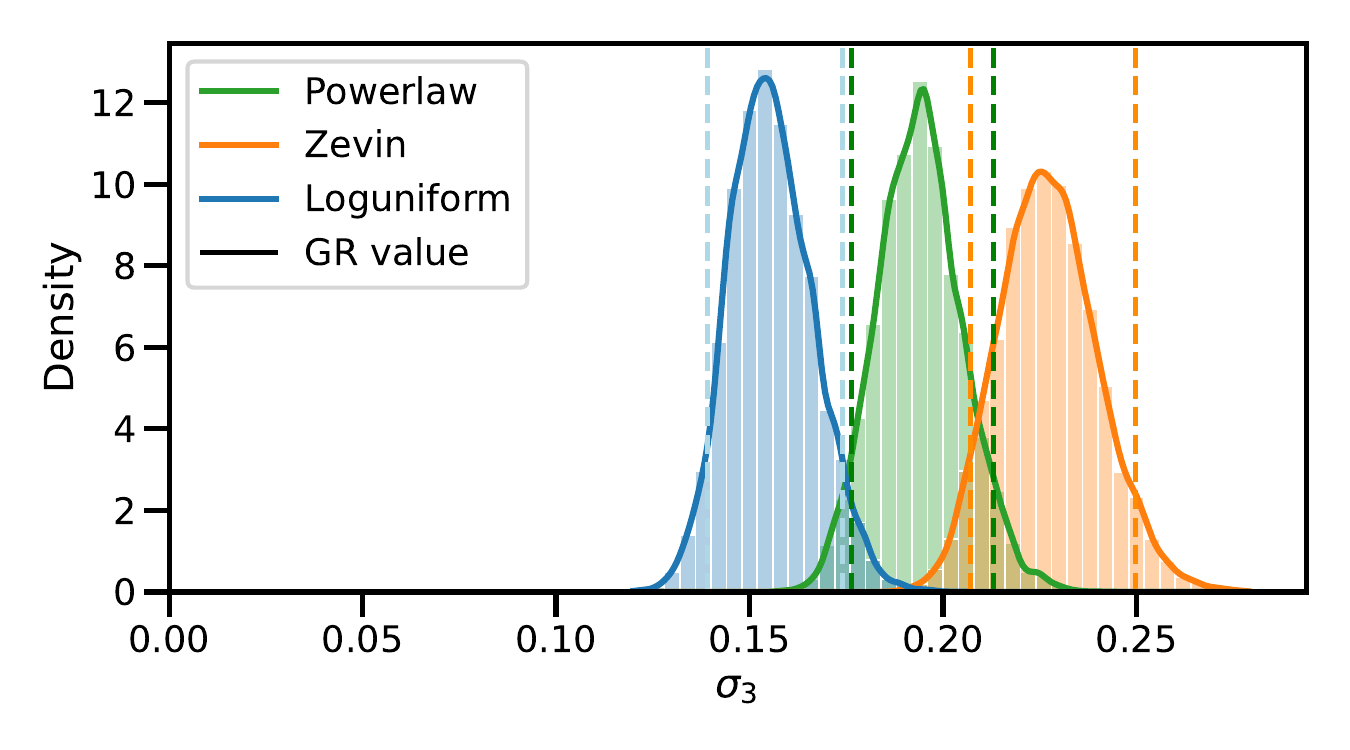}}
    \end{subfigure}

    \begin{subfigure}{\includegraphics[width=0.49\textwidth]{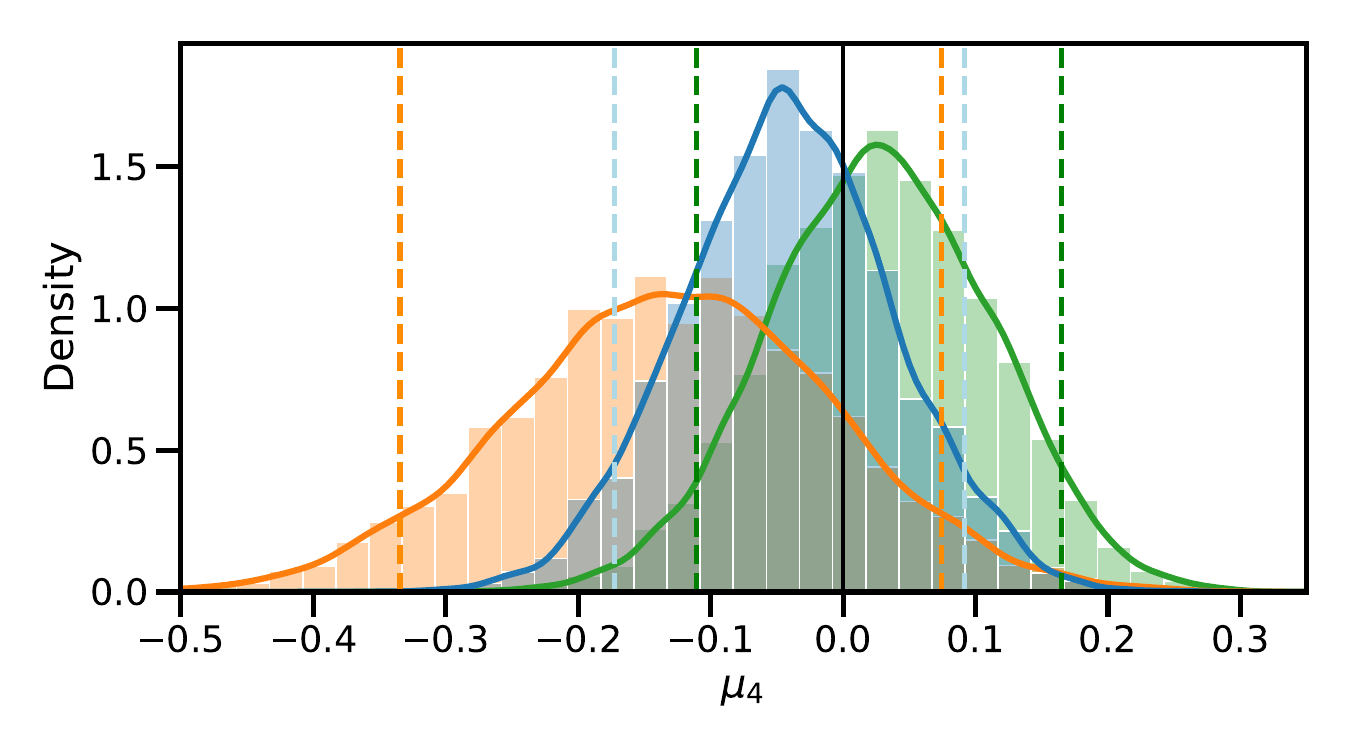}}
    \end{subfigure}
    \vspace{-0.6cm}
    \begin{subfigure}{\includegraphics[width=0.49\textwidth]{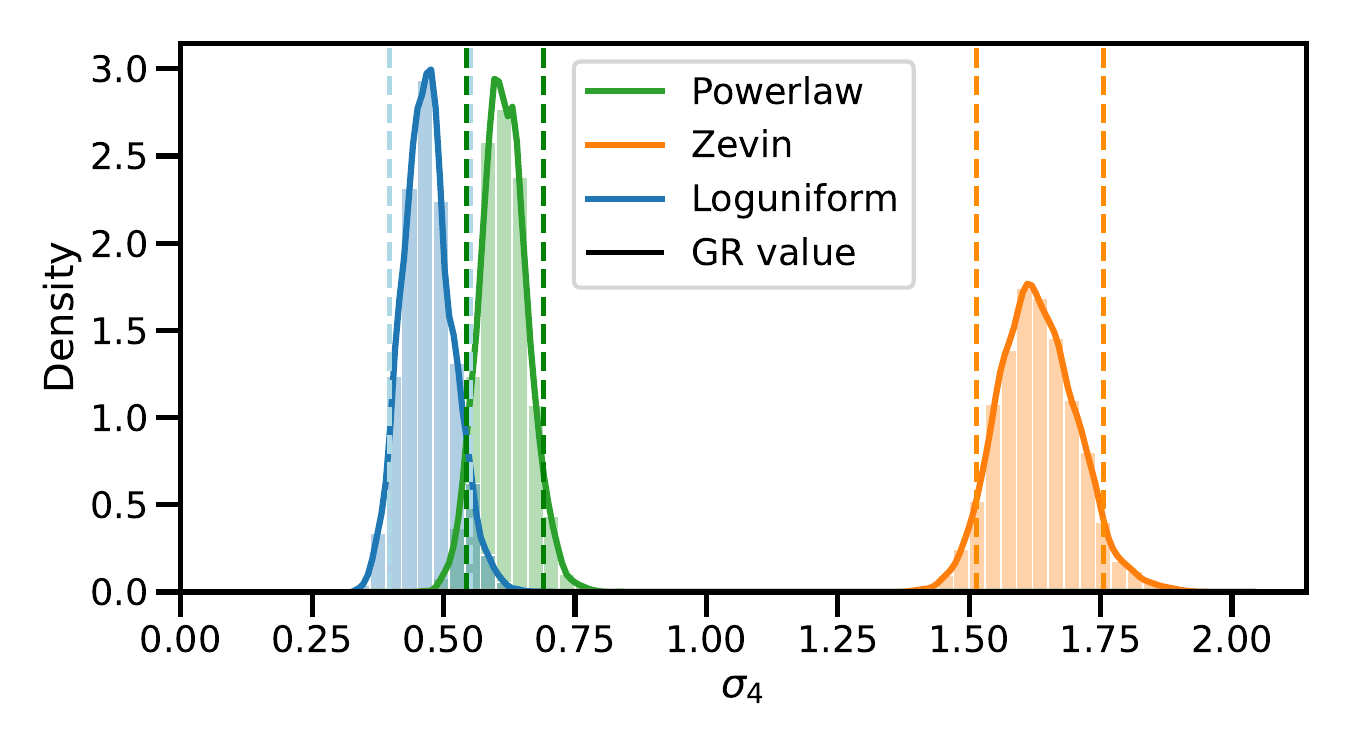}}
    \end{subfigure}
    \caption{Posteriors on the hyperparameters $\mu_i$ (left panels) and $\sigma_i$ (right panels) for the leading-order TGR parameters $\delta\hat{\varphi}_i$ as measured with a single LIGO detector. Panels refer to the 0PN, 1PN, 1.5PN, and 2PN deviation parameters (top to bottom).  The three colors represent the three eccentricity distributions considered in Fig.~\ref{eccentricity_distribution}. The vertical black line indicates the GR value (zero). The dashed (colored) lines represent the $90\%$ credible interval for each posterior. Most posteriors exclude the GR value, indicating a (false) deviation from GR. For each eccentricity distribution we combine data from $303$ sources with $\text{SNR}\geq 12$ and $M\leq 60 M_{\odot}$.}
    \label{ligo1}
\end{figure*}
%%%%%%%%%%%%%%%%%%%%%%%%%%%%%%%%%%%%%%%%%%%%%%%%%%%%%%%%%%%%%%%%%%%%%%%%%%%%%
\begin{figure*}[hpt!]
    \centering
    \begin{subfigure}{\includegraphics[width=0.49\textwidth]{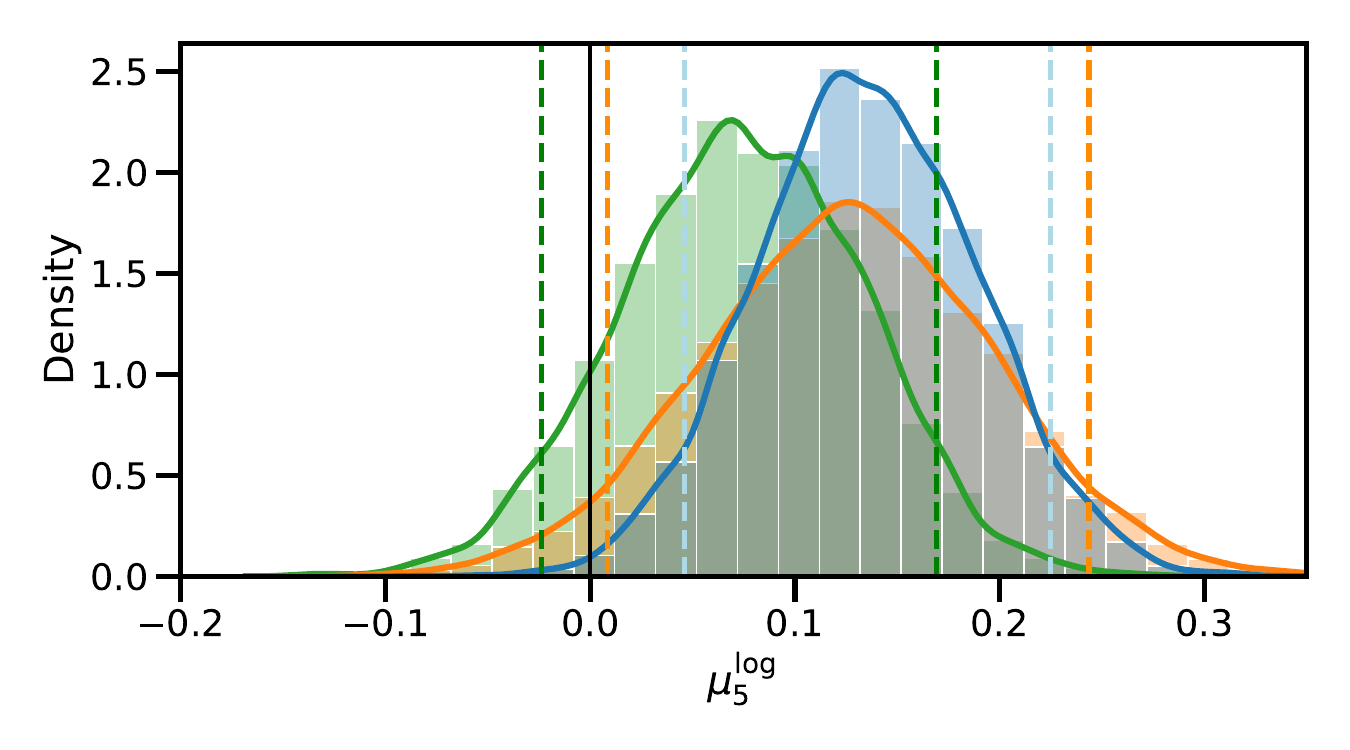}}
    \end{subfigure}
    \vspace{-0.6cm}
    \begin{subfigure}{\includegraphics[width=0.49\textwidth]{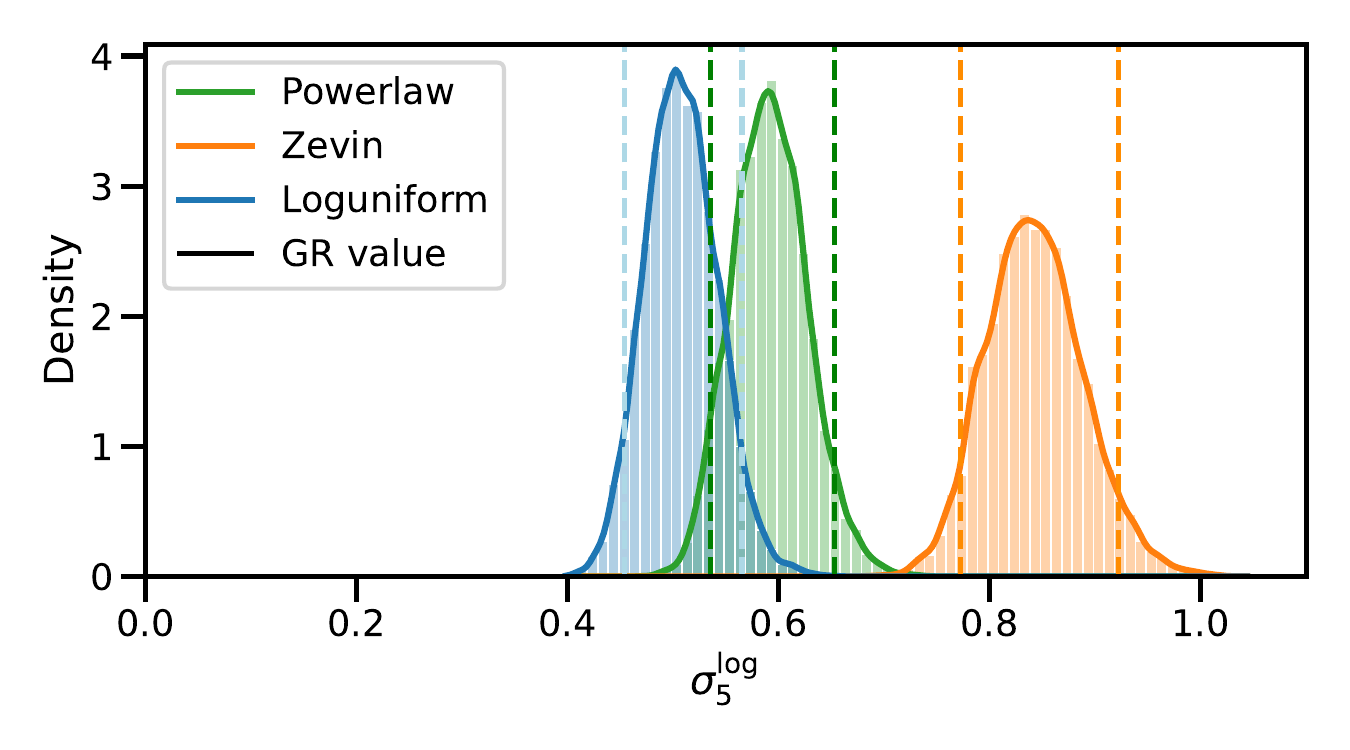}}
    \end{subfigure}
    \begin{subfigure}{\includegraphics[width=0.49\textwidth]{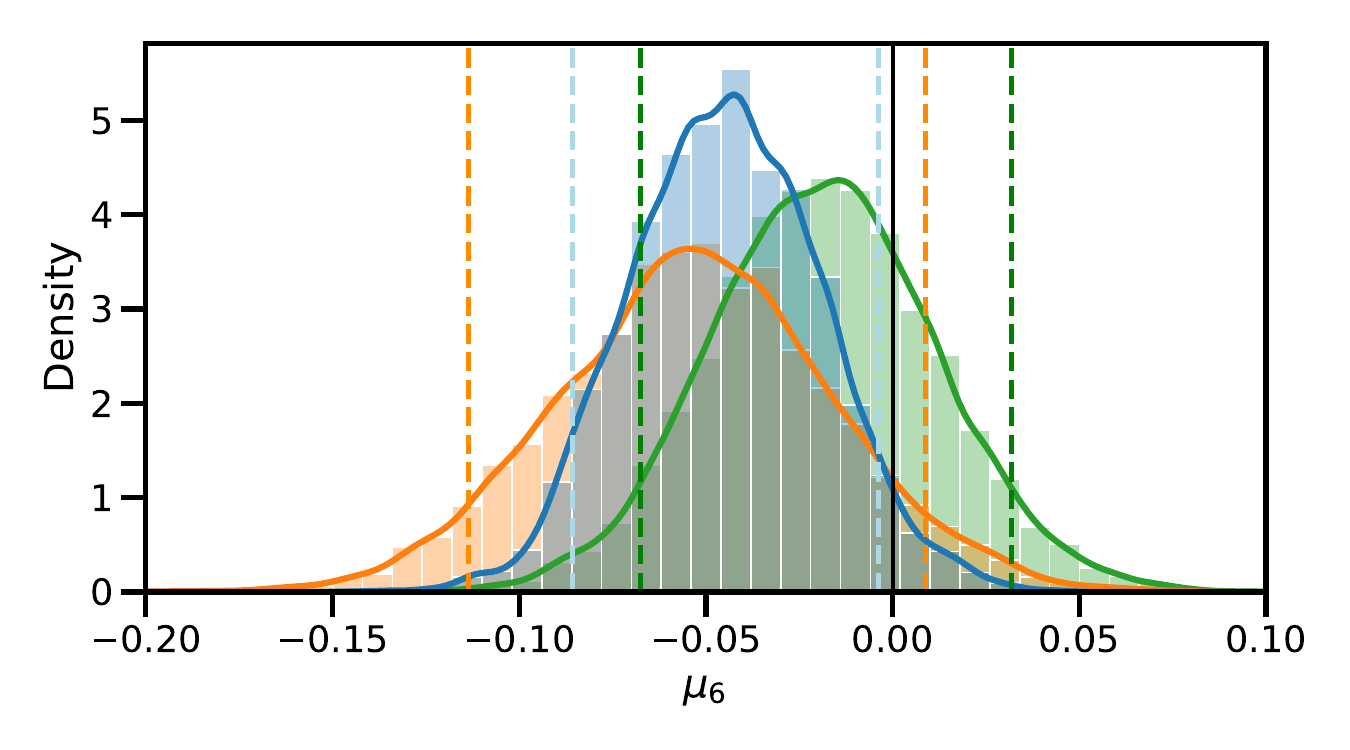}}
    \end{subfigure}
    \vspace{-0.6cm}
    \begin{subfigure}{\includegraphics[width=0.49\textwidth]{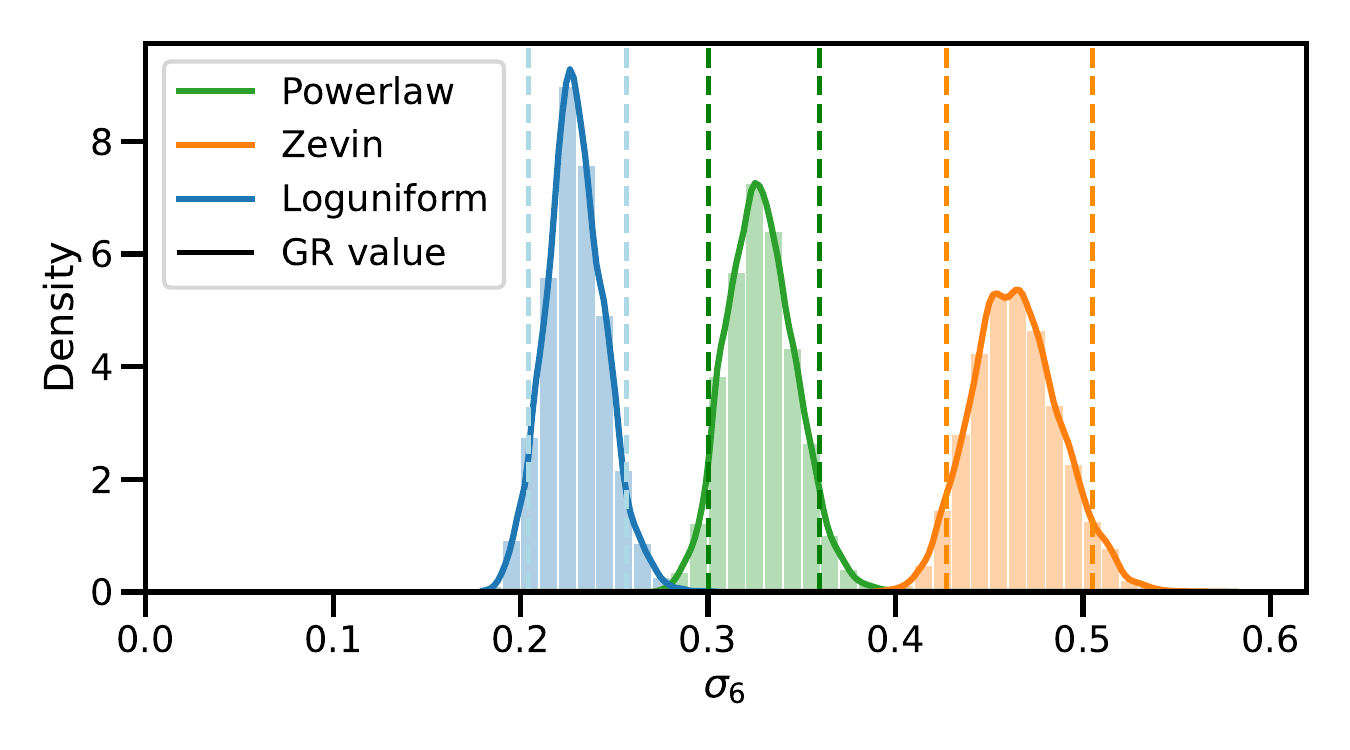}}
    \end{subfigure}
    \begin{subfigure}{\includegraphics[width=0.49\textwidth]{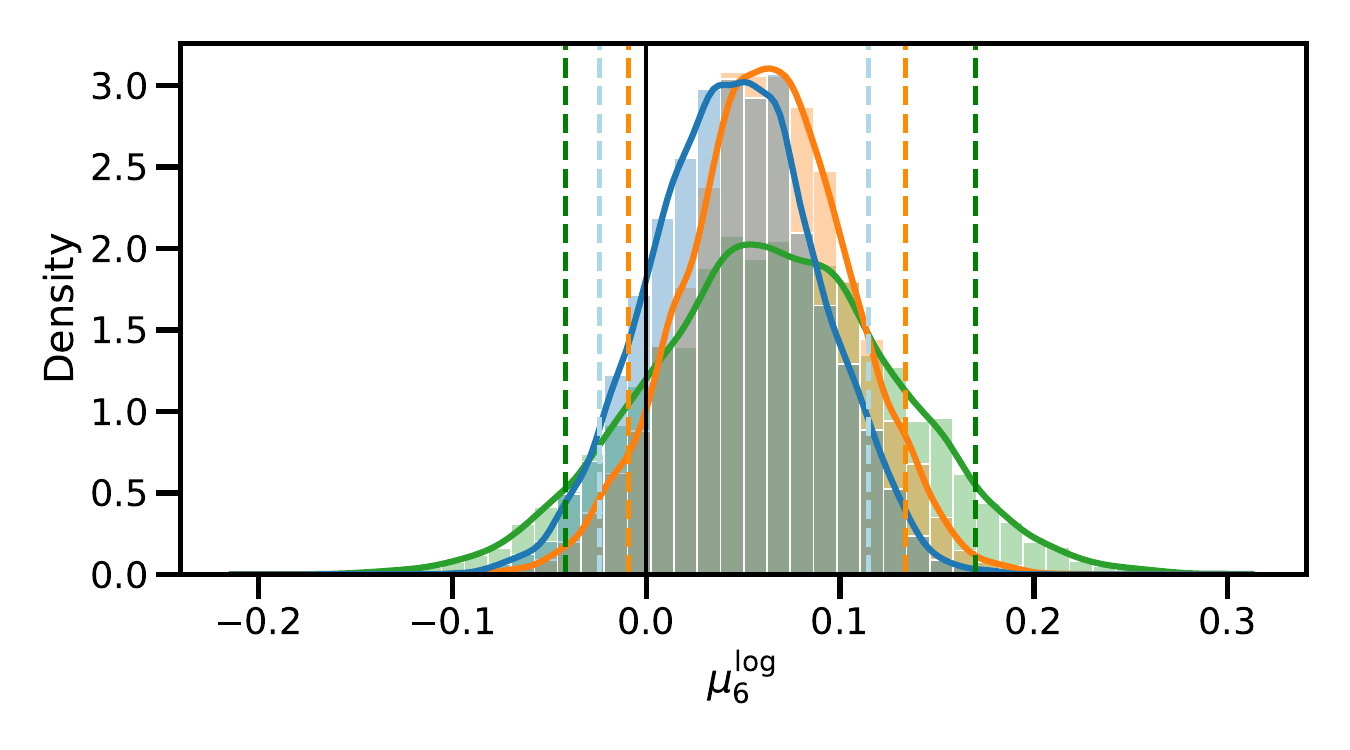}}
    \end{subfigure}
    \vspace{-0.6cm}
    \begin{subfigure}{\includegraphics[width=0.49\textwidth]{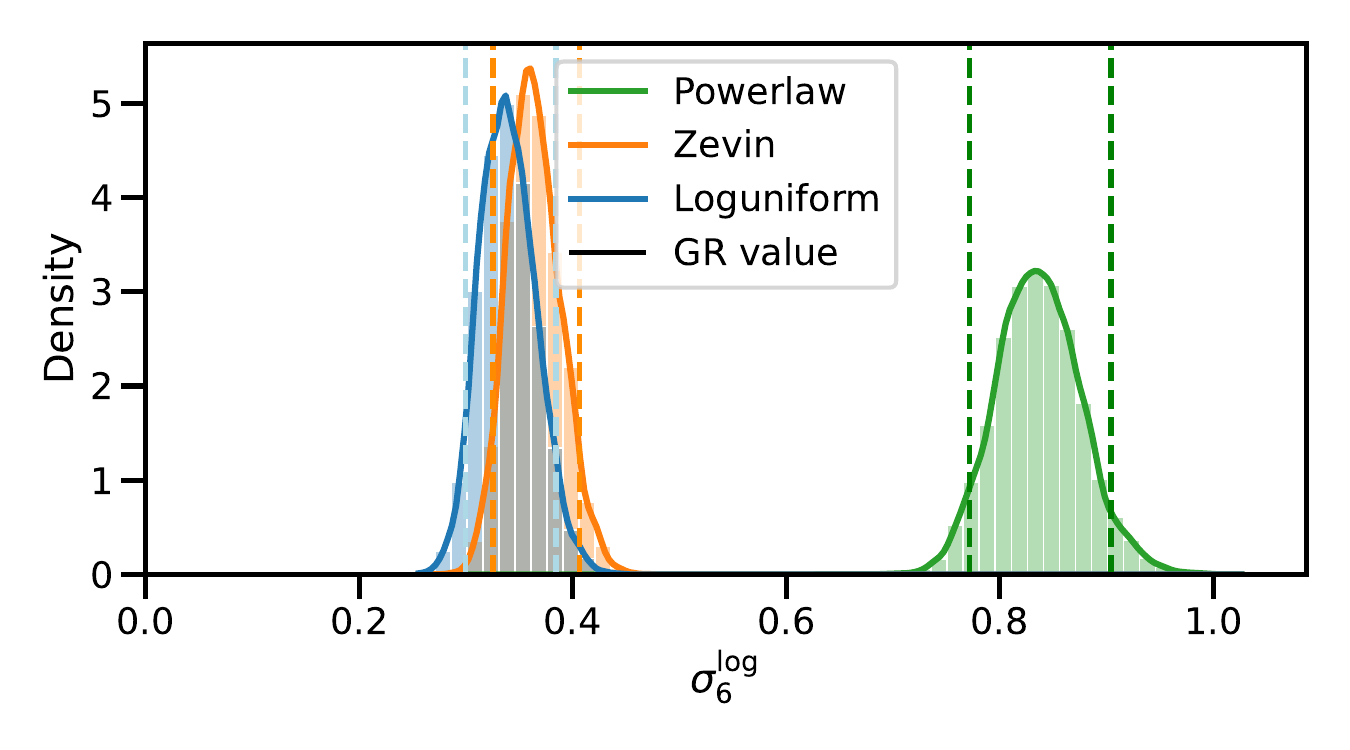}}
    \end{subfigure}
    \begin{subfigure}{\includegraphics[width=0.49\textwidth]{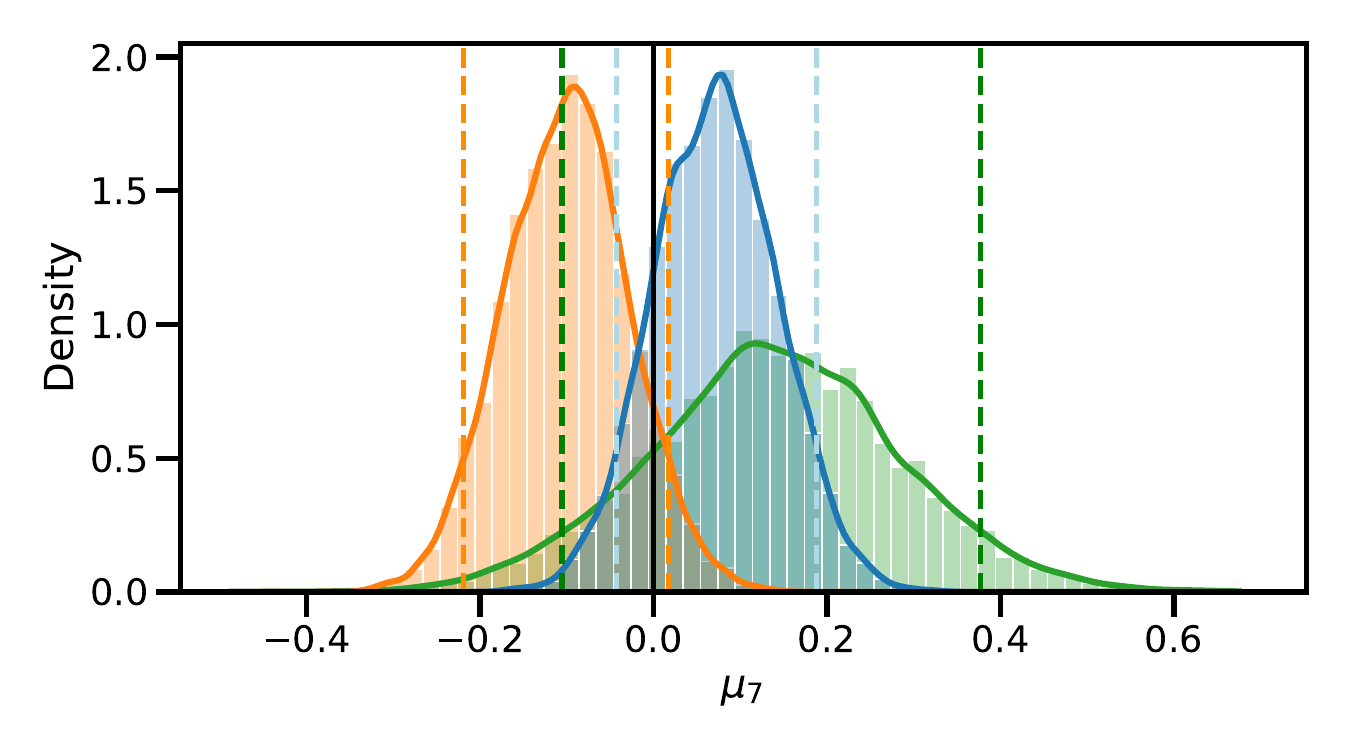}}
    \end{subfigure}
    \vspace{-0.6cm}
     \begin{subfigure}{\includegraphics[width=0.49\textwidth]{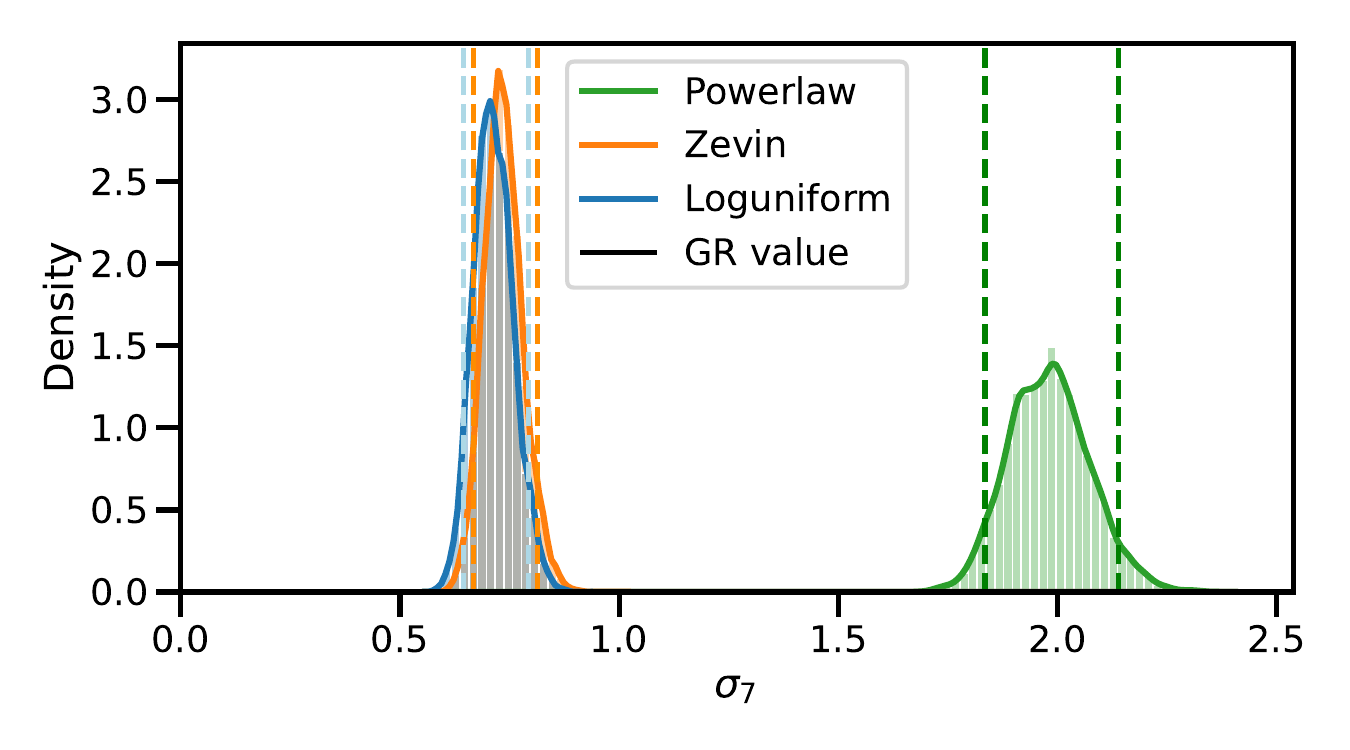}}
    \end{subfigure}
    \begin{subfigure}{\includegraphics[width=0.49\textwidth]{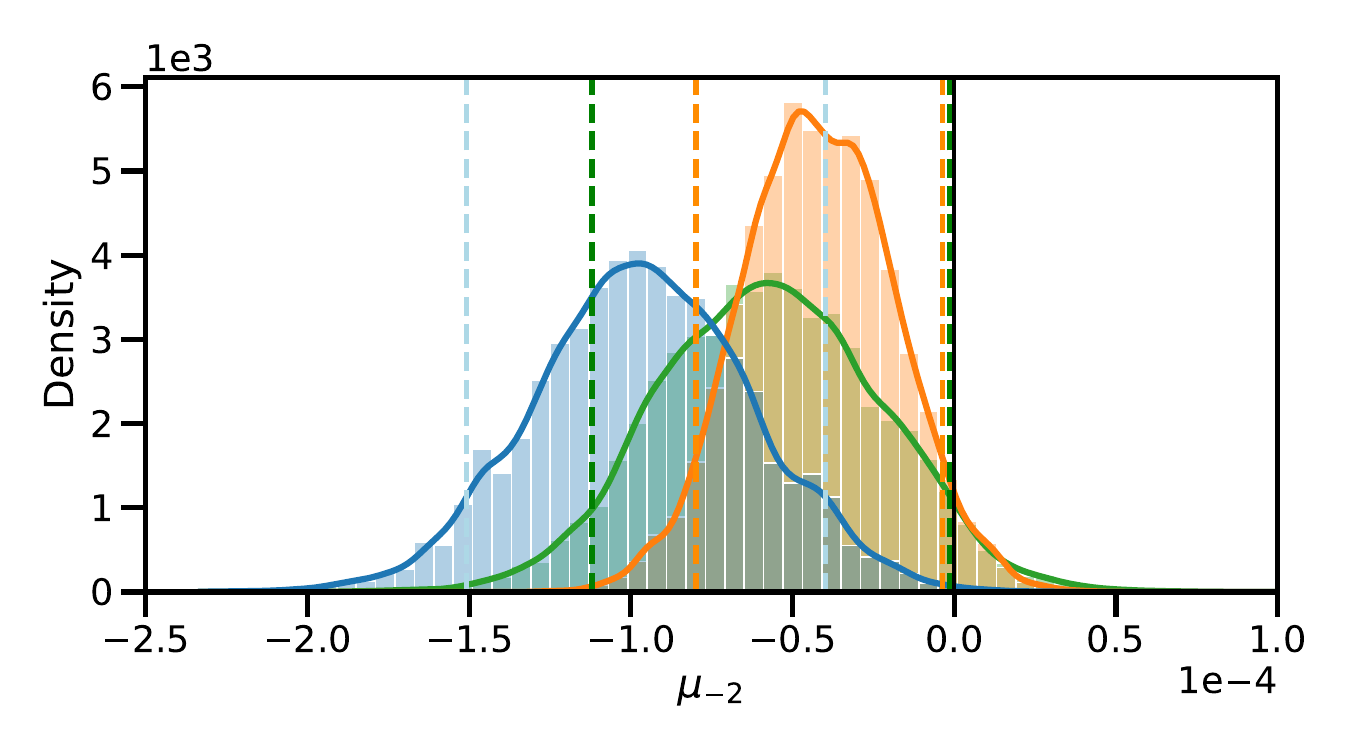}}
    \end{subfigure}
    \vspace{-0.6cm}
    \begin{subfigure}{\includegraphics[width=0.49\textwidth]{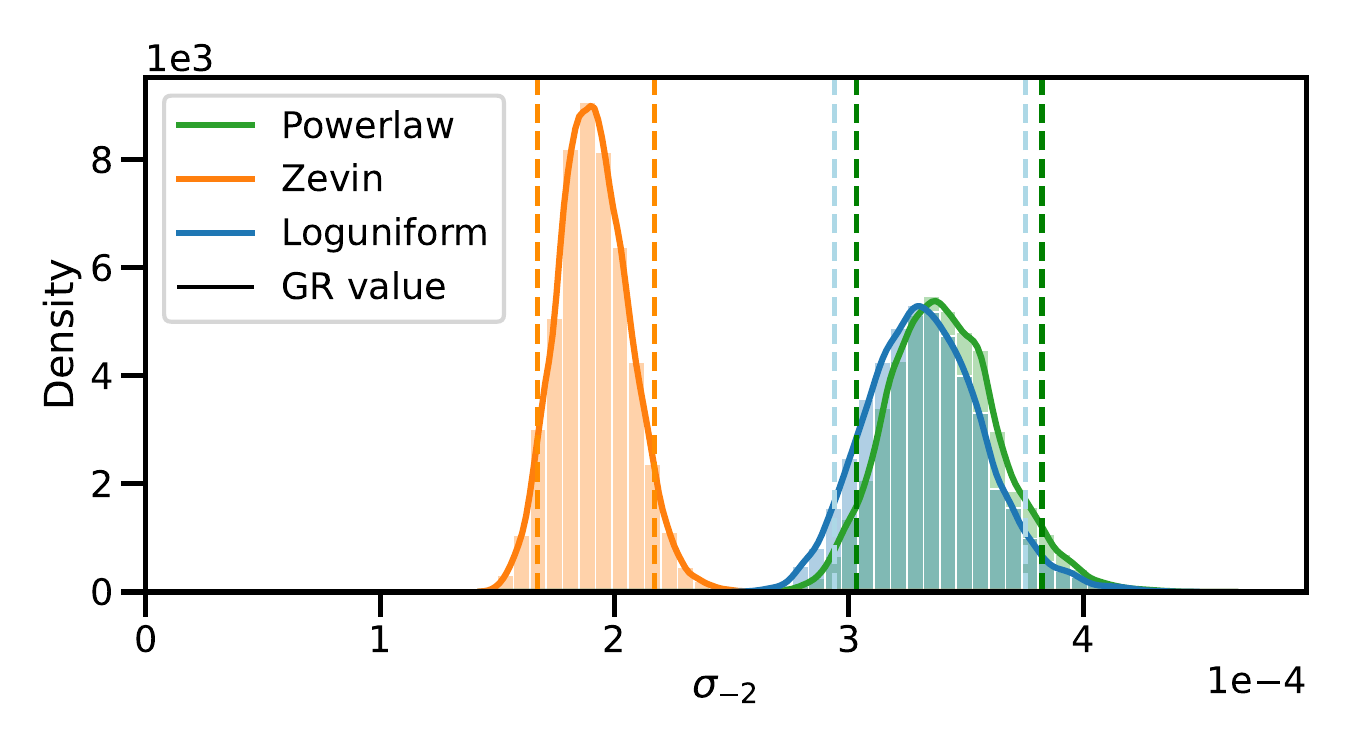}}
    \end{subfigure}
    \caption{Posteriors on hyperparameters $\mu_i$ and $\sigma_i$ for the higher-order TGR parameters $\delta\hat{\varphi}_i$ (2.5PN log, 3PN, 3PN log, 3.5PN) as measured by LIGO. The bottom row shows the -1PN order (dipole) parameter. Conventions are as in Fig.~\ref{ligo1}.}
    \label{ligo2}
\end{figure*}
%%%%%%%%%%%%%%%%%%%%%%%%%%%%%%%%%%%%%%%%%%%%%%%%%%%%%%%%%%%%%%%%%%%%%%%%%%%%%
\begin{figure*}[hpt!]
    \centering
    \begin{subfigure}{\includegraphics[width=0.49\textwidth]{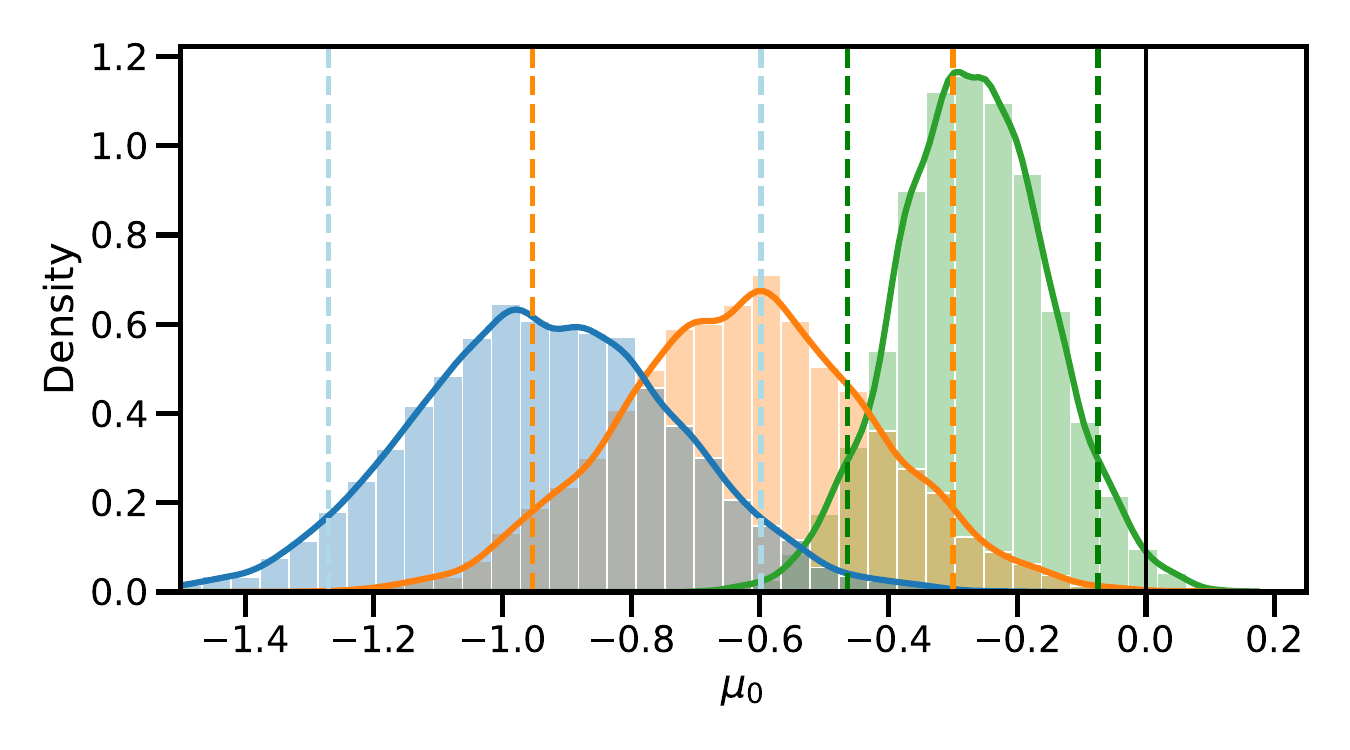}}
    \end{subfigure}
    \vspace{-0.6cm}
    \begin{subfigure}{\includegraphics[width=0.49\textwidth]{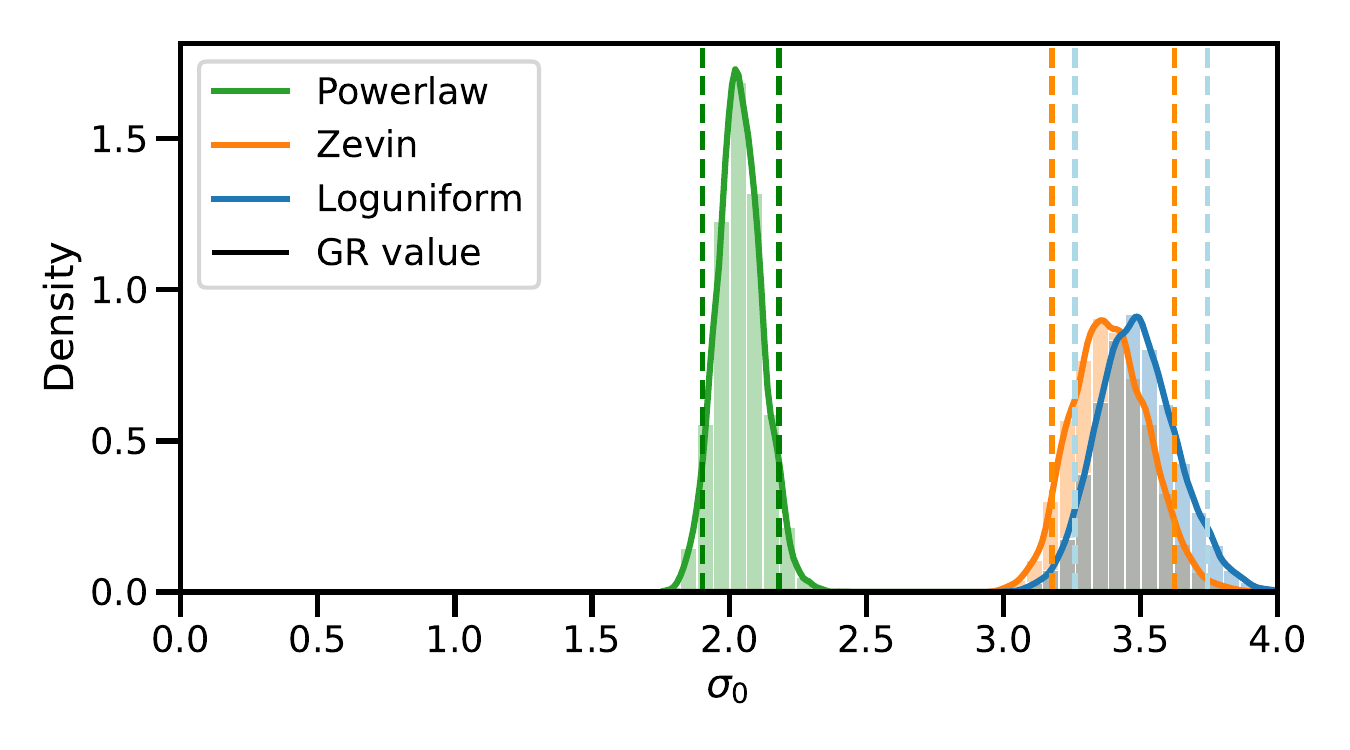}}
    \end{subfigure}

    \begin{subfigure}{\includegraphics[width=0.49\textwidth]{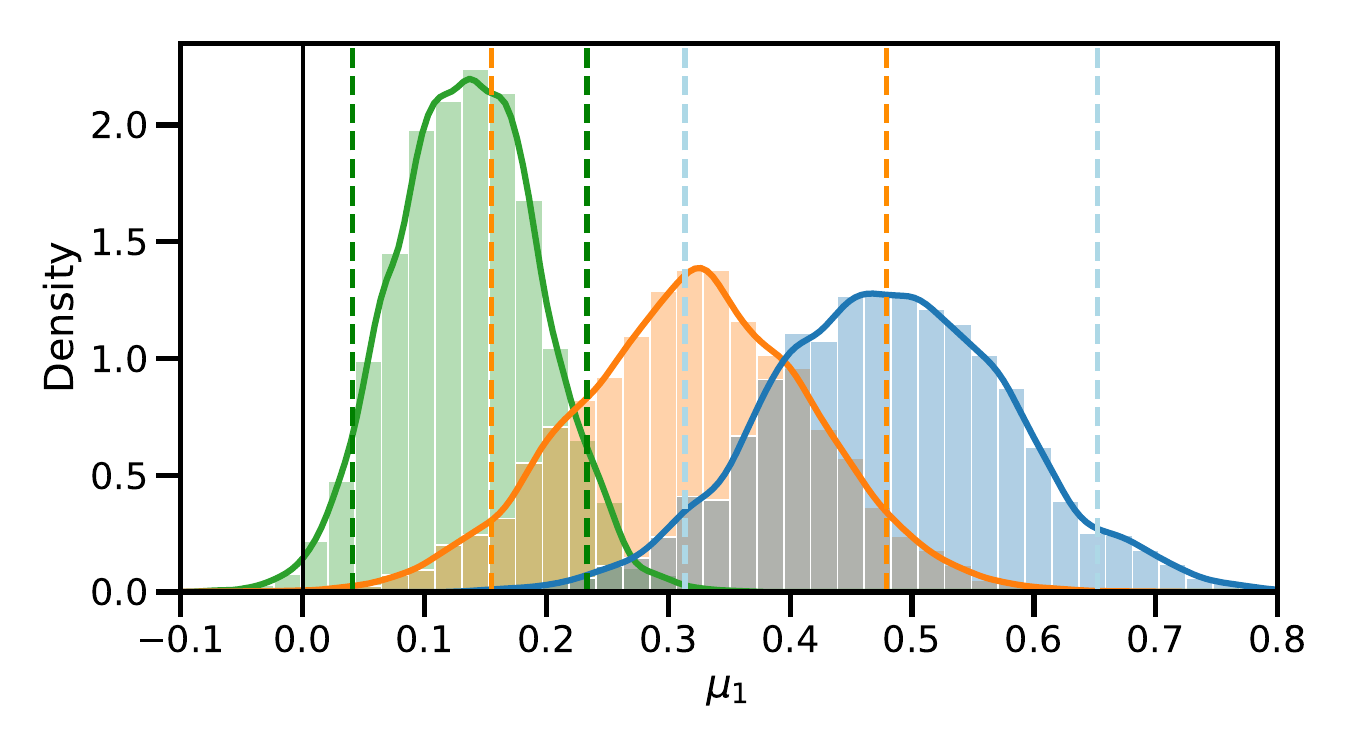}}
    \end{subfigure}
    \vspace{-0.6cm}
    \begin{subfigure}{\includegraphics[width=0.49\textwidth]{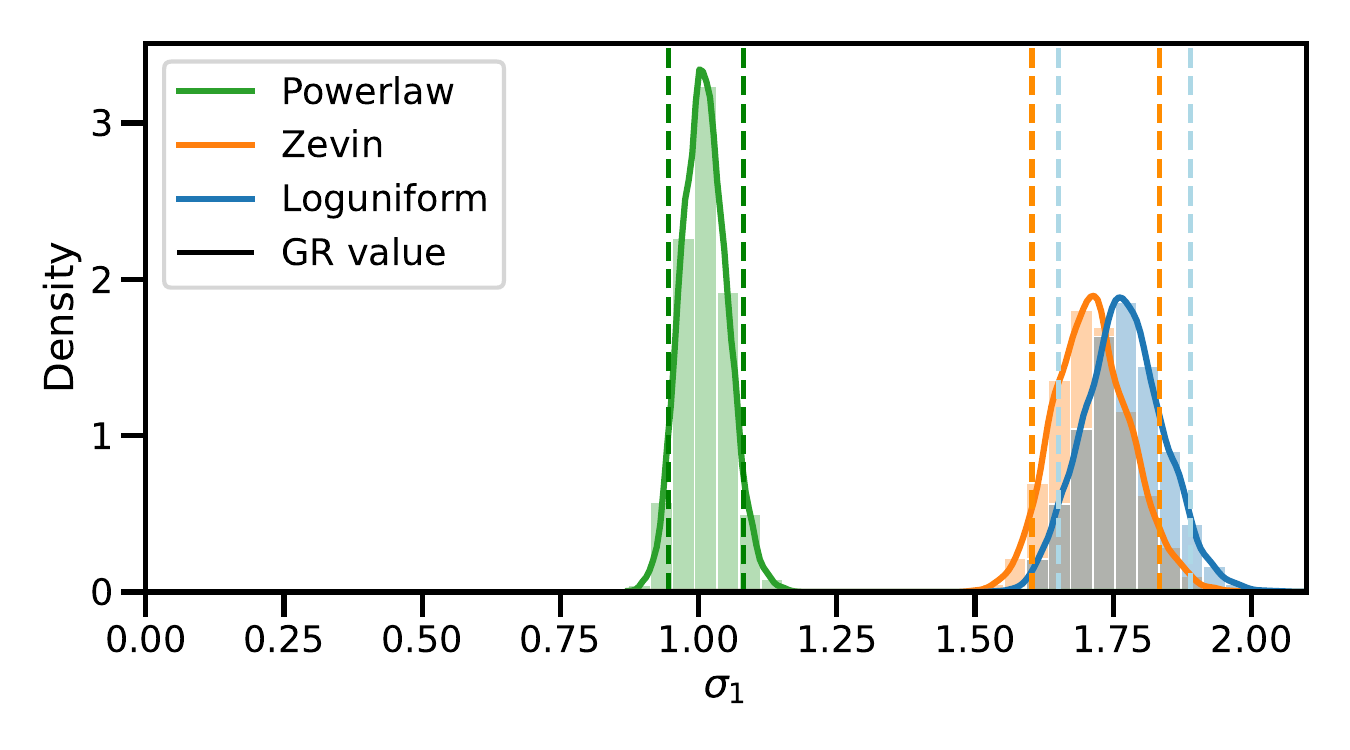}}
    \end{subfigure}

    \begin{subfigure}{\includegraphics[width=0.49\textwidth]{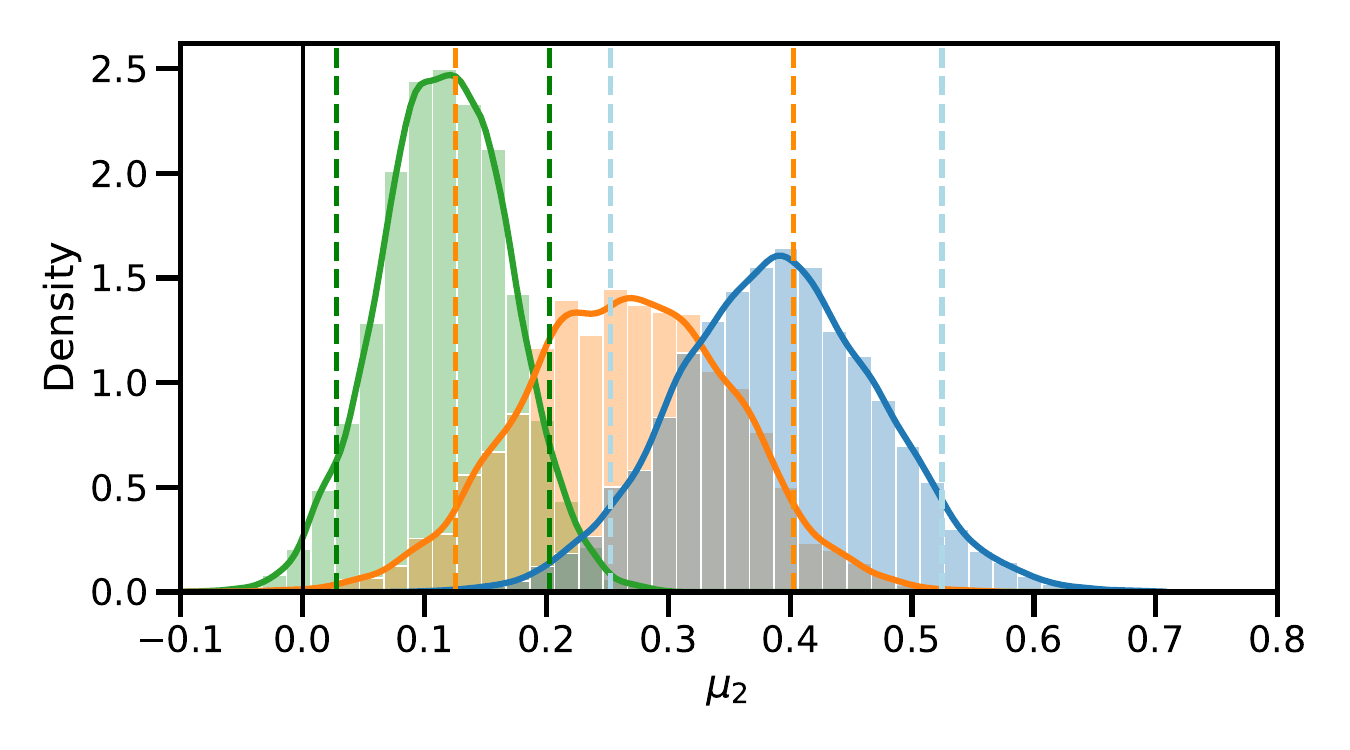}}
    \end{subfigure}
    \vspace{-0.6cm}
    \begin{subfigure}{\includegraphics[width=0.49\textwidth]{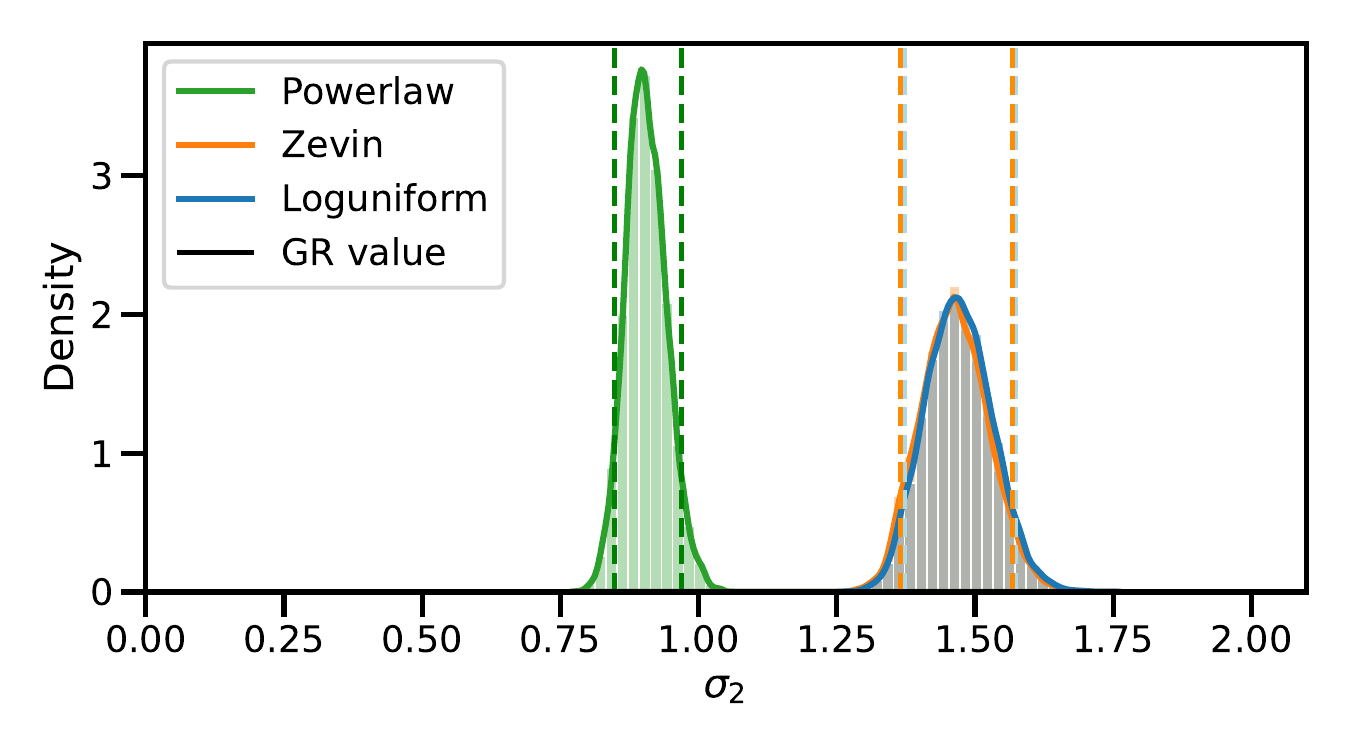}}
    \end{subfigure}

    \begin{subfigure}{\includegraphics[width=0.49\textwidth]{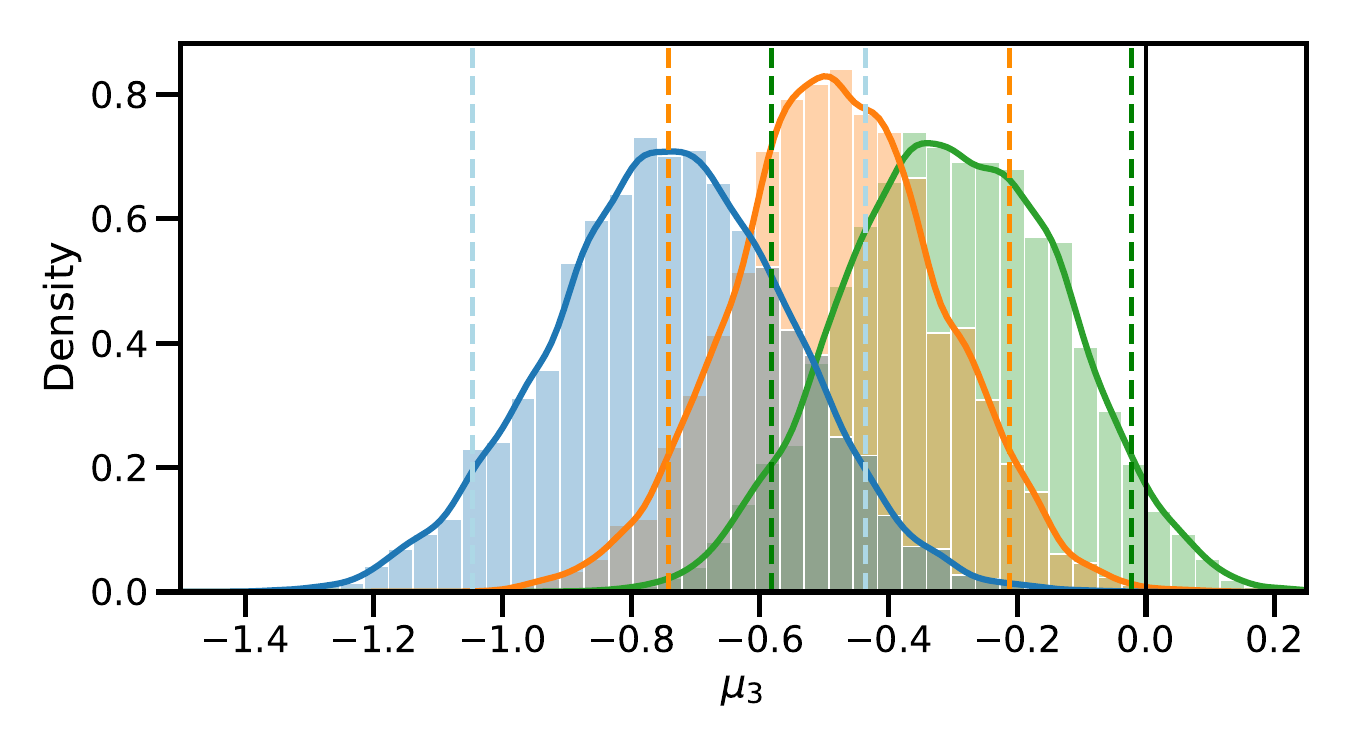}}
    \end{subfigure}
    \vspace{-0.6cm}
    \begin{subfigure}{\includegraphics[width=0.49\textwidth]{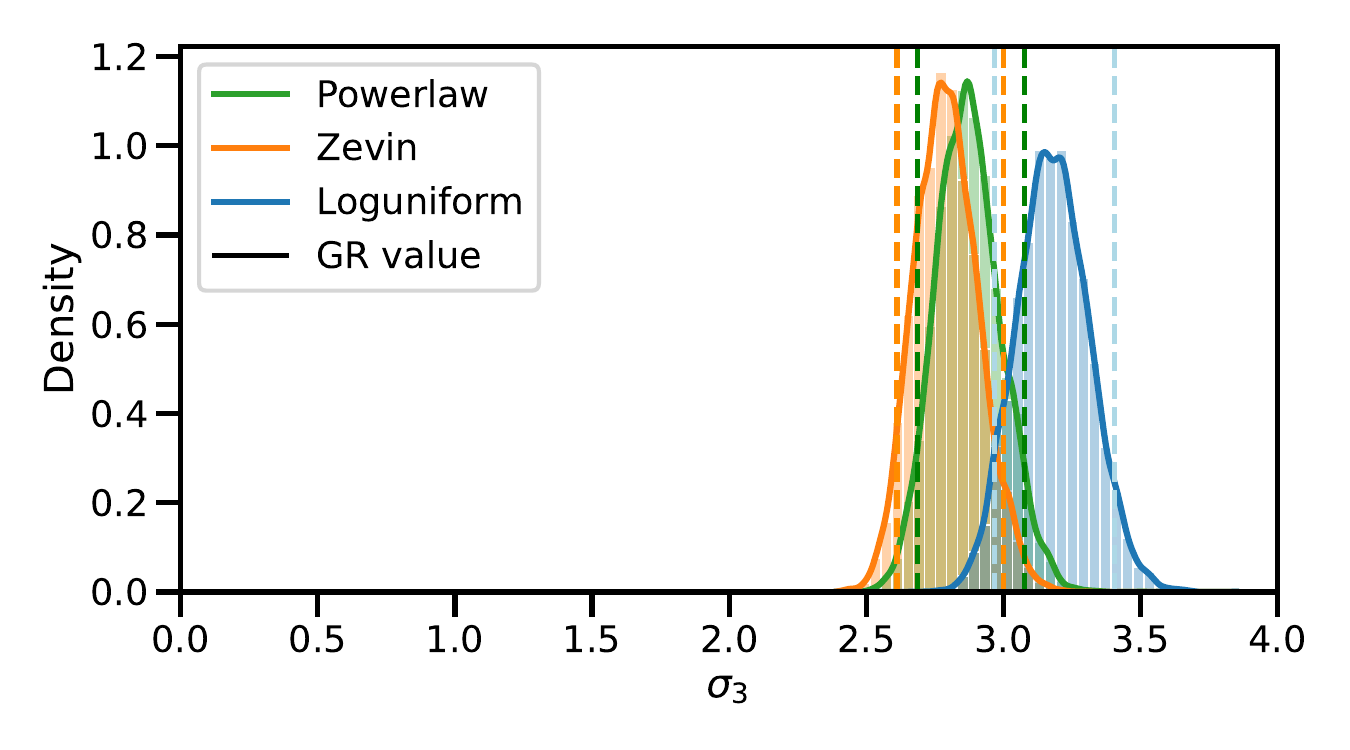}}
    \end{subfigure}

    \begin{subfigure}{\includegraphics[width=0.49\textwidth]{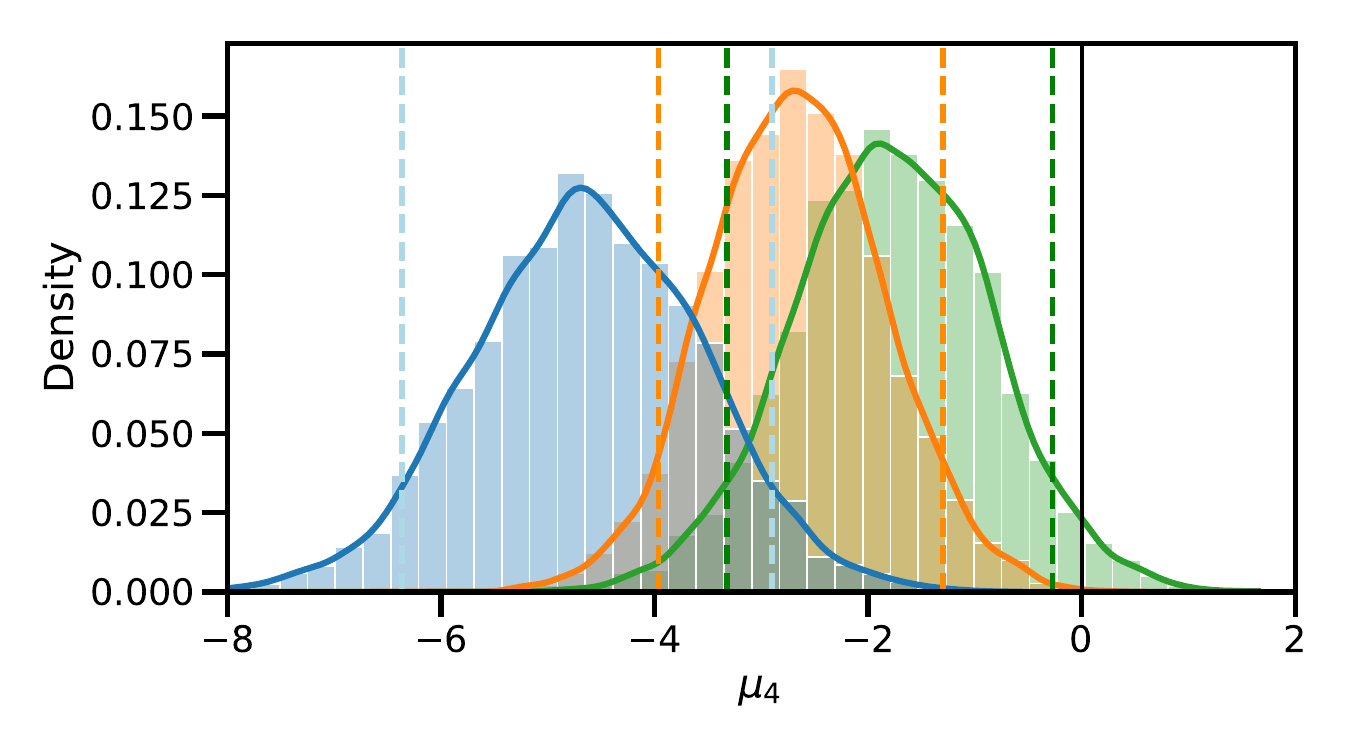}}
    \end{subfigure}
    \vspace{-0.6cm}
    \begin{subfigure}{\includegraphics[width=0.49\textwidth]{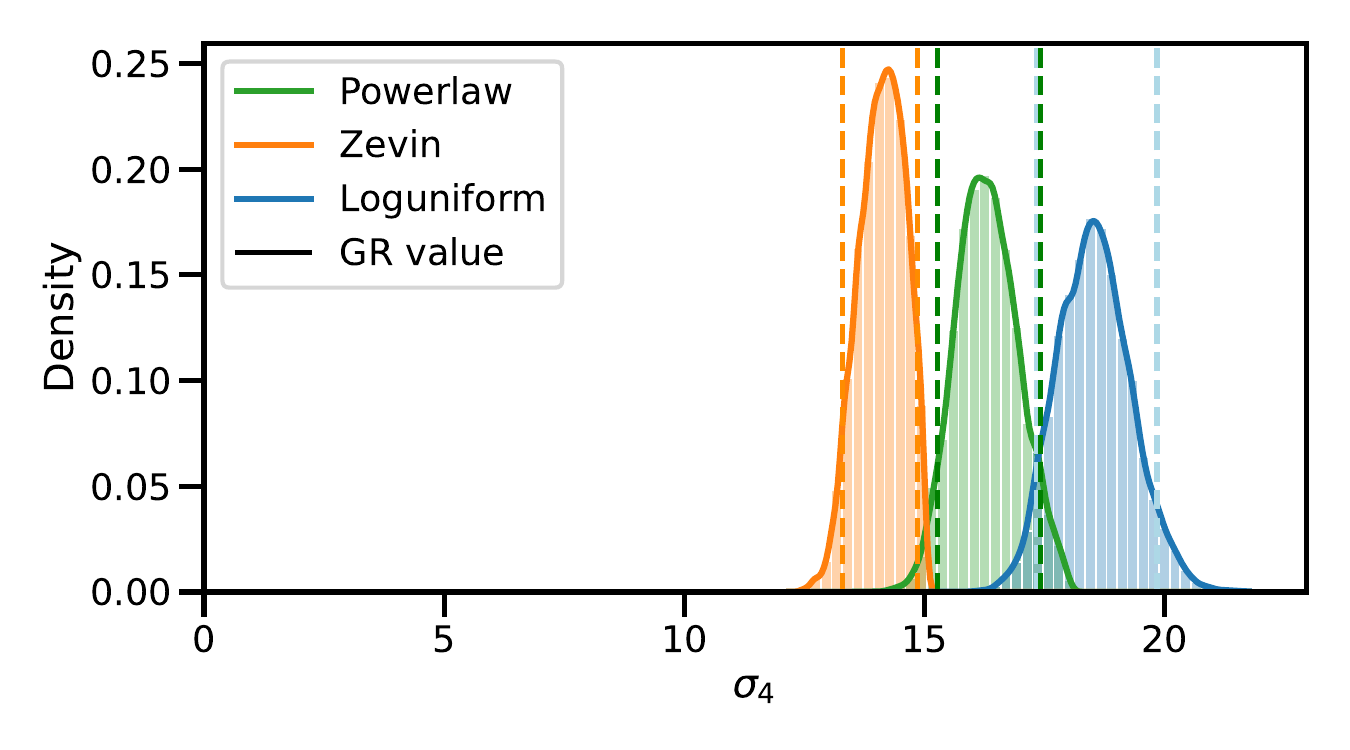}}
    \end{subfigure}
    \caption{Posteriors on the hyperparameters $\mu_i$ (left panels) and $\sigma_i$ (right panels) for the leading-order $\delta\hat{\varphi}_i$ parameters as measured with a single CE detector. Conventions are as in Fig.~\ref{ligo1}.}
    \label{ce1}
\end{figure*}
%%%%%%%%%%%%%%%%%%%%%%%%%%%%%%%%%%%%%%%%%%%%%%%%%%%%%%%%%%%%%%%%%%%%%%%%%%%%%
\begin{figure*}[hpt!]
    \centering
    \begin{subfigure}{\includegraphics[width=0.49\textwidth]{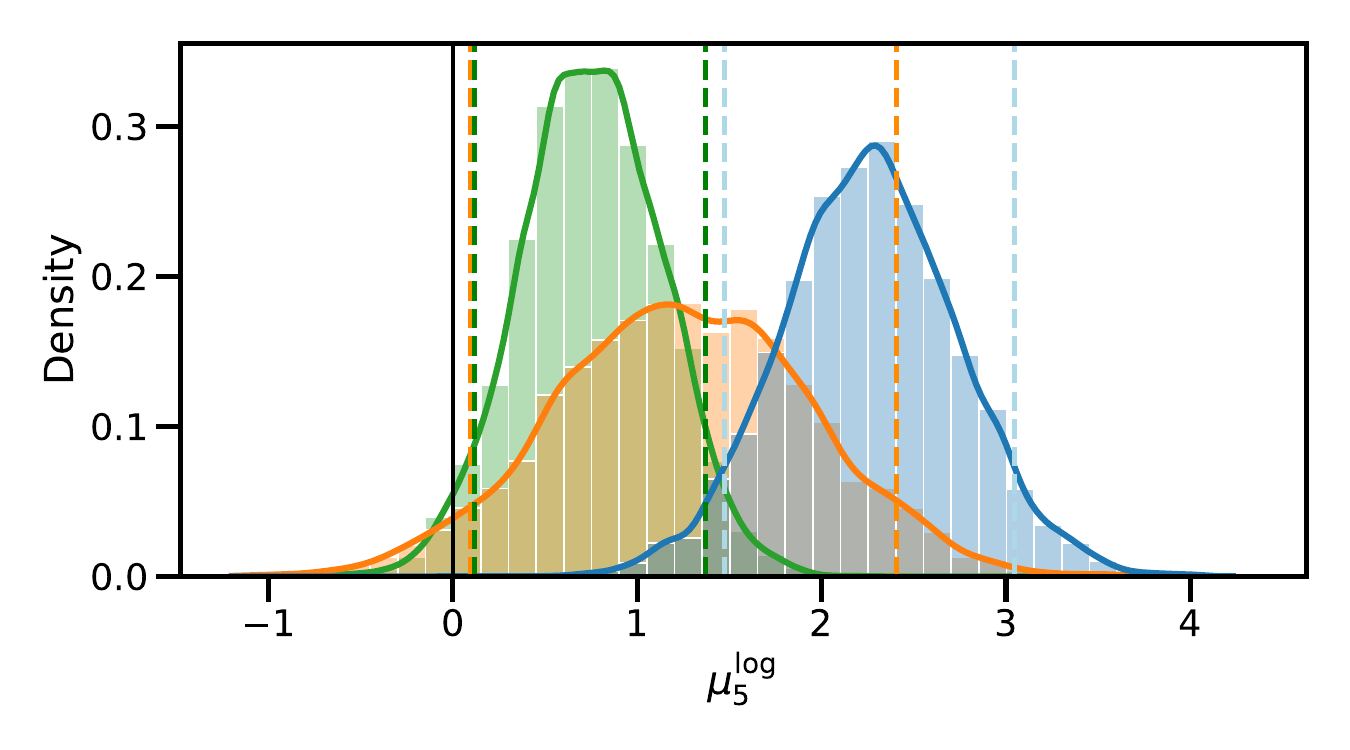}}
    \end{subfigure}
    \vspace{-0.6cm}
    \begin{subfigure}{\includegraphics[width=0.49\textwidth]{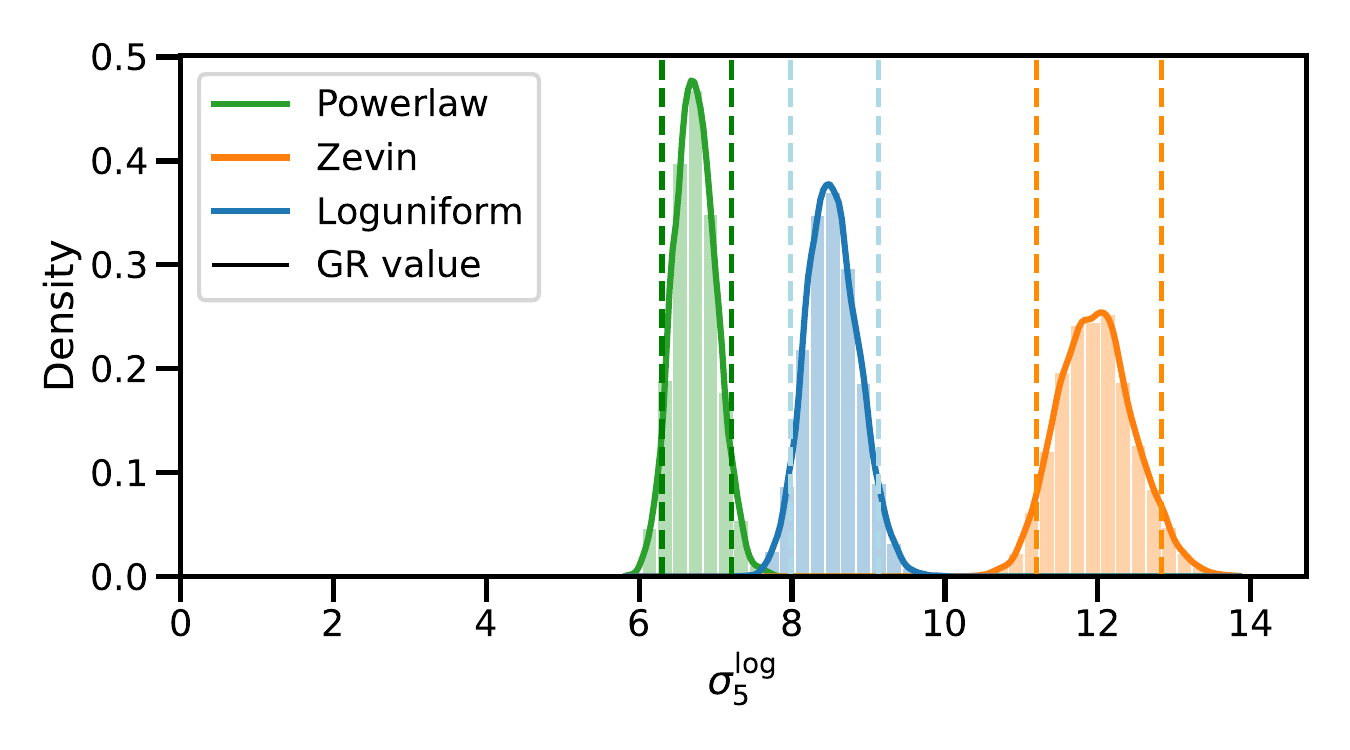}}
    \end{subfigure}

    \begin{subfigure}{\includegraphics[width=0.49\textwidth]{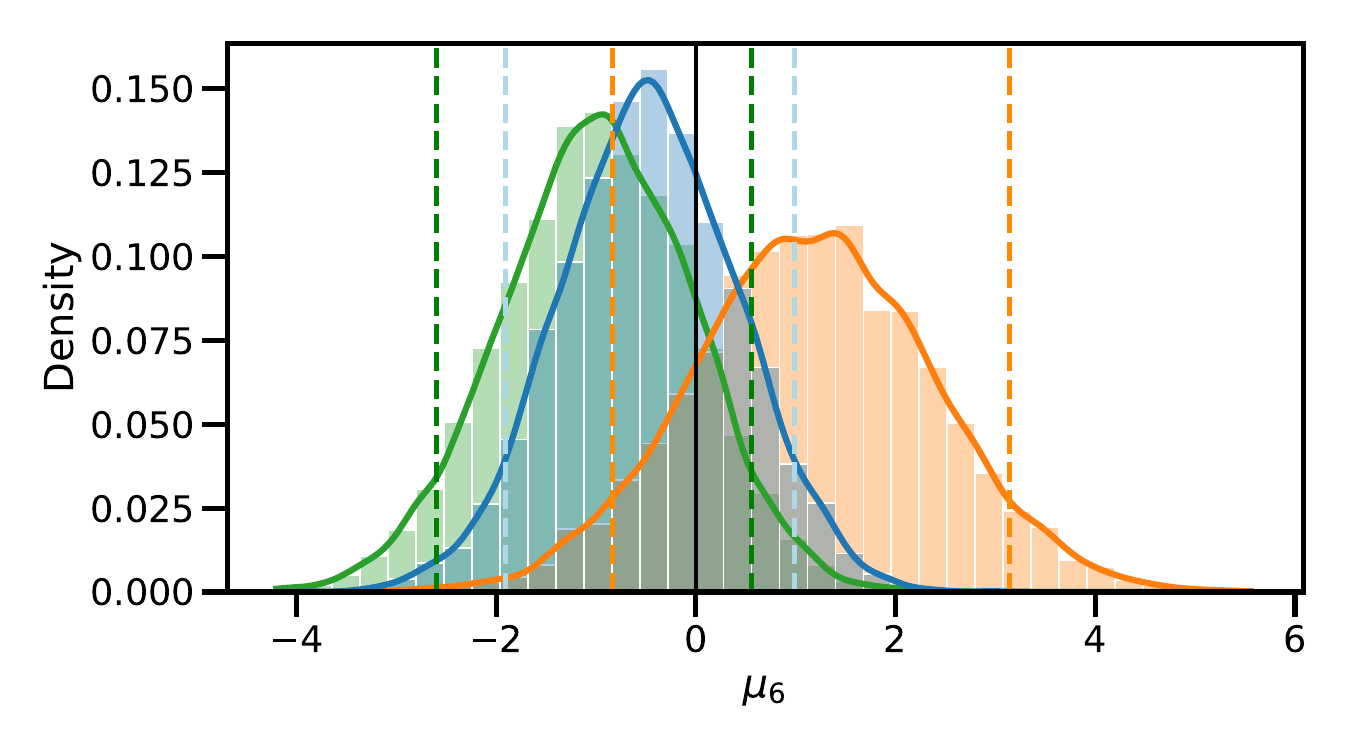}}
    \end{subfigure}
    \vspace{-0.6cm}
    \begin{subfigure}{\includegraphics[width=0.49\textwidth]{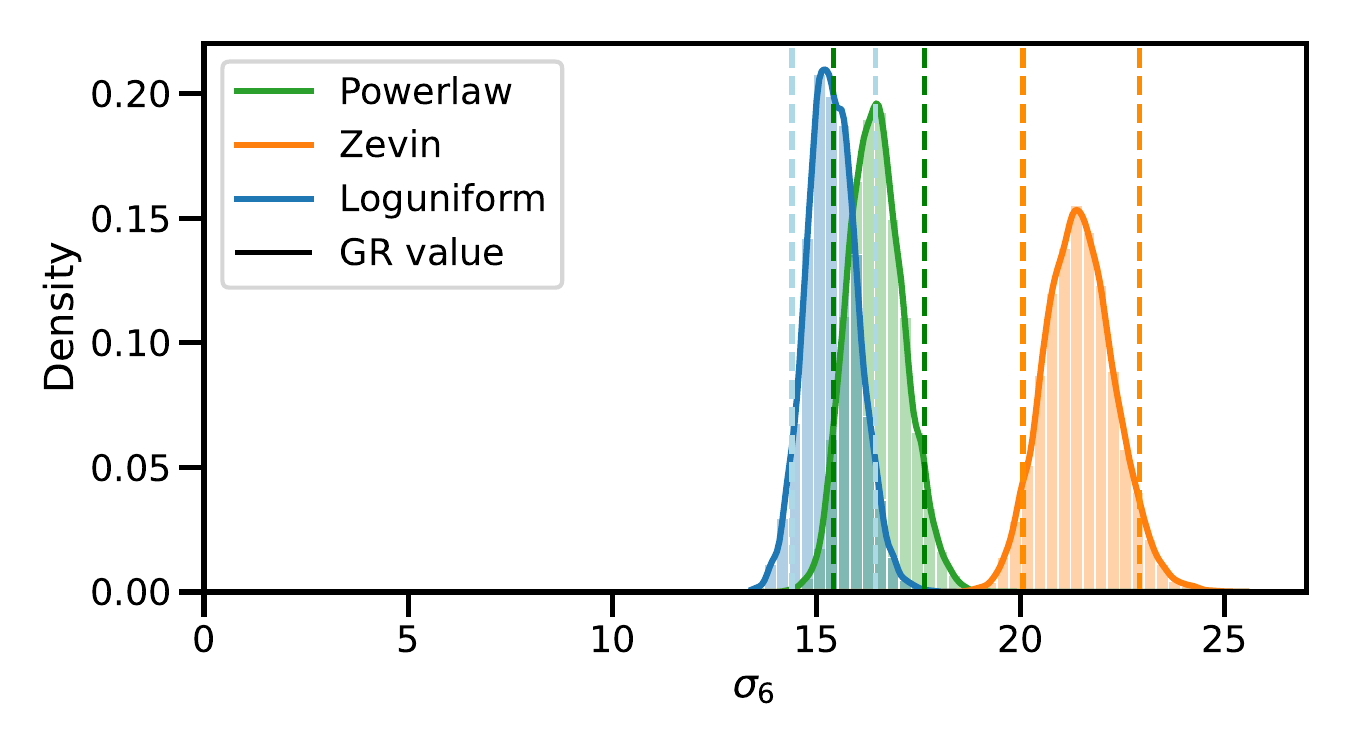}}
    \end{subfigure}

    \begin{subfigure}{\includegraphics[width=0.49\textwidth]{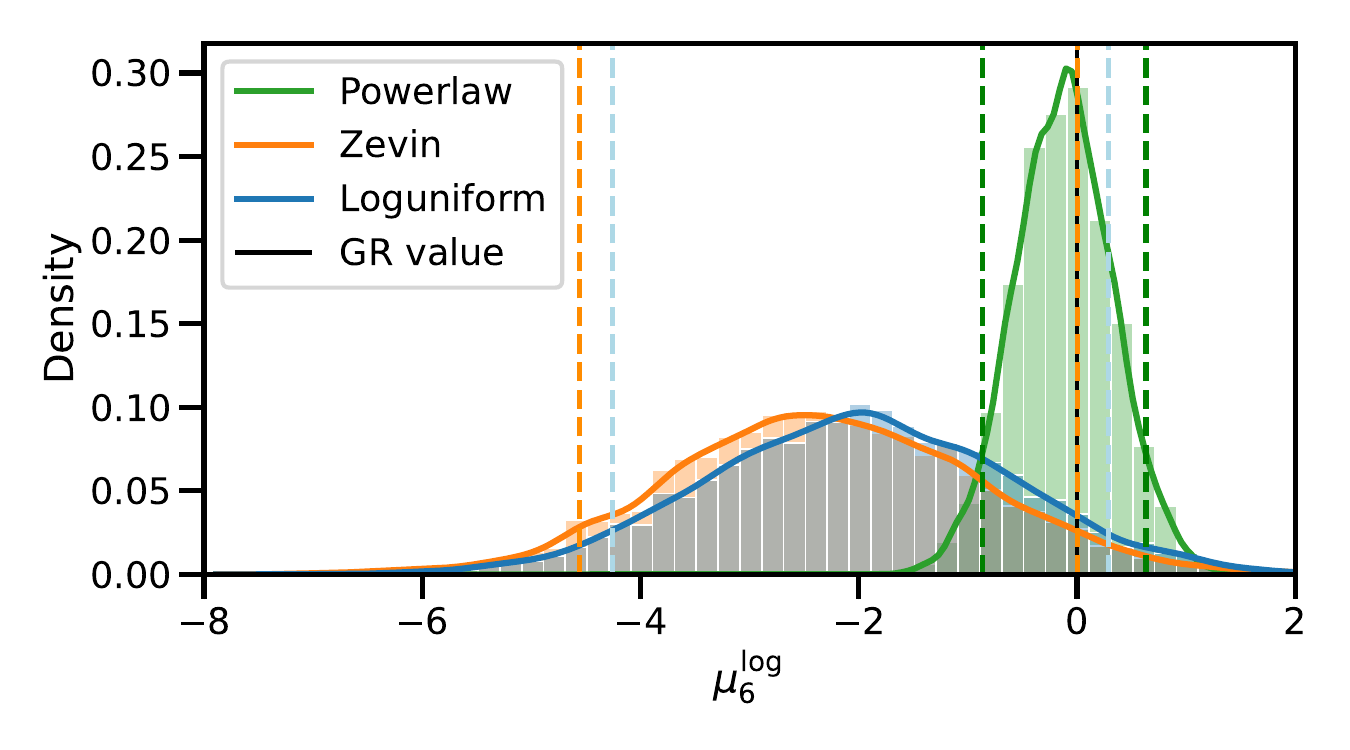}}
    \end{subfigure}
    \vspace{-0.6cm}
    \begin{subfigure}{\includegraphics[width=0.49\textwidth]{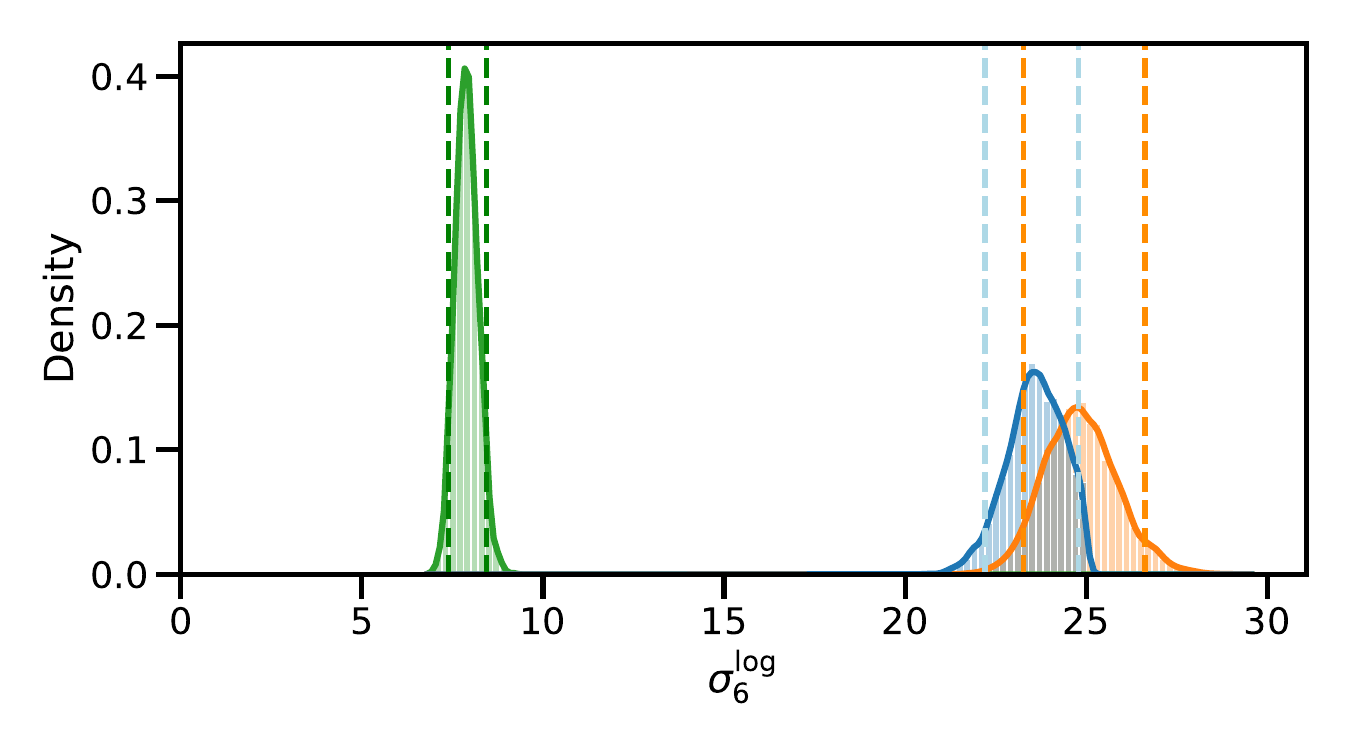}}
    \end{subfigure}

    \begin{subfigure}{\includegraphics[width=0.49\textwidth]{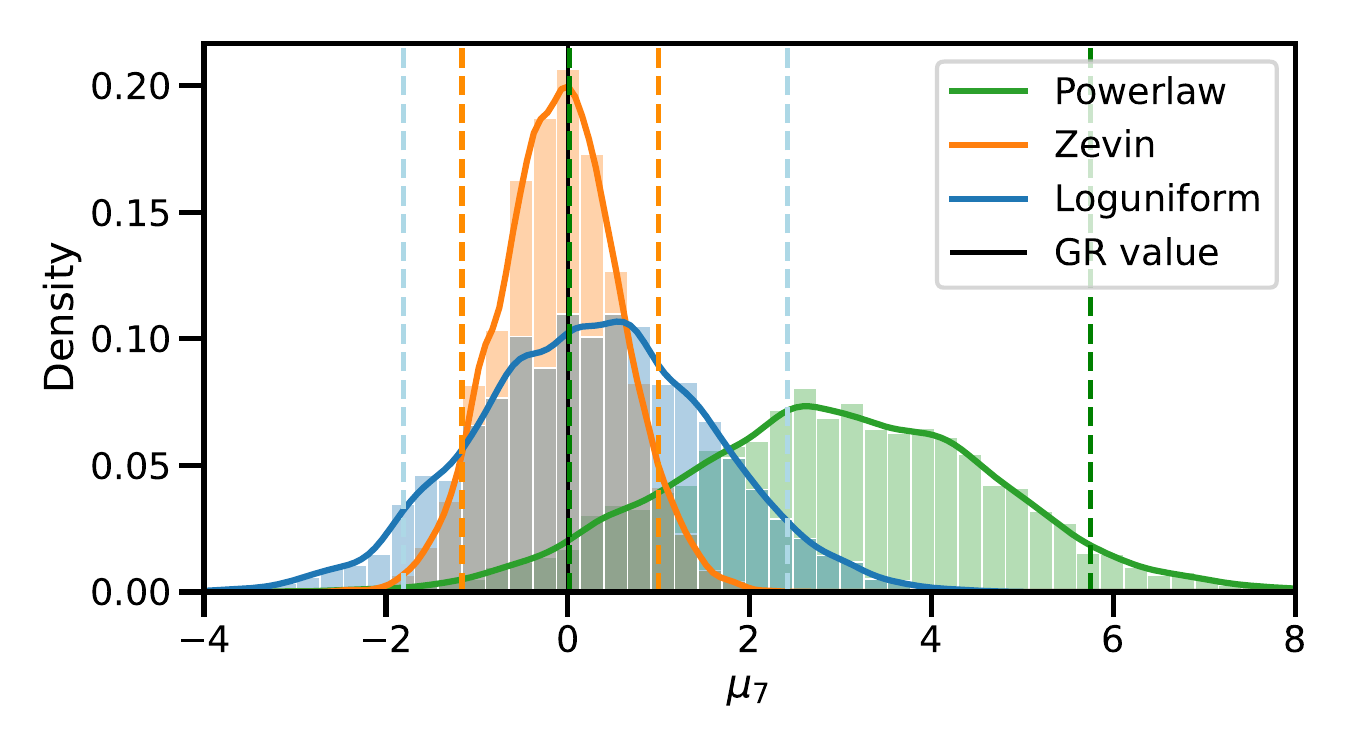}}
    \end{subfigure}
    \vspace{-0.6cm}
    \begin{subfigure}{\includegraphics[width=0.49\textwidth]{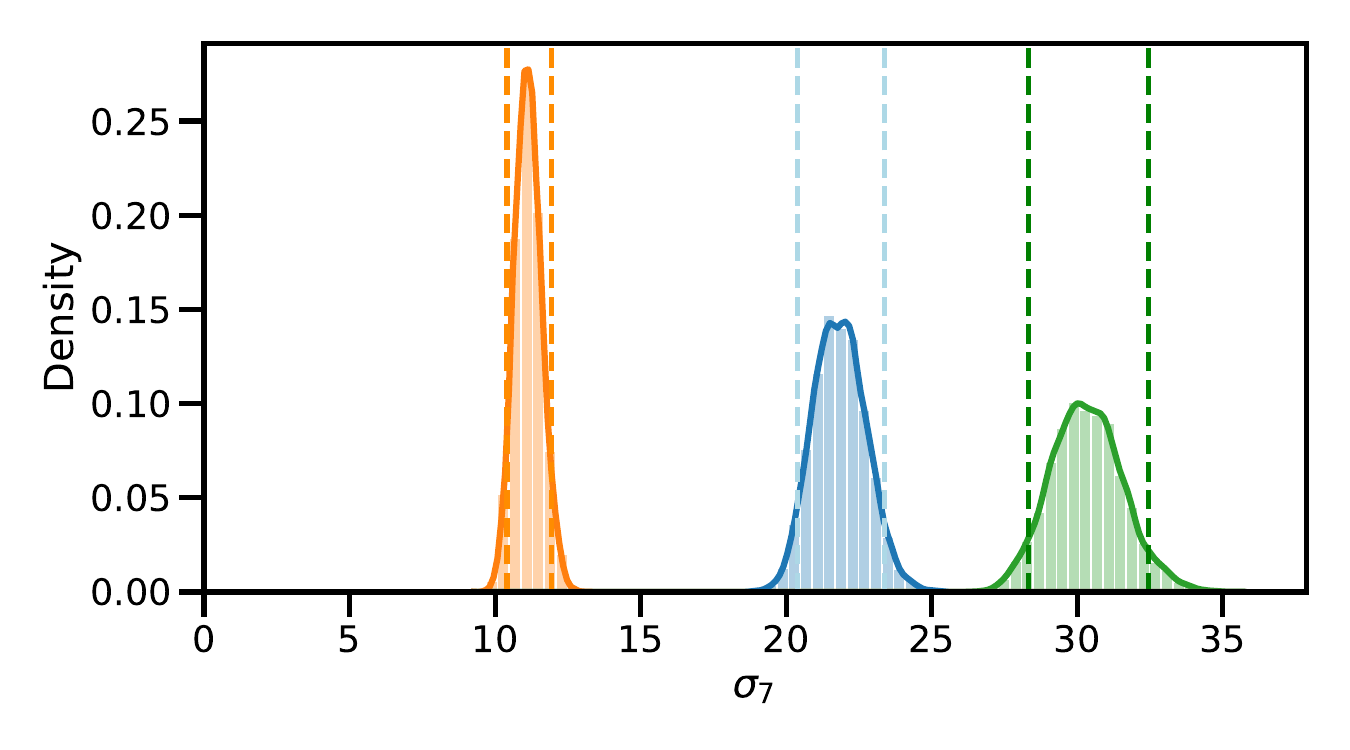}}
    \end{subfigure}

    \begin{subfigure}{\includegraphics[width=0.49\textwidth]{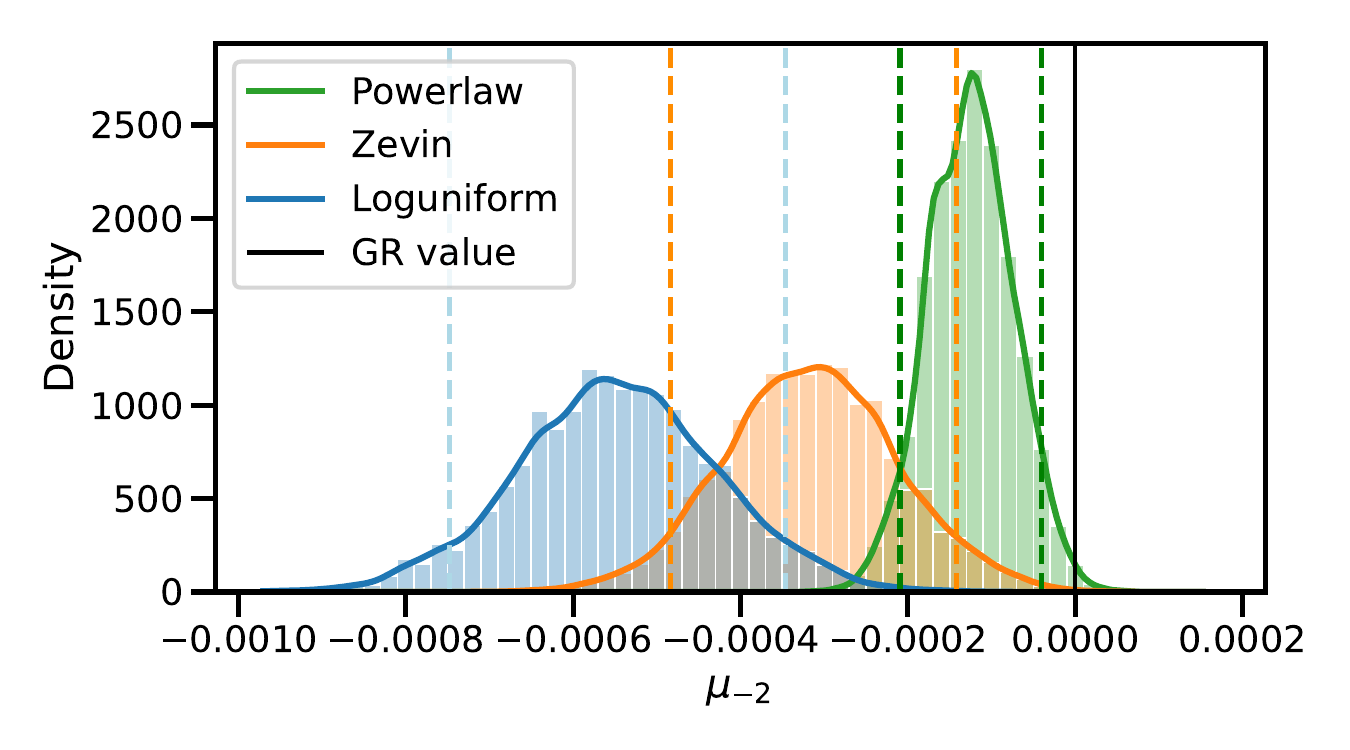}}
    \end{subfigure}
    \vspace{-0.7cm}
    \begin{subfigure}{\includegraphics[width=0.49\textwidth]{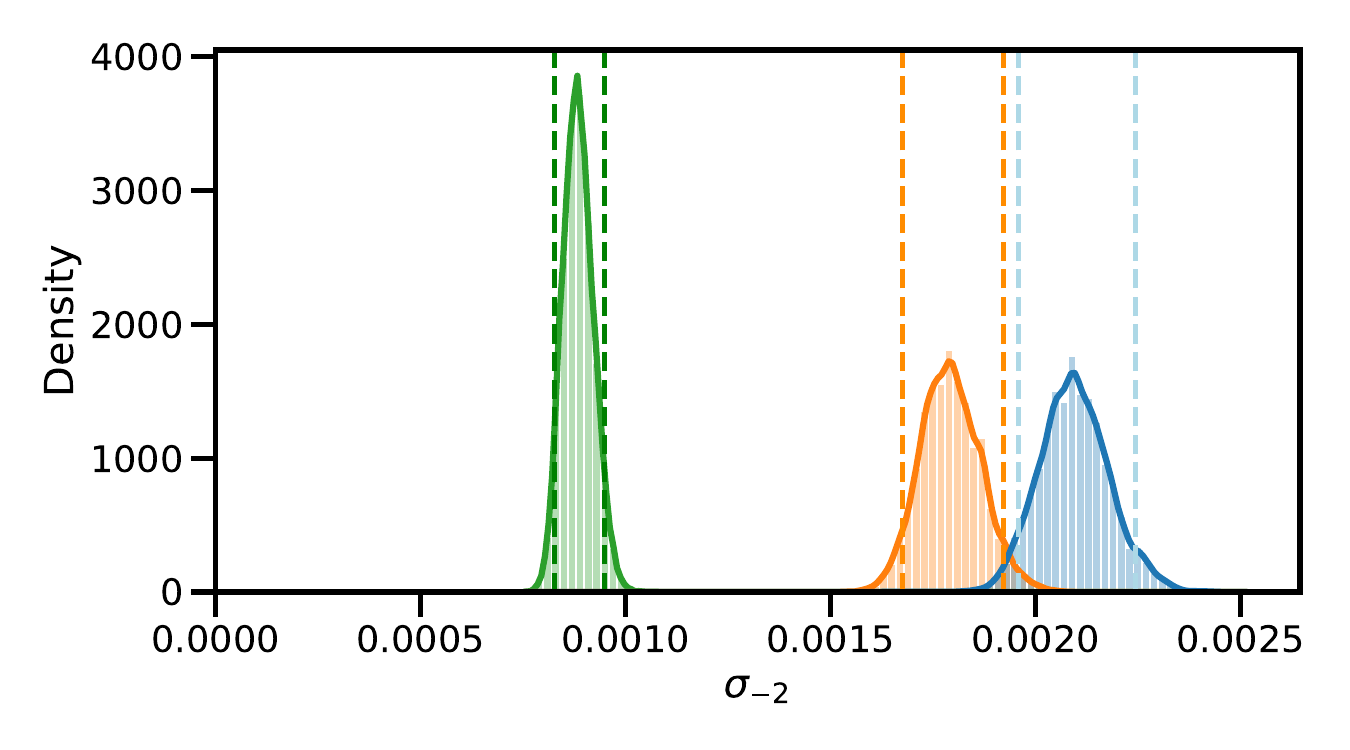}}
    \end{subfigure}
    \caption{Posteriors on hyperparameters $\mu_i$ and $\sigma_i$ for the higher-order and -1PN order TGR parameters $\delta\hat{\varphi}_i$ as measured by a single CE detector. Conventions are as in Fig.~\ref{ligo2}.}
 \label{ce2}
\end{figure*}
%%%%%%%%%%%%%%%%%%%%%%%%%%%%%%%%%%%%%%%%%%%%%%%%%%%%%%%%%%%%%%%%%%%%%%%%%%%%%
%%%%%%%%%%%%%%%%%%%%%%%%%%%%%%%%%%%%%%%%%%%%%%%%%%%%%%%%%%%%%%%%%%%%%%%%%%%%%

%%%%%%%%%%%%%%%%%%%%%%%%%%%%%%%%%%%%%%%%%%%%%%%%%%%%%%%%%%%%%%%%%%%%%%%%%%%%%
%%%%%%%%%%%%%%%%%%%%%%%%%%%%%%%%%%%%%%%%%%%%%%%%%%%%%%%%%%%%%%%%%%%%%%%%%%%%%
\begin{table*}[htp!]
\centering
\renewcommand{\arraystretch}{1.3}
\setlength{\tabcolsep}{4pt}
\begin{tabular}{| c | c | c@{\hspace{0.25cm}} | c@{\hspace{0.25cm}} | c@{\hspace{0.25cm}} | c@{\hspace{0.25cm}} | c@{\hspace{0.25cm}} | c@{\hspace{0.25cm}} | c@{\hspace{0.25cm}} | c@{\hspace{0.25cm}} | c@{\hspace{0.25cm}} | c@{\hspace{0.25cm}} | c@{\hspace{0.25cm}} |} 
\hline
  & $\mu_i$ & $\mu_0$ & $\mu_1$ &  $\mu_2$ &  $\mu_3$  &  $\mu_4$   &  $\mu_{\rm 5}^{\rm log}$  &  $\mu_6$  &  $\mu_{\rm 6}^{\rm log}$  &  $\mu_7$  &  $\mu_{-2}$ \\ 
\hline 
\multicolumn{1}{|c|}{}&\multicolumn{1}{c|}{\bf{eccentricity distribution}} & \multicolumn{10}{c|}{credible level (\%)}\\
\hline
\multirow{3}{3em}{\bf{LIGO}} & Powerlaw & 88& 87 &90 & 91 & 26 & 78& 46& 68 & 66 & 91 \\
 & Zevin & 97 & 95 & 95 & 96 & 71 & 92 & 85 & 85 & 84 & 93\\
& Loguniform & 99& 99& 99 & 99 & 41 & 99 & 92& 70 & 69 & 99\\
 \hline
\multirow{3}{3em}{\bf{CE}} & Powerlaw & 98 & 98 & 97& 93 & 95& 95& 70& 22 & 90&98\\
& Zevin & 99 & 99& 99& 99& 99 & 93 & 67 & 90&10 &99\\
& Loguniform & 99 & 99 & 99 & 99& 99 & 99& 41&85&18&99\\
 
 \hline
\end{tabular}
 \caption{The credible levels at which the $\mu_i$ posteriors in Figures \ref{ligo1}, \ref{ligo2}, \ref{ce1}, and \ref{ce2} exclude the GR values (zero). (Those figures indicate the $90\%$ credible intervals by dashed vertical lines.) We do not show the credible levels for the $\sigma_i$ posteriors as they all exclude the GR values (zero) at a $>99.99\%$ credible level. The credible levels are rounded to the nearest percent.}
 \label{table}

\end{table*}
%%%%%%%%%%%%%%%%%%%%%%%%%%%%%%%%%%%%%%%%%%%%%%%%%%%%%%%%%%%%%%%%%%%%%%%%%%%%%
%%%%%%%%%%%%%%%%%%%%%%%%%%%%%%%%%%%%%%%%%%%%%%%%%%%%%%%%%%%%%%%%%%%%%%%%%%%%

\subsection{Results and discussion}\label{results}
When applying hierarchical Bayesian inference in the testing GR context, it is worth noting that $\mu_i$ and $\sigma_i$ are theoretical parameters that serve as proxies for the (unknown) variation in the $\delta \hat{\varphi}_i^j$ across the BBH population. In the GR limit $\mu_i=0$ and $\sigma_i=0$. Therefore, a GR violation can be indicated by \emph{either} an offset in the $\mu_i$ from zero \emph{or} nonzero values for the $\sigma_i$.

Figures~\ref{ligo1} and \ref{ligo2} show the posterior probability densities for the hyperparameters $\mu_i$ (left panels) and $\sigma_i$ (right panels) corresponding to the different $\delta\hat{\varphi}_i$ for observations in the LIGO band. The three different posteriors correspond to the three eccentricity distributions shown in different colors. The vertical black lines represent the GR values (zero). The vertical lines in the respective colors bound the $90\%$ credible interval. 

For example, the posteriors for the 0PN hyperparameter $\mu_0$ exclude zero at the $\gtrsim88\%$, $\gtrsim97\%$, and $\gtrsim99\%$ credible level for the {\tt Powerlaw}, {\tt Zevin}, and {\tt Loguniform} eccentricity distributions (respectively), indicating a clear deviation from GR (but actually due to the eccentricity-induced bias). Additionally, the $\sigma_0$ posterior excludes the GR value (zero) at $\gtrsim 99.99\%$ credibility for all three eccentricity distributions. This indicates that the scatter in the mean values of the individually measured $\delta\hat{\varphi}_0^j$ cannot be explained by detector noise alone. In our case (where the injected values of the individual $\delta\hat{\varphi}_i^j$ are taken to be zero), it is evident that the eccentricity-induced systematic bias is showing up as a false deviation from GR. As a summary, Table~\ref{table} lists the credible level at which the GR value (zero) is excluded for the $\mu_i$ posteriors shown in Figs.~\ref{ligo1}, ~\ref{ligo2}, ~\ref{ce1}, and ~\ref{ce2}. We do not show the credible levels for the $\sigma_i$ posteriors as they all exclude zero with $\geq 99.99\%$ credible level. 

The three eccentricity distributions show different shifts in the $\mu_i$ and $\sigma_i$ posteriors from zero. The shift in the $\mu_i$ posteriors depends on both the magnitude and the sign of the systematic biases in the individual $\delta\hat{\varphi}_i^j$ posteriors; these in turn depend on the number of eccentric sources, the eccentricities of those sources, and the correlations of the $\delta\hat{\varphi}_i$ with the eccentricity and other source parameters. The shift in the $\sigma_i$ posteriors depends on the spread of the systematic biases. 

The ratio of systematic to statistical errors for an individual event scales as~\cite{Saini:2022igm}
\begin{equation}\label{eq:ratio}
    \frac{\Delta(\delta\hat{\varphi}_i)}{\sigma_{\delta\hat{\varphi}_i}} \sim \frac{e_0^2}{M^{5/6}}\,.
\end{equation}
Hence, for low-mass sources with high values of eccentricity the ratio of systematic to statistical errors is large. Moreover, low-mass sources have more GW cycles in the detector's frequency band, yielding larger systematic errors and small statistical errors. 

\begin{figure*}[ht!]
    \centering
    \begin{subfigure}{\includegraphics[width=0.25\textwidth]{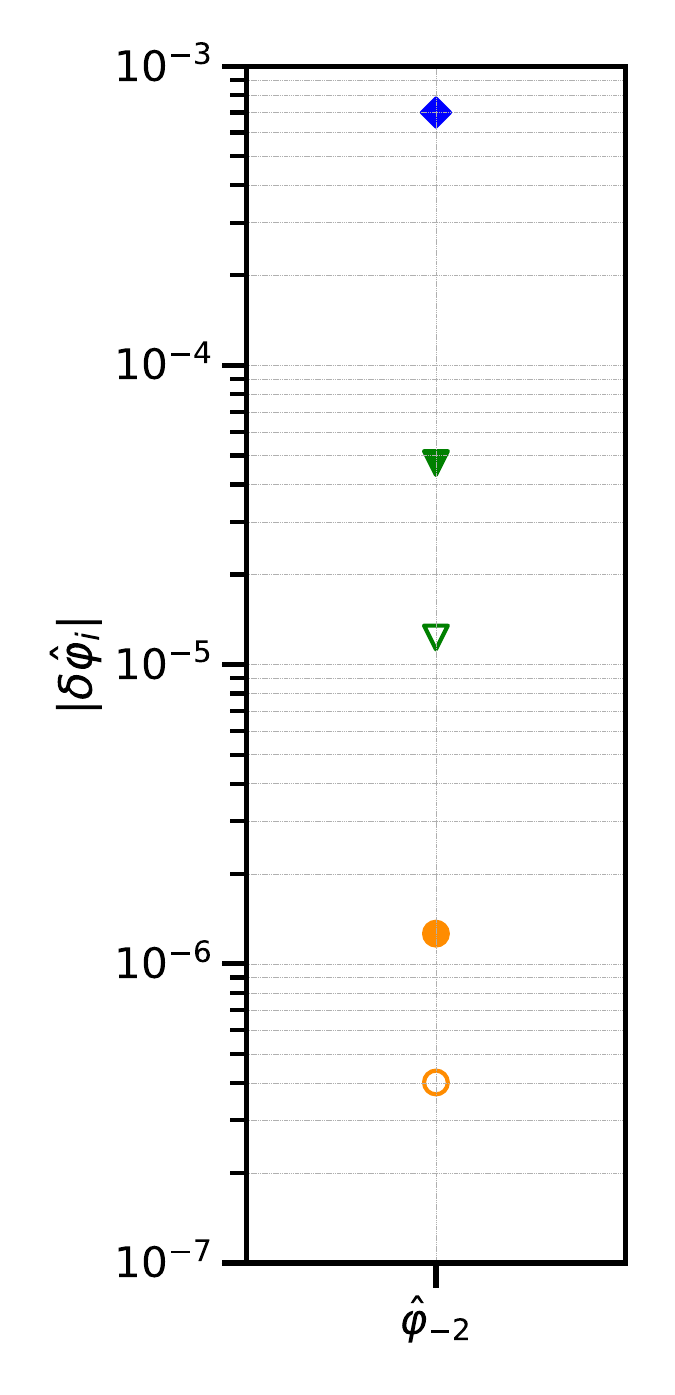}}
    \end{subfigure}
   \begin{subfigure}{\includegraphics[width=0.70\textwidth]{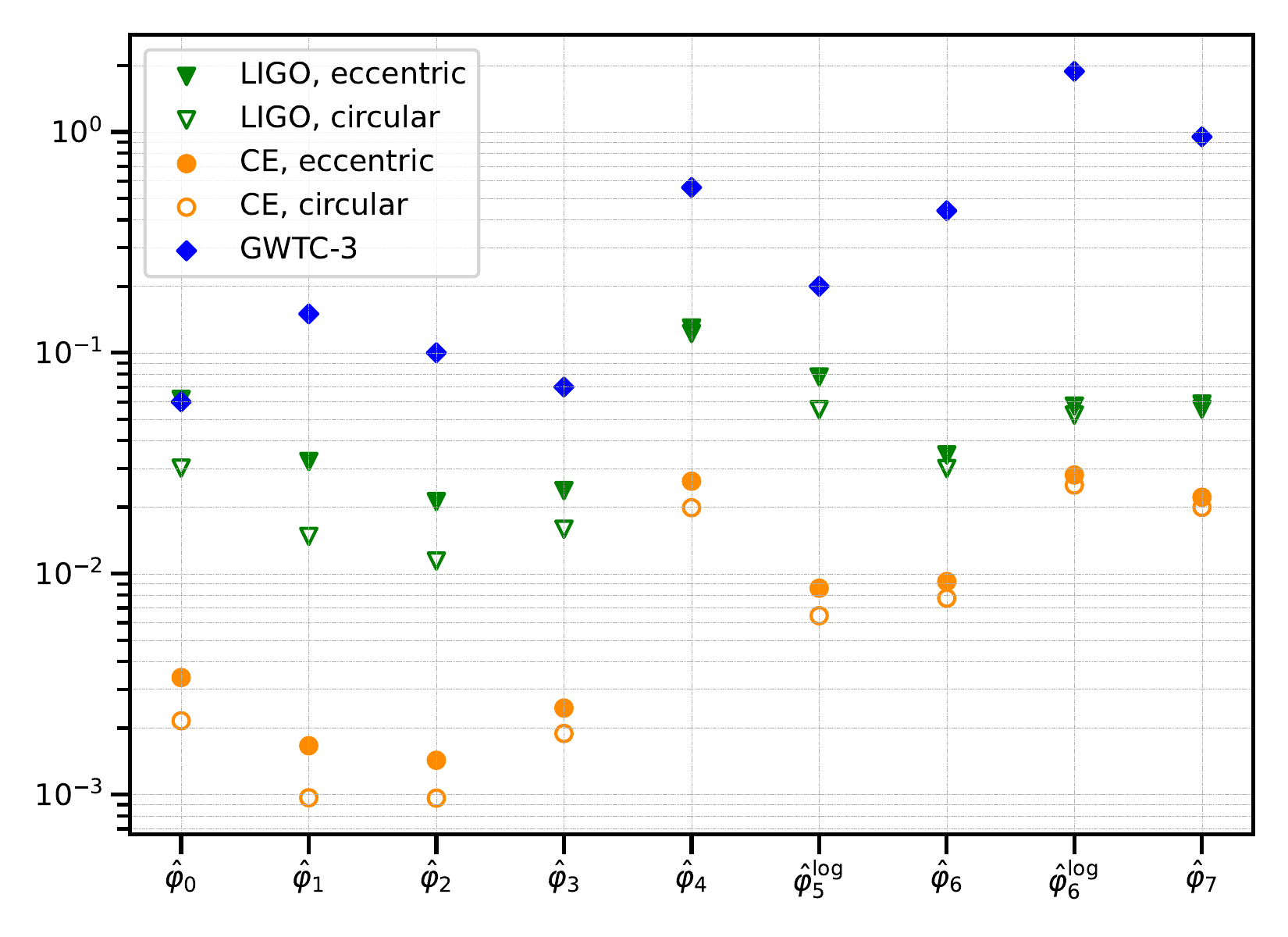}}
   \end{subfigure}
   
    \caption{$90\%$ upper bound on the testing GR parameters $\delta\hat{\varphi}_i$ from $303$ BBHs in our simulated population. The bounds are obtained by assuming that the $\delta\hat{\varphi}_i$ take the GR values (zero) for all events. The green triangles represent the projected bounds for LIGO at design sensitivity, while the orange circles show the projected bounds for the 3G CE detector. The filled symbols correspond to bounds obtained with eccentric waveforms (including $e_0$ as a parameter). Unfilled symbols show bounds in the circular case. The eccentricity samples for $10^5$ BBHs are drawn from the {\tt Loguniform}  distribution with $e_0\in (10^{-7},0.2)$. The blue diamonds represent the $90\%$ upper bound from the GWTC-3~\cite{GWTC3:2021sio} analysis.}
    \label{PN_bounds}
\end{figure*}

The visibility of systematic bias in the $\mu_i$ or $\sigma_i$ posteriors depends on the width of these posteriors. The width of the $\mu_i$ and $\sigma_i$ posteriors depends on the spread in the systematic biases ($\tilde{\mu}_i^j$) and the statistical uncertainties ($\tilde{\sigma}_i^j$) in the individual posteriors $\delta\hat{\varphi}_i^j$. The overall statistical uncertainty scales as $\sim1/\sqrt{N}$ with the number of sources, where $N=303$ in our BBH population. For example, the combined statistical error is of the order of $0.02$ on $\delta\hat{\varphi}_0$ (see Fig.~\ref{PN_bounds}); this is roughly $\sim 5$ and $\sim 20$ times smaller than the $\mu_0$ and $\sigma_0$ values (respectively). Therefore, the contribution of the combined statistical errors from all events is negligible compared to the width of the $\mu_0$ and $\sigma_0$ posteriors. The primary contribution to the width of the $\mu_i$ and $\sigma_i$ posteriors comes from the scatter in the systematic biases. Note that the peak of the $\sigma_i$ posterior is proportional to the scatter in the systematic biases. That means the farther from zero that the $\sigma_i$ posterior peaks, the wider the corresponding $\mu_i$ and $\sigma_i$ posteriors become. This can be observed from Figs.~\ref{ligo1}, ~\ref{ligo2}, ~\ref{ce1}, and ~\ref{ce2}. 

The {\tt Loguniform} eccentricity distribution has the highest number of eccentric sources ($\sim 20$) with measurable eccentricity in the LIGO band ($e_0 \gtrsim 0.05$); thus $\mu_0$,  $\mu_1$, $\mu_2$, $\mu_3$, $\mu_5^{\rm log}$, and $\mu_6$ show inconsistency with GR at a higher credible level compared to the {\tt PowerLaw} and {\tt Zevin} distributions (see Table~\ref{table}). Posteriors for the $\sigma_i$ show inconsistency with GR at a higher credible level than that of the $\mu_i$ posteriors. This is due to the larger scatter in the systematic biases. For LIGO, all $\mu_i$ posteriors contain the mean of the injected systematic biases within a $90\%$ credible interval. However, not all $\sigma_i$ posteriors recover the standard deviation of systematic biases. This is expected because the peak of $\sigma_i$ posteriors also depends on the width of individual posterior $\Tilde{\sigma}_{i}^j$, and the individual event posteriors $\delta\hat{\varphi}_i^j$ have larger statistical errors for LIGO. Therefore, some $\sigma_i$ posteriors are not able to recover the true spread in systematic biases. 

The posteriors of $\mu_4$, $\mu_6^{\rm log}$, and $\mu_7$ do not exclude zero at a very high credible level. This is because some systematic biases are positive and some are negative, resulting in a mean close to zero due to cancellations. However, the corresponding $\sigma_i$ posteriors show a significant offset from zero, indicating an inconsistency with GR for all eccentricity distributions. The posteriors on $\mu_{-2}$ and $\sigma_{-2}$ (bottom row of Fig.~\ref{ligo2}) characterize constraints on the dipole term; this PN coefficient is zero in GR but may be nonzero in alternative theories of gravity. Currently, the dipole term is the most precisely measured $\delta\hat{\varphi}_i$ coefficient in the GWTC-3 dataset~\cite{GWTC3:2021sio}. The $\mu_{-2}$ and $\sigma_{-2}$ posteriors show a clear inconsistency with GR. Note that all $\sigma_i$ posteriors exclude the GR values at a significant credible interval. This is due to the large scatter in the values of the systematic bias.

Figures~\ref{ce1} and \ref{ce2} show the corresponding posteriors on the $\mu_i$ and $\sigma_i$ hyperparameters as measured by a single CE detector. Since CE is $\sim 10$ times more sensitive than LIGO and also sensitive to lower frequencies ($\geq 5$ Hz), the statistical errors on the $\delta\hat{\varphi}_{i}$ are $\sim 10\mbox{--}100$ times smaller compared to those measured by LIGO. Since the individual posteriors are narrower for CE, all $\mu_i$ and $\sigma_i$ posteriors recover the mean and standard deviation of the injected values of the systematic biases, respectively. The posteriors on ($\mu_0$, $\sigma_0$), ($\mu_1$, $\sigma_1$), ($\mu_2$, $\sigma_2$), ($\mu_3$, $\sigma_3$), ($\mu_4$, $\sigma_4$), ($\mu_{5}^{\rm log}$, $\sigma_{5}^{\rm log}$), and ($\mu_{-2}$, $\sigma_{-2}$) exclude zero for all three eccentricity distributions with $\gtrsim90\%$ credible level. Posteriors corresponding to the {\tt Loguniform} distribution exclude zero with a higher credible level, followed by the {\tt Zevin} and {\tt Powerlaw} distributions. The explanation is the same as discussed above: the {\tt Loguniform} distribution has the most sources with relatively higher eccentricity values. However, this is not always the case for all $\delta \hat{\varphi}_i$, as the systematic bias also depends on other source parameters like the masses and spins. 

Since the systematic biases on the higher-order deviation parameters ($\delta\hat{\varphi}_6$, $\delta\hat{\varphi}_{6}^{\rm log}$, $\delta\hat{\varphi}_7$) are distributed on both sides of zero, the $\mu_i$ posteriors for these higher PN-order parameters do not exclude zero with a high credible level (Fig.~\ref{ce2}). However, the $\sigma_i$ posteriors for these parameters exclude zero with high credible levels for all three eccentricity distributions; this is due to the sufficiently large scatter in the systematic biases. Note that the $\mu_i$ and $\sigma_i$ posterior for CE peak at larger values compared to LIGO. CE and LIGO can measure only a fraction of a particular eccentricity distribution. CE can measure eccentricities $\gtrsim 5\times 10^{-3}$ at a $10$ Hz GW frequency~\cite{Saini:2023wdk}, whereas LIGO-type detectors can measure eccentricities $\gtrsim 0.05$ at $10$ Hz~\cite{PhysRevD.98.083028,Favata:2021vhw, Saini:2023wdk}. That means CE can detect more sources for which eccentricity can be measured. Therefore, eccentricity-induced systematic bias has a larger magnitude and larger spread for CE. This causes the $\mu_i$ and $\sigma_i$ posteriors to peak farther from zero compared to the corresponding LIGO posteriors.

In summary, Figs.~\ref{ligo1}--\ref{ce2} show that if there are a few measurably eccentric sources $[\mathcal{O}(10\mbox{--}20)]$ in the observed population, unmodeled eccentricity will bias the hyperparameter inferences. This will become increasingly severe as the catalog size increases, since the posteriors will become narrower and the chances that a few sources will be eccentric will also increase. Though there is no consensus on the rates of eccentric sources in the community, our results give insight into how the present test of GR would respond to a catalog of BBH containing a few eccentric sources. Current constraints on PN parametrized tests from GWTC-3 were obtained by analyzing around $35$ events~\cite{GWTC3:2021sio}. Therefore, it is likely that the current catalog of sources that pass the criteria for inspiral tests lacks measurably eccentric sources. Hence, the current GWTC-3 results do not show any apparent deviations.

\section{Combined bounds on the TGR parameters including the effect of eccentricity} \label{sec:combined bound}
We next consider the projected bounds set by LIGO and CE on the deformation coefficients $\delta \hat{\varphi}_i$ when eccentricity is included as a waveform parameter (rather than treated as an unmodeled bias). As one would expect, inclusion of another dynamical parameter degrades the overall parameter estimation precision. However, the resulting bounds---though slightly worse---are more realistic and free of eccentricity-induced systematic errors. Further, combining information from multiple events should further improve the bounds. We use two methods of combining the events: (a) multiplying individual likelihoods under the assumption that the individual $\delta\hat{\varphi}_i$ take the same value for all events; and (b) hierarchically combining the events assuming $\delta\hat{\varphi}_i$ across events follow a Gaussian distribution characterized by hyperparameters $\mu_i$ and $\sigma_i$.  

%%%%%%%%%%%%%%%%%%%%%%%%%%%%%%%%%%%%%%%%%%%%%%%%%%%%%%%%%%%%%%%%%%%%%%%%%%%%%
%%%%%%%%%%%%%%%%%%%%%%%%%%%%%%%%%%%%%%%%%%%%%%%%%%%%%%%%%%%%%%%%%%%%%%%%%%%%%
%%%%%%%%%%%%%%%%%%%%%%%%%%%%%%%%%%%%%%%%%%%%%%%%%%%%%%%%%%%%%%%%%%%%%%%%%%%%%
%%%%%%%%%%%%%%%%%%%%%%%%%%%%%%%%%%%%%%%%%%%%%%%%%%%%%%%%%%%%%%%%%%%%%%%%%%%%%

\subsection{Multiplication of likelihoods}
For the population of $303$ BBHs synthesized in the previous section, we compute the bounds on the different TGR parameters $\delta \hat{\varphi}_i$. Each $i^{\rm th}$ TGR parameter is considered one at a time, along with the system parameters (including the eccentricity $e_0$). The eccentricities for $10^5$ sources are drawn from the {\tt Loguniform} eccentricity distribution as discussed above. This gives us $303$ Gaussian posteriors for each of the $\delta\hat\varphi_i$. These posteriors are free from any eccentricity-induced systematics; i.e., the mean of these posteriors is centered around zero. Since we assume that each $\delta\hat{\varphi}_i$ takes the same value (zero) for all events, we multiply the likelihoods of the individual events to obtain the combined bounds. Since the likelihood function of the $j^{\rm th}$ source is a Gaussian of the form $\propto e^{-\frac{1}{2}(\delta \hat{\varphi}_i-0)^2/(\tilde{\sigma}_i^j)^2}$, the combined statistical error on the parameter $\delta \hat{\varphi}_i$ from $N$ individual sources in the BBH population is 
\begin{equation}\label{combined bound}
        \frac{1}{\sigma_{\delta \hat{\varphi}_i}^{2}} = \sum_{j=1}^{N}\frac{1}{\big(\tilde{\sigma}_{i}^{j}\big)^{2}}\;.
\end{equation}
Note that in the limiting case where all $N$ sources are identical, the combined constraint is expected to improve by a factor of $\sqrt{N}$.   
Figure~\ref{PN_bounds} shows $90\%$ upper bounds on the $\delta\hat{\varphi}_i$ from the $303$ BBHs as measured by a single LIGO or CE detector. The filled markers show the measured bounds on $\delta\hat{\varphi}_i$ when the eccentricity is also considered as a parameter. The unfilled markers represent the bounds on $\delta\hat{\varphi}_i$ for a population of otherwise identical binaries in circular orbits. The blue diamonds show the $90\%$ upper bounds from the GWTC-3~\cite{GWTC3:2021sio} analysis performed by the LVK Collaboration. For LIGO, the bounds on the TGR parameters with $i\geq 0$ are $\mathcal{O}(10^{-2} \mbox{--} 10^{-1})$. For CE, the bounds on the $i\geq 0$ TGR parameters are $\mathcal{O}(10^{-3}\mbox{--}10^{-2})$. The leading-order parameter bounds for CE show a greater improvement (relative to the LIGO bounds) compared to the bounds on the higher-order TGR parameters. For CE, the bounds on the leading-order TGR parameters ($\delta\hat{\varphi}_0$, $\delta\hat{\varphi}_1$, $\delta\hat{\varphi}_2$, $\delta\hat{\varphi}_3$, $\delta\hat{\varphi}_4$, $\delta\hat{\varphi}_{5}^{\rm log}$) are $\sim 8$--$20$ times better than the LIGO bounds for both the eccentric and circular cases. For the higher-order PN deviation parameters ($\delta\hat{\varphi}_6$, $\delta\hat{\varphi}_{6}^{\rm log}$, $\delta\hat{\varphi}_7$), the improvement in the CE bounds (relative to LIGO) is mild (improving by a factor of $\sim 2\mbox{--}4$). The dipole (-1PN, $i=-2$) term is the most tightly constrained. The bound on the dipole term for circular binaries is $|\delta\hat{\varphi}_{-2}|\lesssim10^{-5}$ for LIGO and $|\delta\hat{\varphi}_{-2}|\lesssim 4 \times 10^{-7}$ for CE. The dipole bound in CE improves by a factor of $\sim 30\mbox{--}40$ with respect to LIGO bound. Note that the GWTC-3 analysis considered $\mathcal{O}(35)$ events, while this study obtained combined bounds from $303$ events. 

When the $\delta\hat{\varphi}_i$ are measured along with the eccentricity parameter, the bounds on the leading-order TGR parameters become weaker by a modest factor of $\sim 1\mbox{--}2$ for both LIGO and CE. Inclusion of eccentricity in the parameter space degrades the dipole bound by a factor of $\sim 3\mbox{--}4$. For the higher-order deviation parameters, the eccentric bounds are comparable to their circular counterparts. This is because the higher-PN-order TGR parameters become dominant at higher GW frequencies, while the orbital eccentricity is a low-frequency effect. Hence, the eccentricity affects the bounds on the dipole term and leading-order TGR parameters most but has only a mild impact on the higher-order TGR parameters.

%%%%%%%%%%%%%%%%%%%%%%%%%%%%%%%%%%%%%%%%%%%%%%%%%%%%%%%%%%%%%%%%%%%%%
%%%%%%%%%%%%%%%%%%%%%%%%%%%%%%%%%%%%%%%%%%%%%%%%%%%%%%%%%%%%%%%%%%%%%%
\begin{figure*}
    \centering
    {\includegraphics[width=1.0\textwidth]{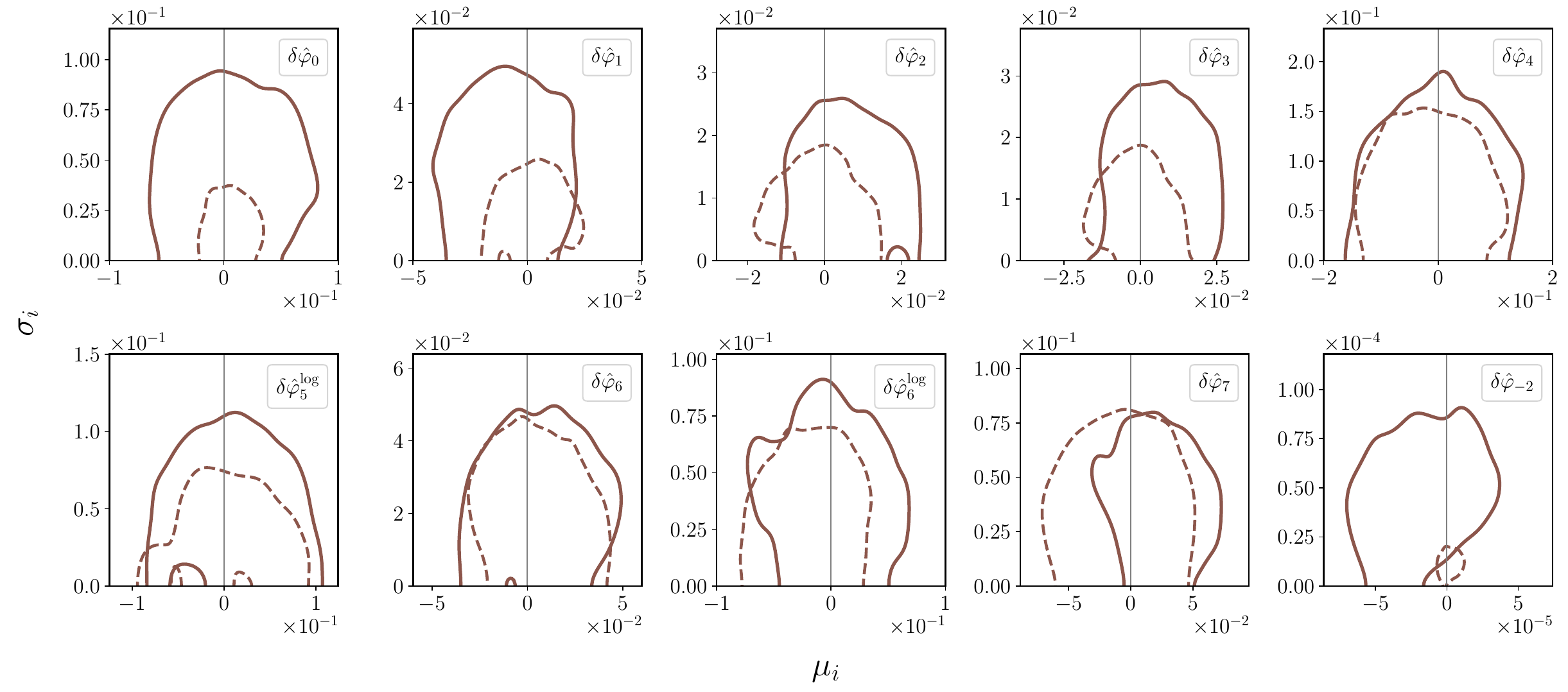}}
\caption{Joint probability distributions for the hyperparameters $\mu_i$ and $\sigma_i$ that characterize the $10$ TGR parameters. The contours represent the $90\%$ credible region set by measurements with a LIGO-like detector operating at design sensitivity. We combine the $303$ BBH events as discussed in Sec.~\ref{population analysis results}. The solid contours consider the case of eccentric binaries; dashed contours represent the circular case.} 
    \label{fig:Hier_bounds_LIGO}
\end{figure*}
%%%%%%%%%%%%%%%%%%%%%%%%%%%%%%%%%%%%%%%%%%%%%%%%%%%%%%%%%%%%%%%%%%%%%%%%%%%%%
%%%%%%%%%%%%%%%%%%%%%%%%%%%%%%%%%%%%%%%%%%%%%%%%%%%%%%%%%%%%%%%%%%%%%%%%%%%%%
\begin{figure*}
    \centering
    {\includegraphics[width=1.0\textwidth]{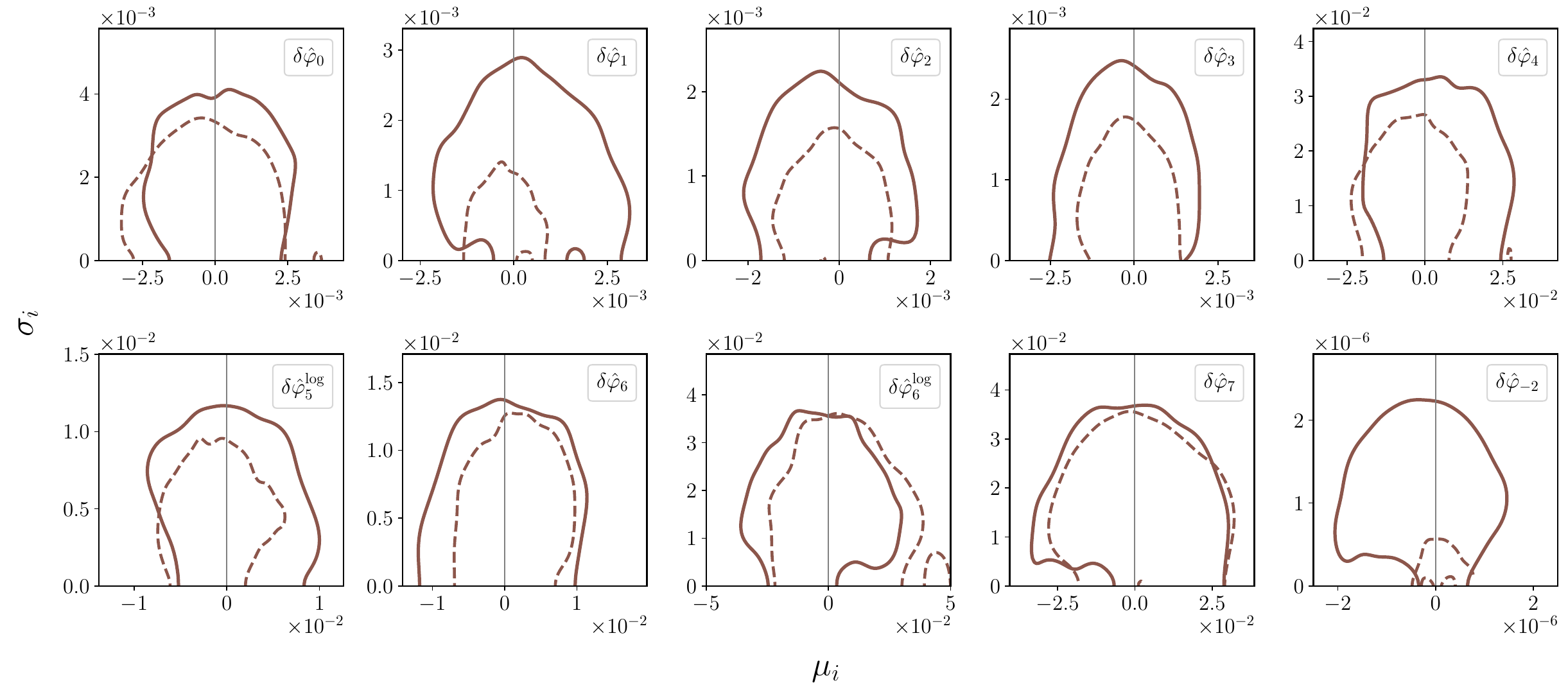}}
\caption{Joint probability distributions for the $\mu_i$ and $\sigma_i$ hyperparameters in the case of a single CE detector. Conventions are the same as Fig.~\ref{fig:Hier_bounds_LIGO}.}
    \label{fig:Hier_bounds_CE}
\end{figure*}
%%%%%%%%%%%%%%%%%%%%%%%%%%%%%%%%%%%%%%%%%%%%%%%%%%%%%%%%%%%%%%%%%%%%%%%%%%%%%
%%%%%%%%%%%%%%%%%%%%%%%%%%%%%%%%%%%%%%%%%%%%%%%%%%%%%%%%%%%%%%%%%%%%%%%%%%%%%

\subsection{Hierarchical inference}
In this section, we obtain constraints on the TGR parameters $\delta\hat{\varphi}_i$ by hierarchically combining the individual events. We consider the same population of $303$ BBHs as discussed in the previous section. The individual posteriors are free of any systematics, and the hyperparameters take the values $\mu_i=0$, $\sigma_i=0$. Hence, the measurement accuracy of the $\mu_i$ and $\sigma_i$ depends only on the statistical uncertainties of the individual $\delta\hat{\varphi}_i$ posteriors. We perform a hierarchical Bayesian analysis on this BBH population as explained in Sec.~\ref{hierarchical inference}. This is done for LIGO and CE, with each of the ten TGR parameters considered one at a time. We consider separately the cases of circular and eccentric orbits (with $e_0$ as a source parameter in the latter). This involves a total of $40$ independent analyses. Note that the $303$ BBHs are a subset of the $10^5$ BBHs drawn from a population analogous to the observed GWTC-3 population and that also pass the SNR and mass criteria discussed above. In the eccentric case we choose the {\tt Loguniform} distribution (described in Sec.~\ref{subsec:ecc}) with $e_0\in (10^{-7}, 0.2)$.

Figure~\ref{fig:Hier_bounds_LIGO} shows the resulting two-dimensional marginalized posterior distributions for $\mu_i$ and $\sigma_i$ in the LIGO case. The contours represent the $90\%$ credible region. Solid lines represent the probability distribution for eccentric binaries, while dashed lines represent the distribution for their circular counterparts. All contours include $\mu_i=0$, $\sigma_i=0$ (except the dipole contour in the eccentric case). The trends in the constraints are similar to the bounds obtained by multiplying the likelihoods: the leading-order $\delta\hat{\varphi}_i$ are better measured than the higher-order $\delta\hat{\varphi}_i$, and the dipole term $\delta\hat{\varphi}_{-2}$ is the best-measured parameter. 

The inclusion of eccentricity in the parameter space degrades the measurement of $\mu_i$ and $\sigma_i$. The joint posteriors of $\mu_i$ and $\sigma_i$ for the $\delta\hat{\varphi}_i$ with $i=(0,1,2,3,5\log)$ are affected more compared to the other $\delta\hat{\varphi}_i$. The $\delta\hat{\varphi}_{-2}$ term is affected the most due to the inclusion of eccentricity in the parameter space. This is because the eccentricity is a low-frequency effect and has a comparatively stronger correlation with the dipole and leading-order PN coefficients. Therefore, the eccentricity affects those $\delta\hat{\varphi}_i$ which are dominant at lower frequencies. The dipole term is the most precisely measured term; the combined bounds are $\mathcal{O}(10^{-4} \mbox{--} 10^{-5})$ for LIGO. Since the peak of $\mu_i$ and $\sigma_i$ posteriors depends on the width of individual posteriors, few sources with large statistical errors on individual $\delta\hat{\varphi}_{-2}$ can lead $\mu_{-2}$, and $\sigma_{-2}$ posteriors peak away from zero. This is what happens for the dipole contours in the eccentric case. If sources with $\sigma_{\delta\hat{\varphi}_{-2}} > 0.005$ are removed from the population, the ($\mu_{-2}-\sigma_{-2}$) contour includes $(0,0)$. We have checked this explicitly.

Figure~\ref{fig:Hier_bounds_CE} shows the corresponding $90\%$ contours for CE. The constraints for CE improve by $\mathcal{O}(10\mbox{--}100)$ compared to LIGO. The improvement for $\delta\hat{\varphi}_{-2}$ is the largest. The inclusion of eccentricity generally affects the $\delta\hat{\varphi}_i$ constraint as in the LIGO case: the leading PN orders $i=(0,1,2,3,4,5\log)$ are affected most, higher PN orders $i=(6,6\log,7)$ are mildly affected, and the dipole term is affected the most.

Note that we model the distributions of the individual $\delta\hat{\varphi}_i$ by a Gaussian distribution characterized by hyperparameters $\mu_i$ and $\sigma_i$. We obtain joint posteriors on the $\mu_i$ and $\sigma_i$ by sampling over them. Some of the $\mu_i$ and $\sigma_i$ can have correlations with the hyperparameters governing the astrophysical population. Simultaneously inferring the astrophysical population when testing GR can provide more stringent constraints on the $\mu_i$ and $\sigma_i$ parameters and avoids any false deviation from GR due to prior assumptions about the astrophysical parameters~\cite{Payne:2023kwj}.

\section{Conclusions}\label{conclusions}
We studied how the presence of eccentricity in a detected population of BBHs affects parametrized tests of GR. Assuming a quasicircular waveform model for an eccentric BBH introduces a systematic bias on the TGR parameters $\delta\hat{\varphi}_i$. These biases on the individual $\delta\hat{\varphi}_i$ posteriors become a more serious issue for a catalog of multiple events. Combining the $\delta\hat{\varphi}_i$ posteriors from multiple events reduces the statistical uncertainty on the population level parameters by $\sim 1/\sqrt{N}$, where $N$ is the number of events. Hence, even small systematic biases (which might otherwise have been within the statistical errors for individual events) can become significant.

We modeled a BBH population analogous to that observed in GWTC-3, while also considering three eccentricity distributions. We used hierarchical Bayesian inference to produce combined constraints from the $303$ simulated BBH events that pass our detection criteria. This assumes that the TGR parameters $\delta\hat{\varphi}_i$ follow a common underlying distribution. This distribution is taken to be a Gaussian characterized by hyperparameters $\mu_i$ and $\sigma_i$ corresponding to the means and standard deviations. 

We show that the systematic biases in the individual event posteriors cause the posteriors for $\mu_i$ and $\sigma_i$ to exclude the GR values (zeros) for the TGR parameters; this exclusion indicates a false GR deviation induced by the neglect of eccentricity in the waveform model. The shift in the $\mu_i$ posteriors depends on the magnitude and sign of the systematic biases---which in turn depends on the eccentricity and other source parameters, as well as correlations between the TGR parameters and source eccentricity. The peak of the $\mu_i$ posteriors is given by the mean of the systematic biases from the individual event posteriors (provided the individual posteriors have similar widths). The peak of the $\sigma_i$ posteriors is roughly proportional to the standard deviation in the individual systematic biases. For those TGR parameters that have systematic biases that are almost evenly distributed about zero, the $\sigma_i$ posteriors provide a better indicator of the bias. In comparison to LIGO, the $\mu_i$ and $\sigma_i$ posteriors for CE typically exclude the GR values at a greater credible level. Hence, systematic biases are an even bigger concern for 3G detectors like CE and ET. 

In addition to considering the eccentricity-induced systematic bias, we also used our simulated BBH population to compute the projected combined bounds on the TGR parameters including eccentricity as a source parameter. We compare these bounds on TGR parameters with the bounds obtained from the same BBH population with circular orbits. We obtained these combined constraints by multiplying the individual likelihoods assuming that each $\delta\hat{\varphi}_i$ takes a common value (zero) across all events. For both LIGO and CE, including eccentricity in the parameter space degrades the combined bounds on the leading-order TGR parameters by a modest factor of $\lesssim 2$; eccentricity has a mild effect on the higher-order TGR parameters. The dipole term has the most tightly constrained bound. The inclusion of eccentricity degrades the dipole bound by a factor of $\sim 3\mbox{--}4$. We also computed the hierarchically combined bound, assuming the TGR parameters can take independent values for each event across the BBH population. The hierarchically combined constraints on the TGR parameters follow a trend similar to the case where the likelihoods are multiplied. 

LIGO (CE) at design sensitivity is expected to detect $220\mbox{--}360$ ($8.6\times10^4 \mbox{--}5.4 \times 10^5$) BBHs per year~\cite{Baibhav:2019gxm}. In LIGO's current observing run (O4), it is plausible that a few of the detected BBHs could have non-negligible eccentricity. Previous TGR analyses used quasicircular waveform models~\cite{GWTC3:2021sio}. If the same analyses are used for O4 or future observing runs, there is a risk of falsely claiming a GR violation when combining information from multiple events. Our study attempts to improve the understanding of eccentricity-induced systematic errors when performing parametrized tests of GR using a population of detected binary black holes.

\section*{Acknowledgments}   
We thank Aaron Zimmerman for a careful reading of the draft and Michael Zevin for providing the data used to model the eccentricity distribution in \cite{Zevin:2021rtf}. K.G.A., P.S., and S.A.B.~acknowledge support from the Department of Science and Technology and the Science and Engineering Research Board (SERB) of India via  Swarnajayanti Fellowship Grant No.~DST/SJF/PSA-01/2017-18 and support from Infosys foundation. K.G.A also acknowledges Core Research Grant No.~CRG/2021/004565 and MATRICS Grant (Mathematical Research Impact Centric Support) No.~MTR/2020/000177 of the SERB. M.F.~was supported by NSF (National Science Foundation) Grant No.~PHY-1653374. The authors are grateful for computational resources provided by the LIGO Laboratory and supported by the National Science Foundation Grants No. PHY-0757058 and No. PHY-0823459. This paper has been assigned the LIGO Preprint No.~P2300370. This study made use of the following software packages: {\tt NumPy} \cite{2020Natur.585..357H}, {\tt Scipy}\cite{2020NatMe..17..261V}, {\tt Matplotlib}~\cite{4160265}, {\tt Seaborn}~\cite{Waskom2021}, {\tt jupyter}~\cite{soton403913}, {\tt pandas}~\cite{mckinney-proc-scipy-2010}, {\tt GWPopulation} \cite{Talbot:2019okv}, and {\tt Dynesty}~\cite{speagle2020dynesty}.

\bibliographystyle{apsrev}
\bibliography{bibliography}%

\end{document}